\documentclass[11pt]{article}
\usepackage{verbatim,amsmath,amssymb}
\usepackage{epsfig,float,color}
\usepackage{geometry}
\usepackage{setspace}
\usepackage{natbib}
\usepackage{amsmath,amssymb,amsthm,mathrsfs,amsfonts,dsfont} 
\usepackage{hyperref}
\usepackage{framed}

\definecolor{darkblue}{rgb}{0,0,1}
\hypersetup{pdftex=true, colorlinks=true, breaklinks=true, linkcolor=darkblue, menucolor=darkblue, pagecolor=darkblue, citecolor=darkblue, urlcolor=darkblue}

\newcommand{\bitm}{\begin{itemize}}
\newcommand{\eitm}{\end{itemize}}
\newcommand{\bnumr}{\begin{enumerate}}
\newcommand{\enumr}{\end{enumerate}}

\newcommand{\mrT}{\mathrm{T}}

\newcommand {\eqb}[1]{\begin{equation}\begin{array}{#1}}
\newcommand {\eqe}{\end{array}\end{equation}}

\newcommand {\esb}[1]{\begin{equation*}\begin{array}{#1}}
\newcommand {\ese}{\end{array}\end{equation*}}
\newcommand {\ds}{\displaystyle}

\newcommand {\pa}[2]{\frac{\partial{#1}}{\partial{#2}}}

\newcommand {\back}{\! \! \!}
\newcommand {\is}{\back &=& \back}
\newcommand {\dis}{\back &:=& \back}

\newcommand {\plus}{\back &+& \back}

\newcommand {\norm}[1]{\|#1\|}
\newcommand {\tr}{\mathrm{tr}\,}
\newcommand {\sign}{\mathrm{sign}\,}

\newcommand {\dif}{\mathrm{d}}

\newcommand {\II}{{I\kern-.3em I}}
\newcommand {\III}{{I\kern-.3em I\kern-.3em I}}

\newcommand {\mrc}{\mathrm{c}}

\newcommand {\mre}{\mathrm{e}}

\newcommand {\mrk}{\mathrm{k}}

\newcommand {\mrm}{\mathrm{m}}
\newcommand {\mrn}{\mathrm{n}}
\newcommand {\mro}{\mathrm{o}}
\newcommand {\mrp}{\mathrm{p}}

\newcommand {\mrs}{\mathrm{s}}

\newcommand {\me}{\mathbf{e}}
\newcommand {\mf}{\mathbf{f}}

\newcommand {\mk}{\mathbf{k}}

\newcommand {\mx}{\mathbf{x}}

\newcommand {\ba}{\boldsymbol{a}}

\newcommand {\bc}{\boldsymbol{c}}
\newcommand {\bd}{\boldsymbol{d}}

\newcommand {\bff}{\boldsymbol{f}}
\newcommand {\bg}{\boldsymbol{g}}

\newcommand {\bm}{\boldsymbol{m}}
\newcommand {\bn}{\boldsymbol{n}}

\newcommand {\bt}{\boldsymbol{t}}

\newcommand {\bx}{\boldsymbol{x}}

\newcommand {\bxi}{\mbox{\boldmath$\xi$}}

\newcommand {\mB}{\mathbf{B}}
\newcommand {\mC}{\mathbf{C}}

\newcommand {\mM}{\mathbf{M}}
\newcommand {\mN}{\mathbf{N}}

\newcommand {\bA}{\boldsymbol{A}}

\newcommand {\bP}{\boldsymbol{P}}

\newcommand {\bT}{\boldsymbol{T}}

\newcommand {\bsig}{\mbox{\boldmath$\sigma$}}

\newcommand {\btau}{\mbox{\boldmath$\tau$}}

\newcommand {\bone}{\mathbf{1}}

\newcommand {\IR}{{\rm\kern.24em
   \vrule width.02em height1.53ex depth-.05ex
   \kern-.3em R}}
\newcommand {\ic}{{\rm\kern.20em
   \vrule width.02em height1.0ex depth-.05ex
   \kern-.22em c}}
\newcommand {\ia}{{\rm\kern.20em
   \vrule width.02em height1.05ex depth-.0ex
   \kern-.25em a}}
\newcommand {\IC}{{\rm\kern.24em
   \vrule width.02em height1.4ex depth-.05ex
   \kern-.26em C}}
\newcommand {\ID}{{\rm\kern.34em
   \vrule width.02em height1.5ex depth-.05ex
   \kern-.36em D}}
\newcommand {\IS}{{\rm\kern.24em
   \vrule width.02em height1.6ex depth.05ex
   \kern-.26em S}}
\newcommand {\IT}{{\rm\kern.50em
   \vrule width.02em height1.55ex depth-.05ex
   \kern-.52em T}}

\newcommand {\IE}{{\rm\kern.24em
   \vrule width.02em height1.55ex depth-.05ex
   \kern-.33em E}}
\newcommand {\IEa}{{\rm\kern.24em
   \vrule width.02em height1.55ex depth-.05ex
   \kern-.33em E}^{1}_{ijkl}}
\newcommand {\IEb}{{\rm\kern.24em
   \vrule width.02em height1.55ex depth-.05ex
   \kern-.33em E}^{2}_{ijkl}}

\newcommand {\sB}{\mathcal{B}}
\newcommand {\sC}{\mathcal{C}}
\newcommand {\sD}{\mathcal{D}}
\newcommand {\sE}{\mathcal{E}}

\newcommand {\sK}{\mathcal{K}}
\newcommand {\sL}{\mathcal{L}}

\newcommand {\sP}{\mathcal{P}}

\newcommand {\Ass}[2]{\kern 0.9ex \vrule width0.45em height0.2ex depth0ex \kern -2.1ex \bigwedge_{#1}^{#2}}
\newcommand {\ASS}[2]{\kern 1.45ex \vrule width0.5em height0.2ex depth0ex \kern -2.65ex \bigwedge_{#1}^{#2}}

\begin{document}

\begin{center}
\Large{\bf{A concise frictional contact formulation based on surface potentials and isogeometric discretization}}\\

\end{center}

\begin{center}
\large{Thang X. Duong\footnote{corresponding author, email: duong@aices.rwth-aachen.de} and Roger A. Sauer}\\
\vspace{4mm}

\small{\textit{Aachen Institute for Advanced Study in Computational Engineering Science (AICES), RWTH Aachen
University, Templergraben 55, 52056 Aachen, Germany}}

\vspace{3mm}
Published\footnote{This pdf is the personal version of an article whose final publication is available at \href{http://dx.doi.org/10.1007/s00466-019-01689-0}{http://link.springer.com/}} 
in \textit{Computational Mechanics},
\href{http://dx.doi.org/10.1007/s00466-019-01689-0}{DOI: 10.1007/s00466-019-01689-0} \\
Submitted on 31 August 2018, Revised on 18 December 2018, Accepted on 20 January 2019
\end{center}


\rule{\linewidth}{.15mm}
{\bf Abstract:}
This work presents a concise theoretical and computational framework for the finite element formulation of frictional contact problems with arbitrarily large deformation and sliding. The aim of this work is to extend the contact theory based on surface potentials \citep{spbc} to account for friction. Coulomb friction under  isothermal conditions is considered here. For a consistent friction formulation, we start with the first and second laws of thermodynamics and derive the governing equations at the contact interface. A so-called \textit{interacting gap} can then be defined as a kinematic variable unifying both sliding/sticking and normal/tangential contact.
A variational principle for the frictional system can then be formulated based on a purely kinematical constraint. The direct elimination approach  applied to the tangential part of this constraint leads to the so-called \textit{moving friction cone} approach of \citet{Wriggers2003}. 
Compared with existing friction formulations, our approach reduces the theoretical and computational complexity. Several numerical examples are presented to demonstrate the accuracy and robustness of the proposed friction formulation.
 

\vspace{-4mm}
\rule{\linewidth}{.15mm}
%


\section{Introduction}\label{s:intro}

The computation of contact problems has made substantial progress due to \textcolor{black}{two recent developments: New} constraint enforcement techniques, such as mortar methods -- among others by \citet{puso04a,yang05,Gitterle10,Apop2012,Kim12,Lorenzis2012,Temizer2013,Hiermeier2018}-- and  isogeometric discretization methods \citep{hughes05} for contact problems \citep{Lu2011, temizer2011,temizer12, Laura2011,Dittmann14,nece,Buffa15,Apop2016,dimitri2017,Duong2018,weeger_2018}. For the latter development, we also refer to the comprehensive review paper of  \citet{Laura2014} and references therein. Mortar methods increase the robustness by weakening the contact constraint enforcement over the contact surfaces. Isogeometric discretization methods can provide smooth contact surfaces, which enhance the robustness of  both Gauss-point-to-segment (GPTS) and novel mortar contact formulations \citep{Laura2014}.  This is because the smoothness of isogeometric surfaces helps avoiding all issues associated with discontinuities (e.g.~kinks) at element boundaries as they appear in classical Lagrange discretization.

Most of the existing contact formulations employ a phenomenological approach that considers the contact problem as a numerical contact constraint, foregoing the underlying complex interactions at  atomistic scales.
 Normal and tangential contact are thus usually treated independently \citep{krstulovic02}.  For tangential contact computation in particular, the algorithms of elastoplasticity are usually considered.  Accordingly, the concepts of associated/non-associated flow and plastic slip criteria have been adopted to friction (see \citet{wriggers-contact} and references therein). A large number of references have applied elastoplasticity algorithms to nonlinear sliding problems, e.g.~\citet{krstulovic02,laursen,spbf,neto16}.  In this paper, we will refer to this approach as the \textit{standard} formulation.

Although the above-mentioned approach is usually appropriate for most engineering (macroscopic scale) problems, it exhibits the following two main drawbacks  \citep{spbc}: First, the independent treatment of  normal and tangential contact may lead to physical inconsistencies. For instance, for normal contact, a slave point interacts with  the closest projection point on the master surface, while for tangential contact, the slave point interacts with the sliding point,  which in case of a penalty regularization is different from the projection point.  Second, the algorithmic treatment for determining the sliding point, which corresponds to the plastic strain, is complicated due to relying on the tangential slip in the parameter space. This means that a finite element implementation for 3D friction needs special attention as the sliding point crosses an element boundary. This issue makes  friction formulations complicated and usually difficult to implement.

In order to alleviate those issues, the so-called \textit{moving friction cone} (MFC) method has been proposed by \citet{Wriggers2003}. The idea of MFC is to use a single gap vector for both normal and tangential contact instead of the two independent ones in the standard formulation. The first issue of inconsistency for normal and tangential contact is thus avoided. Further, to fix the second issue, the sliding point is determined by the condition that the gap vector is orthogonal to the surface normal of the Coulomb friction cone. This approach enables to formulate a contact formulation that is more elegant, easier to implement, as well as facilitates a compact finite element code. The MFC  method has been extended successfully to the three dimensional \textit{node-to-segment method} \citep{Wriggers04}, and  the \textit{GPTS  method} \citep{fischer06}.
 
Apart from the phenomenological approach discussed above, physically-motivated contact interaction models (see e.g.~\citet{argento97,sauer07b,sauer08b}) become desirable at small length scales. 
 \textcolor{black}{An example are coupled adhesion and friction models that are motivated from biological or bio-inspired adhesive systems \citep{adhesionfric}.}
 In this case, physical interactions -- such as van-der-Waals adhesion, electrostatic interactions, cohesive-zone contact, or atomistic interactions -- are dominating so that  macroscopic contact models are no longer suitable.  For an overview of these interactions see e.g.~\citet{shadowitz,raous99,persson,delpiero10,sauer-phd,Temizer2016,kilic2016} and references therein. 

In order to incorporate both phenomenological and physically-motivated approaches, \citet{spbc} provide a unified formulation  based on the concept of surface potentials. According to this formulation, a potential that fully characterizes surface interactions between two bodies is constructed as a function of the gap vector. Depending on the definition of the gap vector, three classes of interactions are identified: point interaction, short-range, and long-range surface interactions.  An advantage of the  formulation of \citet{spbc} is, that the surface potential can be merely numerical, but also allows for physically-motivated interactions such as van-der-Waals adhesion, electrostatic interactions, cohesive-zone contact, and atomistic interactions.  However, the existing framework is restricted to the frictionless case. 

In this contribution, we provide an extension of \textit{the surface potential-based contact  formulation} to friction.  Point interactions  and penalty-based constraint enforcement are particularly considered here. The application to physically-motivated interactions with friction are subject of future work.

Besides, we also aim at providing an advancement of the MFC method by an alternative and concise theoretical framework that has a clear connection with a variational principle and that is consistent with the laws of thermodynamics. For the latter purpose, we will systematically derive the basic equations for the friction problem by starting from the first and the second laws of thermodynamics. We restrict ourselves here to Coulomb friction  although our approach can be extended to other friction laws.

Unlike adopted elastoplasticity algorithms, the present work  formulates the variational principle for friction problems based on a purely kinematical constraint function by defining a new gap vector, called the \textit{interacting gap}.  For the determination of the sliding point, we use the direct elimination of the kinematical constraint function. Therefore, normal and tangential contact are treated in a consistent manner, and the sliding point can be determined by solving a local equation that does not  rely on the tangential slip in the parameter space. This direct elimination approach turns out to be identical to the MFC concept. Our framework here, however, can also recover  adopted elastoplasticity algorithms by expressing the contact potential as an equivalent force constraint instead of a kinematical constraint. 

Furthermore, this work presents the corresponding finite element implementation using the novel isogeometric discretization technique for \textcolor{black}{frictional} contact problems. Additionally, an unbiased friction formulation is also provided here through the \textit{two-half-pass approach} of \citet{spbf}.

Compared to existing friction formulations, this work has the following novelties: \vspace{-0.2cm}
\bitm
\item The extension of \textit{surface potential-based contact} to friction.
\item The advancement of the \textit{moving friction cone} approach to an alternative but concise theoretical framework.
\item The explicit demonstration of the thermodynamic consistency of the  proposed contact formulation.
\item Accurate determination of the tangential traction direction based on smooth isogeometric  surface discretizations. 
\eitm

%
%
%
%
%

The remaining parts of this paper are structured as follows. Sec.~\ref{s:oneD} studies the contact thermodynamics of an elementary friction system. In this section, a variational principle and a direct elimination approach for the determination of the sliding point are also presented. Sec.~\ref{s:theory} extends the concept to general isothermal 3D friction. In Sec.~\ref{s:fem}, the corresponding finite element formulation is presented.  Sec.~\ref{s:examples} provides several numerical examples to assess the proposed formulation in comparison with existing ones in the literature. Sec.~\ref{s:conclude} concludes the paper.


\section{An elementary friction system }\label{s:oneD}
This section presents  the thermodynamics of an elementary friction system. It provides  restrictions on the form of the governing equations and clarifies the basic concepts of the variational principle for frictional contact problems. The latter is used to formulate a computational model for 3D friction in Sec.~\ref{s:theory}.

\begin{figure}[h]
\begin{center} \unitlength1cm
\begin{picture}(0,6)
\put(-5.2,0.0){\includegraphics[width=0.7\textwidth]{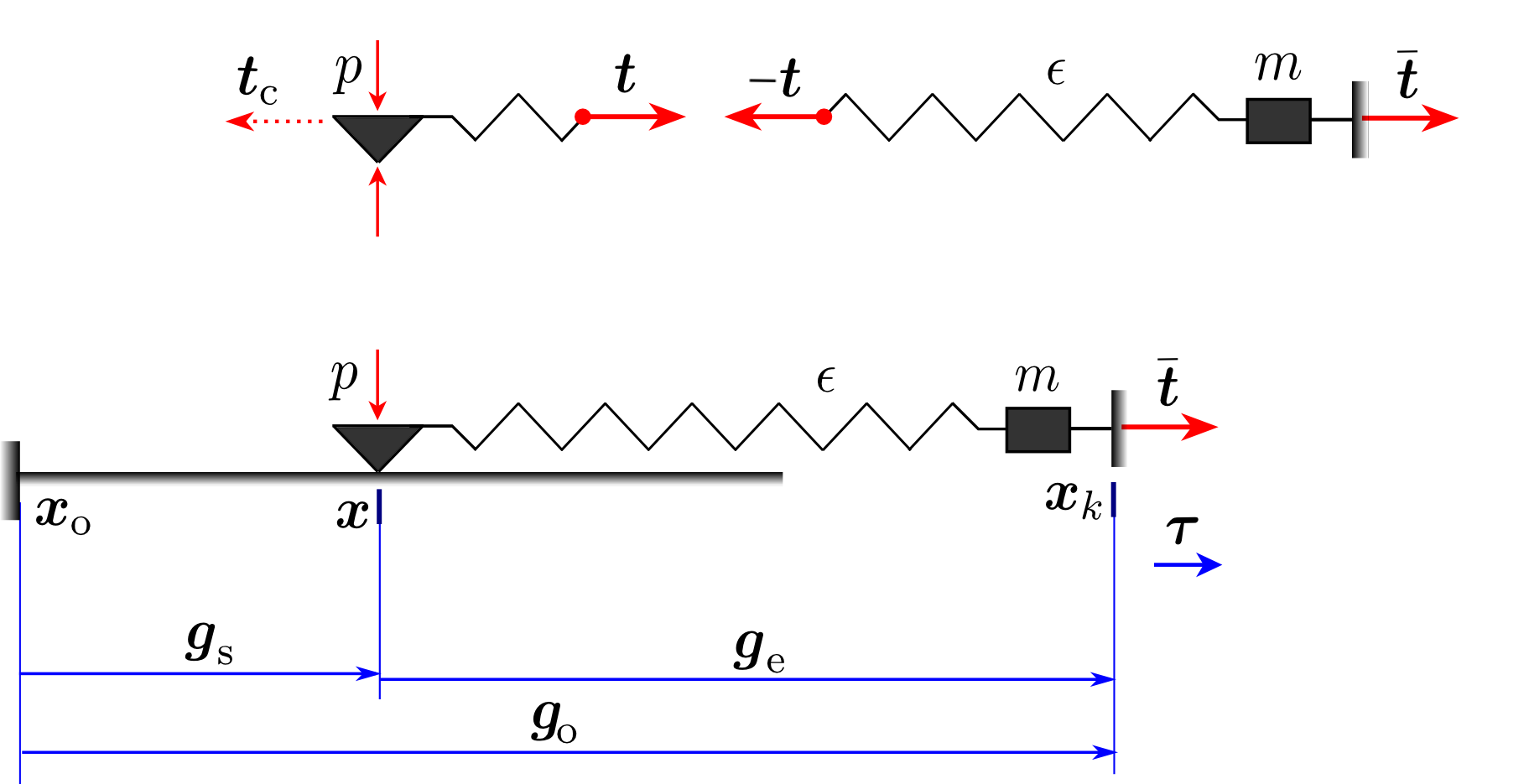}}
\end{picture}
\caption{A rheological model for frictional contact: Under an external force $\bar{\bt}$, sliding takes place from $\bx_\mro$ to $\bx$ (lower figure). The spring,  slider and mass represent potential energy-storing, energy-dissipating, and kinetic energy-storing units. The upper figure depicts the free-body diagram with the contact and spring forces. }
\label{f:friction1D}
\end{center}
\end{figure} 

Consider the conceptual sliding friction model visualized in Fig.~\ref{f:friction1D}. The free energy stored in the system is idealized by the massless spring with stiffness $\epsilon$. The energy dissipated in the form of heat is represented by the (massless) slider unit. Kinetic energy is stored in the mass unit $m$. Fig.~\ref{f:friction1D} also shows the free body diagram where $\bt$ represents the force (per surface area) in the spring and $\bt_\mrc$ denotes the frictional contact force (per surface area) acting on the slider. These forces are induced by the external force, denoted $\bar{\bt}$, which is parallel to
\eqb{l}
\btau:=\ds \frac{\dot{\bg}_\mro}{\norm{\dot{\bg}_\mro}} = \ds \frac{\dot{\bg}_\mrs}{\norm{\dot{\bg}_\mrs}} =  \ds \frac{\dot{\bg}_\mre}{\norm{\dot{\bg}_\mre}} ~.
\label{e:deftau}
\eqe


\textcolor{black}{Here we assume that} the total gap $\bg_\mro:=\bx_k - \bx_\mro$ can be split into the elastic part $\bg_\mre:=\bx_k - \bx$ and the sliding part $\bg_{\mrs}$. Further, in order to uniquely determine how much energy is stored and dissipated for given $\bg_\mro$, the pair $(\bg_\mre,\,\bg_\mrs)$ is chosen here as the state variables of the system.

It should be noted that Fig.~\ref{f:friction1D} is only conceptual. That is, the entire slider-spring-mass system corresponds to a single material point on the contact surface.
Further, the elastic gap $\bg_\mre$ can be understood as a stretch measure of the spring. 



\subsection{Laws of thermodynamics}\label{s:DISS}

The first law of thermodynamics states that the temporal change of the total energy is equal to the external mechanical power and supplied thermal power. That is,
\eqb{l}
\dot{u} + \dot{\sK} =  P_\mathrm{ext} + r~,
\label{e:Firstlaw}
\eqe

where $u$ and $r$ denote the internal energy and the thermal power, respectively, and where 
\eqb{l}
\sK=\frac{1}{2}\,m\,\dot{\bg}_\mro\cdot\dot{\bg}_\mro~,
\label{e:kin0}
\eqe
is the kinetic energy, and
\eqb{l}
P_\mathrm{ext} = \bar{\bt}\cdot\dot{\bg}_\mro~,
\label{e:Pext0}
\eqe
denotes the power supplied by the external force. \textcolor{black}{Note that all quantities discussed in this section refer to a material point on a continuum surface.}

The mechanical power balance can be obtained by taking the scalar product of the  velocity $\dot{\bg}_\mro$ and the force equilibrium of the spring-mass system (see Fig.~\ref{f:friction1D} (upper right side)),
\eqb{l}
\bar{\bt} - \bt - m\,\ddot{\bg}_\mro = \boldsymbol{0}~.
\eqe
By doing so and taking Eqs.~\eqref{e:kin0} and \eqref{e:Pext0} into account, we get
\eqb{l}
P_\mathrm{ext}= \dot{\sK} + \bt\cdot\dot{\bg_\mro}~,
\label{e:Pext}
\eqe
where the product $\bt\cdot\dot{\bg_\mro}$ expresses the internal power of the system. Eq.~\eqref{e:Pext} implies that the external mechanical power $P_\mathrm{ext}$  leads to a change of  kinetic energy and internal power.
Inserting Eq.~\eqref{e:Pext} into Eq.~\eqref{e:Firstlaw} yields
\eqb{l}
 \bt\cdot\dot{\bg}_\mro= \dot{u} - r~,
\label{e:firstlaw2}
\eqe
which eliminates the change of the kinetic energy. Eq.~\eqref{e:firstlaw2} implies that the change of the internal energy minus the thermal power is equal to the internal power of the system. 

The second law of thermodynamics states that the energy dissipation rate $\sD$ (or dissipation in short) is non-negative. That is,
 \eqb{l}
\sD = T\,\dot{s} - r \ge 0~,
\label{e:secondlaw}
\eqe
where $T$ and $s$ denote the absolute temperature and the entropy of the system, respectively. 

\textcolor{black}{In the following, we are restricting ourselves to isothermal systems, i.e.~$\dot{T}=0$, so that the dissipation inequality follows from Eqs.~\eqref{e:firstlaw2} and \eqref{e:secondlaw} as
 \eqb{l}
\sD  =  \bt\cdot\dot{\bg}_\mro - \dot{\psi} \geq 0~,
\label{e:Dissp}
\eqe
where} 
$\psi := u - T\,s$ denotes the Helmholtz free energy.

\subsection{Constitutive equations of the friction system}\label{s:GOV}
This section presents a derivation of the constitutive equations based on the laws of thermodynamic presented in the previous section. In order to make use of restriction~\eqref{e:Dissp},  the free energy and the dissipation must be specified. Here, Coulomb's friction law will be used for demonstration.
Accordingly, we consider 
\eqb{llll}
\psi  \dis \ds\frac{1}{2}\,\epsilon\, \bg_\mre\cdot\bg_\mre~,\\[3mm]
\sD  \dis -\bt_\mrc\cdot\dot{\bg}_\mrs~,  
\label{e:clD}
\eqe
with
\eqb{l}
\norm{\bt_\mrc} \leq \norm{\bt_\mrc^\mathrm{max}}~, \quad \bt_{\mrc}^\mathrm{max} :=-\mu\,p\,\btau~,
\label{e:CLcondition}
\eqe
where $\btau$ is defined by Eq.~\eqref{e:deftau}, $p>0$ denotes the normal contact pressure, and $\epsilon$ and $\mu$ are the model parameters. 

Eqs.~\eqref{e:clD} and \eqref{e:CLcondition}, together with the choice of state variables  $(\bg_\mre,\,\bg_\mrs)$, fully characterize the system. That is, all governing equations can be derived from them. Indeed, inserting Eq.~\eqref{e:clD} into Eq.~\eqref{e:Dissp} gives
\eqb{lll}
 \left(\bt -  \ds\pa{\psi}{\bg_\mre}\right) \cdot\dot{\bg}_\mre + (\bt + \bt_{\mrc})\cdot\dot{\bg}_\mrs= 0~.
\eqe
Since this equation holds for an arbitrary evolution of the state variables, we obtain the governing equations of the frictional system for sliding as
\eqb{lll}
\bt = \ds\pa{\psi}{\bg_\mre} = \epsilon\,\bg_\mre~,\\[3mm]
\bt + \bt_{\mrc}=\boldsymbol{0}~,
\label{e:gveq}
\eqe
where  $\bt_\mrc$ is subject to condition~\eqref{e:CLcondition}. 
Eq.~(\ref{e:gveq}.2) simply represents the equilibrium between the stress in the spring and the stress in the slider unit (see Fig.~\ref{f:friction1D} (upper left)). The two equations of \eqref{e:gveq} are called the constitutive law and the evolution equation, respectively, following the terminology in material modeling. Eq.~(\ref{e:gveq}.2) reproduces the observation that the friction force always resists the external force $\bar{\bt}$.

%
\textbf{Remark~1.} According to Eq.~(\ref{e:clD}.2), zero dissipation corresponds to one of the following two cases:\vspace{-0.2cm}
\bitm
\item Sticking  ($\dot{\bg}_\mrs=\boldsymbol{0}$): in this case $\dot{\bg}_\mro = \dot{\bg}_\mre +\dot{\bg}_\mrs = \dot{\bg}_\mre$.  This means the whole change of the total gap goes into  stretching the spring.
\item Frictionless slip ($\bt_{\mrc}=\boldsymbol{0}$): this happens when either $\mu=0$ or  $p=0$ (interface separation). \textcolor{black}{In this case, all the external power goes into changing the kinetic energy, as is seen from Eq.~\eqref{e:Pext}, since $\bt=\boldsymbol{0}$ and $\bg_\mre = \boldsymbol{0}$  due to Eq.~\eqref{e:gveq}.}
\eitm
\textbf{Remark~2.} Eq.~\eqref{e:gveq} should be satisfied for both sticking and sliding processes. However, these two cases must be distinguished.
In case of sticking, the state of the system is uniquely defined by only one variable, $\bg_\mre$, which becomes directly observable and controllable from the outside. This means that the spring force $\bt$ is prescribed on the system via Eq.~(\ref{e:gveq}.1). It follows from Eq.~(\ref{e:gveq}.2) that the friction force is \textit{driven} by (or determined from) $\bt$ as $\bt_{\mrc}:=-\bt$. In the sliding case, on the other hand,  $\bg_\mre$ is an internal variable and thus cannot be observed and controlled from the outside. But the friction force is observed to be $\bt_\mrc =  \bt_{\mrc}^\mathrm{max}$.  Thus, to satisfy Eq.~(\ref{e:gveq}.2), the friction force must \textit{drive} the spring force as $\bt:=-\bt_{\mrc}^{\mathrm{max}}$.

\textbf{Remark~3.} The presented model also works for the case $\epsilon\rightarrow \infty$, which corresponds to imposing the inextension constraint $\bg_\mre=\boldsymbol{0}$ on the spring. In this case, the spring potential $\psi$ in Eq.~\eqref{e:clD} is simply replaced by $\psi:= \boldsymbol{\lambda}\cdot\bg_\mre$, where $\boldsymbol{\lambda}$ is the Lagrange multiplier for the inextensibility constraint.



We have derived the two governing equations~\eqref{e:gveq} based on thermodynamical restrictions. In this paper, we will treat the evolution equation as a constraint, so that the governing equations can be recast as a minimization principle. This is particularly convenient for a computational formulation. The variational principle will be discussed in the following.



\subsection{Variational principle and a direct elimination approach}\label{s:varprin}

The governing equations~\eqref{e:gveq} can be also recast into a variational principle.  To this end, the constitutive law~(\ref{e:gveq}.1) is seen to be derived from the free energy $\psi(\bg_\mre)$, while the evolution equation during sliding~(\ref{e:gveq}.2) can be treated as the force constraint
\eqb{lll}
\bff_{\!\bt}: =  \bt + \bt_{\mrc}^\mathrm{max} = \boldsymbol{0}~.
\label{e:fconstraint}
\eqe


Thus, the potential for the friction force unifying both sticking and sliding can be written as
\eqb{lll}
W(\bg_\mre,\boldsymbol{\gamma}):=  \psi(\bg_\me) +  \omega\, \boldsymbol{\gamma}\cdot (\bt  +  \bt_{\mrc}^\mathrm{max})~, 
\label{e:potfricionF}
\eqe
where $\boldsymbol{\gamma}$ denotes the Lagrange multiplier to constraint~\eqref{e:fconstraint}, which carries the physical meaning of the rate of the sliding gap (as seen from Eq.~\eqref{e:physgama}),  and $\omega:=H (\norm{\bt}/\norm{\bt_{\mrc}^\mathrm{max}})$ denotes the Heaviside function of the changing stick-slip criterion. \textcolor{black}{The rear term in Eq.~\eqref{e:potfricionF} is much alike the damage evolution in a bulk material model (see e.g.~\cite{Khiem2017}).} Based on potential \eqref{e:potfricionF}, the stationary condition, $\delta W(\bg_\mre,\boldsymbol{\gamma}) = 0$ for all $\delta\bg_e$ and $\delta\boldsymbol{\gamma}$, recovers the governing equations~\eqref{e:gveq}.

\textcolor{black}{The Lagrange multiplier $\boldsymbol{\gamma}$ in Eq.~\eqref{e:potfricionF} can be treated as an additional unknown of the system. Alternatively, a penalty regularization can be used.}  In this paper, we will employ another approach that eliminates constraint~\eqref{e:fconstraint} directly. To this end, we first recast the force constraint~\eqref{e:fconstraint} into the equivalent kinematic constraint,  since relation~(\ref{e:gveq}.1) is assumed to be a unique function of $\bg_\mre$, as
\eqb{lll}
\bff_{\!\bg}: = \bg_\mre  -  \bg_\mre^\mathrm{max} = \boldsymbol{0}~,
\label{e:gconstraint}
\eqe
where $\bg_\mre^\mathrm{max}$ denotes the critical stretch in the spring during sliding. In particular for Coulomb friction, it can be defined by $\bg_\mre^\mathrm{max} :=- \bt_{\mrc}^\mathrm{max}/\epsilon$. 

Given $\bx_k$, we can find the position $\bx=\bx_\mrm$, called the \textit{sliding point}, that satisfies constraint~\eqref{e:gconstraint} during sliding, so that $\bg_\mre$ becomes $\bg_{\mrm}:=\bg_\mre(\bx_\mrm)$.  Potential~\eqref{e:potfricionF} thus can be simply replaced  by
\eqb{l}
W(\bg_\mre) =  \psi(\hat{\bg}) = \ds\frac{1}{2}\,\epsilon\, \hat{\bg}\cdot\hat{\bg}~,
\label{e:Wge}
\eqe
where $\hat{\bg}$ denotes the so-called \textit{interacting gap} defined by
\eqb{l}
\hat{\bg} := (1-\omega)\,\bg_\mre + \omega\,\bg_{\mrm} = \bx_k - \hat{\bx}~,
\label{e:integap1D}
\eqe
 and $\hat{\bx}$ denotes the so-called \textit{interacting point} defined by
\eqb{l}
\hat{\bx} := (1-\omega)\,\bx_\mro + \omega\,\bx_\mrm~.
\eqe
Therefore, the frictional contact problem in turn can be fully determined by three points: $\bx_k$, $\bx_\mro$, and $\bx_\mrm$. While $\bx_k$ and $\bx_\mro$ are given, $\bx_\mrm$ can be found by solving Eq.~\eqref{e:gconstraint}.


This approach will be extended to general 3D friction problems in Sec.~\ref{s:theory}.

\textbf{Remark 4.}  \textcolor{black}{Compared to classical friction formulations based on elastoplasticity algorithms, the rear term of Eq.~\eqref{e:potfricionF} can be identified as the third Kuhn-Tucker condition for the sliding state. Indeed, considering $\bt = \norm{\bt}\,\btau$ (see Fig.~\ref{f:friction1D} (upper left side)) and Eq.~(\ref{e:CLcondition}.2), we can write
\eqb{lll}
 \boldsymbol{\gamma}\cdot (\bt  +  \bt_{\mrc}^\mathrm{max}) =  (\boldsymbol{\gamma}\cdot\btau)\,f_\mrs = \gamma\,f_\mrs ~, 
\eqe
where $f_\mrs := \norm{\bt} - \mu\,p$ denotes the so-called slip function (i.e.~the friction cone), and thus 
\eqb{l}
\boldsymbol{\gamma} = \gamma\,\btau = \dot{\bg}_\mrs
 \label{e:physgama}
 \eqe
 from the argument of maximum dissipation (see e.g.~\citet{Simo1987,wriggers-contact}).}

\section{A computational model for 3D friction }\label{s:theory}
This section presents a computational formulation for general 3D friction problems following the variational principle with direct elimination presented above.


\subsection{Contact surface description}
The contact surface, denoted by $\partial\sB$,  can be described by the one-to-one mapping of a point $\bxi \,\hat{=}\, (\xi^1\,,\xi^2)$ in parameter space $\sP$ to the point $\bx\in\partial\sB$ as
\eqb{l}
\bx=  \bx(\bxi,t)~.
\eqe
A set of tangent vectors on $\partial\sB$ can then be defined by
\eqb{l} 
 \ba_\alpha:=\ds\pa{\bx}{\xi^\alpha},\quad (\alpha=1,2)~,
 \label{e:tangentsl}
\eqe
and the unit normal vector can be defined by
\eqb{l}
\bn := \ds\frac{\ba_1\times\ba_2}{\norm{\ba_1\times\ba_2}}~.
\label{e:normalmsl}
\eqe
With these,  $\partial\sB$ can be characterized by the surface metric, 
\eqb{l} 
a_{\alpha\beta}=\ba_\alpha\cdot\ba_\beta~.
\label{e:a_ab}
\eqe
%
\textcolor{black}{With this,} the dual tangent vectors,  defined by $\ba^\alpha\cdot\ba_\beta = \delta^\alpha_\beta$, are related to the tangent vectors~\eqref{e:tangentsl} by
\eqb{l} 
\ba_{\alpha}=a_{\alpha\beta}\,\ba^\beta~.
\eqe
\textcolor{black}{Here and in the following, summation is implied on repeated indices}. With the basis $\{ \ba_\alpha,~\bn\}$ and its dual $\{ \ba^\alpha,~\bn\}$, the normal and tangential projection tensors are defined by
\eqb{l} 
\bP_{\!\mrn}:=\bn\otimes\bn~,
\eqe
and
\eqb{l} 
\bP_{\!\tau}:=\ba_\alpha\otimes\ba^\alpha~,
\eqe
respectively. \textcolor{black}{Note that $\bP_{\!\mrn} +\bP_{\!\tau}$  is equal to the 3D identity tensor $\bone$.} Further, in order to track changes of $\partial\sB$ during deformation, one chooses a reference configuration denoted $\partial\sB_0$. On $\partial\sB_0$, tangent vectors $\bA_\alpha$ and surface metric $A_{\alpha\beta}$ can be defined like Eqs~\eqref{e:tangentsl} and \eqref{e:a_ab}, respectively. The area change of the contact surface then reads
\eqb{l} 
 J:=\ds\frac{\sqrt{\det[a_{\alpha\beta}]}}{ \sqrt{\det[A_{\alpha\beta}]}} = \ds\frac{\norm{\ba_1\times\ba_2}}{\norm{\bA_1\times\bA_2}}~.
 \label{e:Jss}
\eqe

\subsection{Contact kinematics}
\begin{figure}[htp]
\begin{center} \unitlength1cm
\begin{picture}(0,7)
\put(-6.5,0.0){\includegraphics[width=0.7\textwidth]{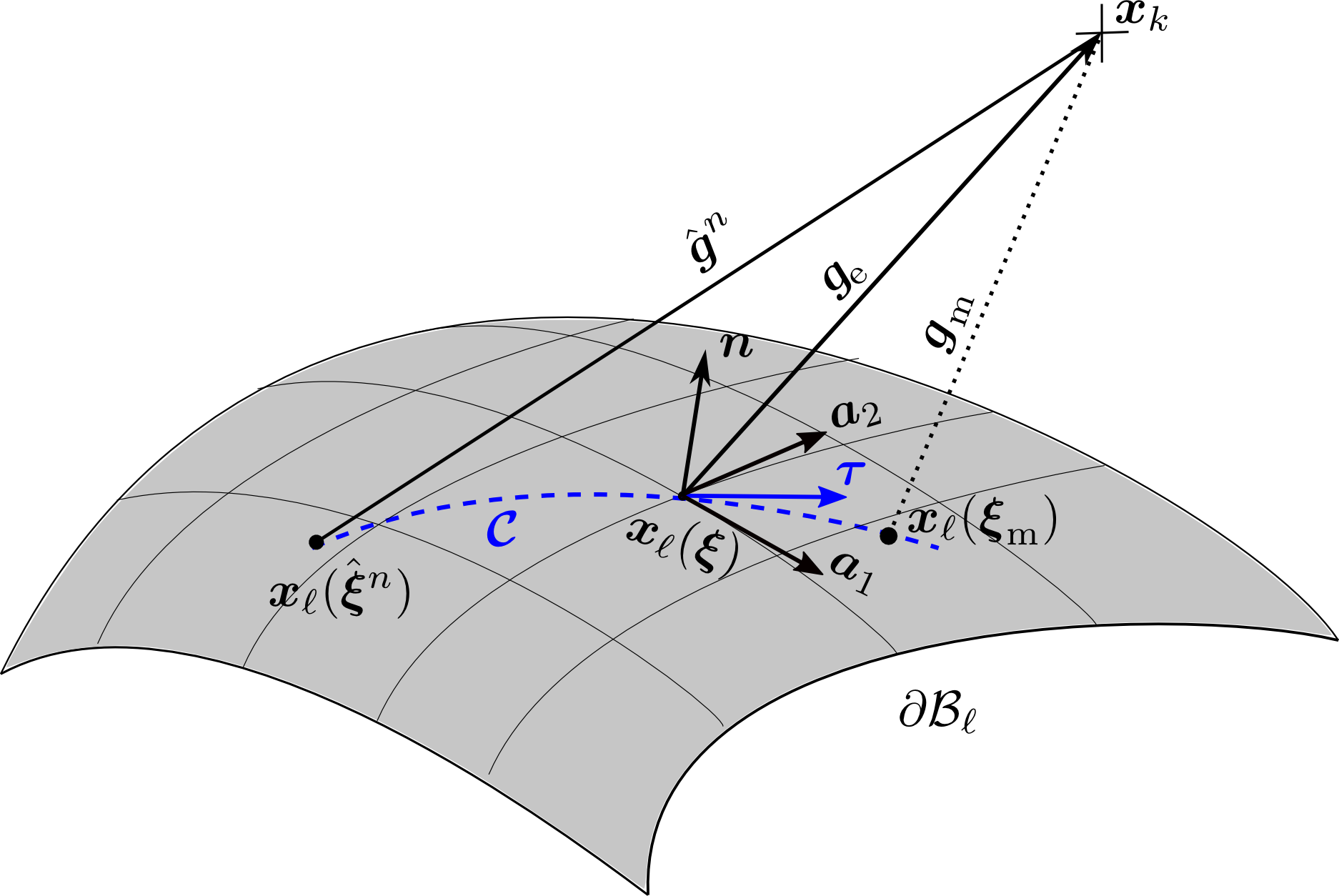}}
\end{picture}
\caption{Frictional contact kinematics:  $\bx_k$  on slave surface $\partial\sB_k$ (not shown) interacts with master surface $\partial\sB_\ell$ along sliding path $\sC$ over the time step $n\rightarrow n+1$. $\bx_\ell(\hat{\bxi}{^n})$, $\bx_\ell(\bxi)$, and $\bx_\ell(\bxi_\mrm)$ on $\sC$ denote the current position of the previous interacting point, current intermediate point, and the current sliding point, respectively.}
\label{f:frictionm}
\end{center}
\end{figure} 
In order to formulate frictional contact between two bodies $\sB_1$ and $\sB_2$,  we consider interactions between a given point $\bx_k\in\partial\sB_k$ called \textit{slave point} ($k=1$ or $2$)  and the neighboring contact surface $\partial\sB_\ell$ ($\ell =2$ or $1$) as shown in Fig.~\ref{f:frictionm}.  Here, one sets $k$ equal to either $1$ or $2$ for the \textit{full-pass} contact algorithm \citep{laursen93}, while $k$ is looped over $1$ and $2$ for the \textit{two-half-pass} algorithm \citep{spbc, spbf}. Further,  point interaction is assumed in this paper. That is, $\bx_k$ can interact with at most one point $\bx_\ell \in \partial\sB_\ell$ at a given time. 
In the following, for the sake of conciseness, all variables without  superscript $n$ are evaluated at the current time $t_{n+1}$ if not stated otherwise.

In order to characterize the interaction, the elastic gap 
vector can be defined as
   (see Fig.~\ref{f:frictionm}) 
\eqb{l}
\bg_\mre(\bxi) :=  \bx_k - \bx_\ell(\bxi) ~,
\label{e:gap}
\eqe
where $\bxi$ is a general point in $\sP$.
Further, the contact gap can be decomposed into tangential and normal contributions as
\eqb{l}
\bg_\mre(\bxi) =  \bg_{\mrn} + \bg_{\tau}~,
\eqe
where 
\eqb{lll}
 \bg_{\mrn}(\bxi) \dis \bP_{\!\mrn} \,\bg_\mre~,\\[2mm]
 \bg_{\tau}(\bxi) \dis \bP_{\!\tau}\,\bg_\mre~.
 \label{e:gtau}
\eqe

%

During sliding, the tangential gap should satisfy the following constraint 
\eqb{l}
\bff_{\!\bg}(\bxi) := \bg_{\tau} - \bg^\mathrm{max}_{\tau} = \boldsymbol{0}~,
\label{e:phiT}
\eqe
where $\bg^\mathrm{max}_{\tau}$ denotes the critical value during sliding, which can be determined by a friction law (see Sec.~\ref{s:coulombfr}). 

In order to obtain a unified expression for both sticking and sliding,  in analogy to Sec.~\ref{s:varprin} we now define the so-called \textit{interacting point} in $\sP$ at time $t_{n+1}$  as
\eqb{lll}
\hat{\bxi} \dis \bxi_\mrp \quad $at initial contact, otherwise$ \\[2mm]
\hat{\bxi} \dis (1 - \omega)\,\hat{\bxi}{^n} + \omega\,  \bxi_\mrm~,
 \label{e:interacxi}
\eqe
where $\bxi_\mrp$ denotes the closest projection point of $\bx_k$, and $\bxi_\mrm:= \{ \bxi \, | \,  \bff_{\!\bg}(\bxi) = \boldsymbol{0} \}$ denotes the so-called \textit{sliding point} that can be found by solving Eq.~\eqref{e:phiT} in the current configuration.  Eq.~(\ref{e:interacxi}.2) implies that the current interacting point $\hat{\bxi}$ is equal to the previous interacting point during sticking ($\omega=0$), and to the sliding point during sliding ($\omega=1$).

With this, the corresponding \textit{interacting gap} at $t_{n+1}$ is defined by 
\eqb{l}
\hat{\bg} := 
 (1-\omega)\,\hat{\bg}{^n} + \omega\,\bg_{\mrm}~,
 \label{e:interacgap}
\eqe
where 
\eqb{lll}
\hat{\bg}{^n} \dis \bx_k - \bx_\ell(\hat{\bxi}{^n})~, \\[2mm]
\bg_{\mrm} \dis \bx_k - \bx_\ell(\bxi_\mrm)~.
\eqe
\textcolor{black}{Here, $\hat{\bg}^n$ denotes the interacting gap vector at time $t_n$ and should not be confused with the normal gap vector $\bg_\mrn$ defined by Eq.~(\ref{e:gtau}.1)}. Eq.~\eqref{e:interacgap} 
implies that during sticking (i.e.~$\omega=0$), the slave point $\bx_k$ interacts with the current position of the previous interacting point $\bx_\ell(\hat{\bxi}{^n})$. On the other hand during sliding (i.e.~$\omega=1$), $\bx_k$ interacts with current sliding point $\bx_\ell(\bxi_\mrm)$.

Further, from Eq.~\eqref{e:interacgap}, the variation of the interacting gap reads
\eqb{l}
\delta\hat{\bg} =  (1-\omega)\,\delta\hat{\bg}{^n} + \omega\,\delta\bg_{\mrm}~,
\label{e:vargapm}
\eqe
where (see e.g.~\citet{wriggers-contact})
\eqb{llll}
\delta\hat{\bg}{^n} \is {\delta\bx_k - \delta\bx_{\ell}}|_\text{\tiny $\bxi = \hat{\bxi}{^n}$}~,
 \\[5mm]
\delta\bg_{\mrm} \is \delta\bx_k - \left.\delta\bx_\ell \right|_\text{\tiny $\bxi=\bxi_\mrm$} - \ba_\alpha \,\delta\xi_\mrm^{\alpha}~.
\eqe





\textbf{Remark~5.} For the 1D case shown in Fig.~\ref{f:friction1D}, Eq.~\eqref{e:interacgap} reduces to Eq.~\eqref{e:integap1D} as $\bx_\ell(\hat{\bxi}{^n})\, \hat{=}\, \bx_\mro$ and $\bx_\ell(\bxi_\mrm)\, \hat{=}\, \bx_\mrm$ since no parametrization has been used for the master surface in this case. 

\subsection{Coulomb friction}\label{s:coulombfr}

For Coulomb friction in particular, $\bg_{\tau}^\mathrm{max}$ in Eq.~\eqref{e:phiT} is given by
\eqb{l}
\bg^\mathrm{max}_{\tau}(\bxi) := \mu\,  \norm{\bg_{\mrn}} \,\btau~, 
\label{e:gtmax}
\eqe
where $\btau$ denotes the unit tangent vector of the sliding direction, which takes the instantaneous direction of the sliding velocity,
\eqb{l}
\btau =  \textcolor{black}{\ds\frac{\sL{\bg}_{\tau}}{\norm{\sL{\bg}_{\tau}}}}~.
\eqe
\textcolor{black}{Here, $\sL{\bg}_{\tau}$ denotes the temporal Lie derivative of ${\bg}_{\tau}$ is equal to the tangential relative velocity between the two bodies.} However, \textcolor{black}{since $\sL{\bg}_{\tau}$ is unknown, for} simplicity, an explicit scheme is usually adopted such that $\btau$ is approximated based on the interacting point at the previous time step.

\begin{figure}[htp]
\begin{center} \unitlength1cm
\begin{picture}(0,4)
\put(-7.9,0.0){\includegraphics[width=1\textwidth]{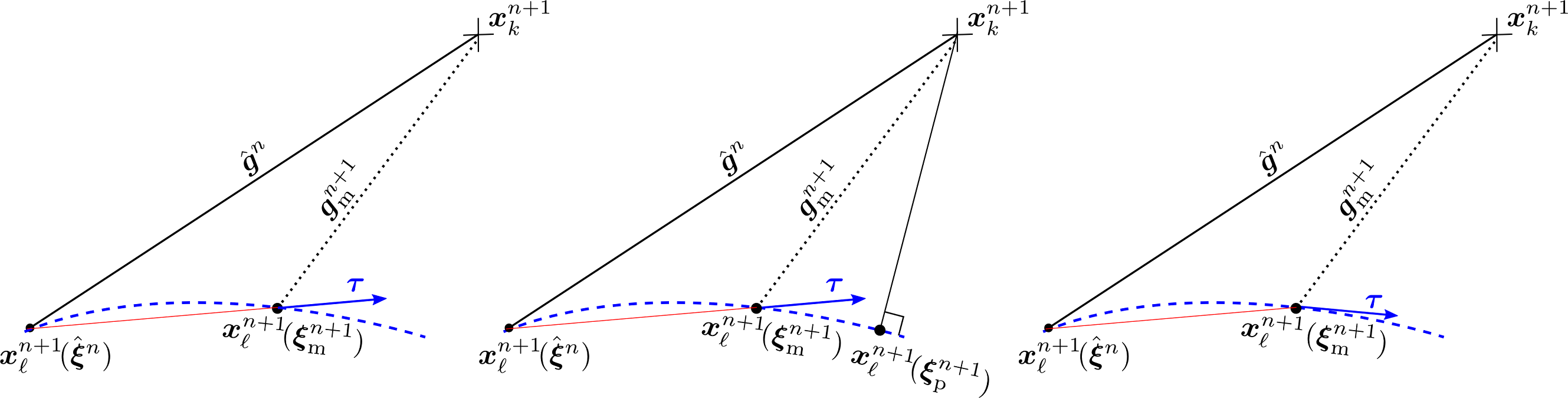}}
\end{picture}
\caption{A comparison of various frictional contact formulations in terms of the tangential traction direction $\btau$ and points involving the computation of frictional forces:  \citet{fischer06} (left), \citet{spbf}(middle), and currently proposed formulation (right). In the left and middle figures, $\btau$ is the secant direction, while in the right figure, it is the tangential direction. In \citet{spbf}, three points on the master surface are used to evaluate contact forces, while only two points are required in \citet{fischer06} and the presented formulation.}
\label{f:kincomp}
\end{center}
\end{figure}

Note that in the context of the predictor-corrector approach, the approximation of  $\btau$  corresponds to the choice for the direction of the trial traction. \textcolor{black}{Fig.~\ref{f:kincomp} (left \& middle)  depicts the choice of the secant direction as it is adopted in the formulations of  \citet{fischer06} and \citet{spbf}.   In this paper here, since we employ a smooth contact surface discretization based on isogeometric analysis, a more  accurate choice for the tangent direction is considered, see Fig.~\ref{f:kincomp} (right). That is,}
\eqb{l}
\btau (\bxi) \approx \ds\frac{ \bP_{\!\tau}\,\hat{\bg}{^n}}{\norm{\bP_{\!\tau}\,\hat{\bg}{^n}}}~,
\label{e:mytau}
\eqe
where $\bP_{\!\tau}$ is evaluated at the current sliding point $\bxi_\mrm$. \textcolor{black}{Eq.~\eqref{e:mytau} implies that $\btau $ results from the projection of the previous interacting gap vector  $\hat{\bg}^n$ onto the tangent plane of the master surface at current sliding point $\bxi_\mrm$ accounting for arbitrary surface deformations.}

%
Furthermore, in order to determine sliding point $\bxi_\mrm$, Eq.~\eqref{e:phiT} is expressed as
\eqb{l}
\bff_{\!\bg} = f_\alpha\,\ba^\alpha= (\bg_\mre\cdot\ba_\alpha)\, \ba^\alpha - (\bg^\mathrm{max}_{\tau}\cdot \ba_\alpha)\, \ba^\alpha= \boldsymbol{0}~.
\eqe
We thus obtain
\eqb{l}
f_\alpha (\xi^1,\xi^2):= (\bg_\mre  - \bg^\mathrm{max}_{\tau})\cdot \ba_\alpha = 0~,
\label{e:phiTa}
\eqe
which is a system of nonlinear equations that can be solved for $\bxi_\mrm$ by a local Newton-Raphson method (see Appendix~\ref{s:localNewton}). As seen, Eq.~\eqref{e:phiTa} unifies both frictionless and frictional cases. When  $\mu=0$, i.e.~$ \bg^\mathrm{max}_{\tau}= \boldsymbol{0}$ due to Eq.~\eqref{e:gtmax}, Eq.~\eqref{e:phiTa} becomes $f_\alpha:= \bg_\mre\cdot\ba_\alpha  = 0$. This implies that  $\bxi_\mrm$ is identical to the closest projection point $\bxi_\mrp$ for the frictionless case.
 \textbf{Remark~6.} For some special contact problems,  Eq.~\eqref{e:phiTa} can be solved analytically for  $\bx_\ell(\bxi_\mrm)$. For instance, for the 2D contact problem of a deformable solid with a rigid plane considered in  example~\ref{s:example1}, the position of the sliding point is
\eqb{l}
x_\mrm = \ds x_\mrk - \mu\,\frac{\sign(\hat{g}^n_\mrn)}{\sign(\hat{g}^n_\tau)}\,\textcolor{black}{\frac{\epsilon_\mrn}{\epsilon_\tau}} \,g_\mrn~,
\label{e:slidepoint2D}
\eqe
where \textcolor{black}{$\epsilon_\mrn$ and $\epsilon_\tau$ denote the penalty parameters in normal and tangential directions, and where}  $\hat{g}^n_\mrn:=\hat{\bg}{^n}\cdot\bn$, $\hat{g}^n_\tau:=\hat{\bg}{^n}\cdot\btau$, and  $g_\mrn:=\bg_\mre\cdot\bn$.

\subsection{A surface potential for frictional contact}

In general, the surface potential for frictional contact is assumed to be a function of the \textit{interacting gap} defined by Eq.~\eqref{e:interacgap}. Here, similar to Eq.~\eqref{e:Wge}, we consider the simple quadratic interaction potential between $\bx_k$ and   $\partial\sB_\ell$,
\eqb{l}
W(\hat{\bg}):=   \ds\frac{1}{2}\,\hat{\bg} \cdot \boldsymbol{\epsilon}\,\hat{\bg}~,
\label{e:WxGen}
\eqe
with the unit \textit{energy per reference area}. In Eq.~\eqref{e:WxGen}, $\boldsymbol{\epsilon}$ is a tensor defined by
\eqb{l}
\boldsymbol{\epsilon} := \epsilon_\mrn\,\bP_{\!\mrn} + \epsilon_\tau\,\bP_{\!\tau}~,
\eqe
where \textcolor{black}{$\epsilon_\mrn(\phi)$ and $\epsilon_\tau(\phi)$ are functions of $\phi:= H(\hat{\bn}\cdot \hat{\bg})$,  with $\hat{\bn}:=\bn(\hat{\bxi})$}. The Heaviside function $H$ is incorporated to account for contact activation/deactivation.\\
In the following, the interaction is considered isotropic for a concise presentation, i.e.~$\epsilon_\mrn=\epsilon_\tau=\epsilon$. In this case, Eq.~\eqref{e:WxGen} reduces to 
\eqb{l}
W(\hat{\bg}):=  \ds\frac{1}{2}\,\epsilon\,\hat{\bg}\cdot\hat{\bg}~.
\label{e:Wx}
\eqe
Given $W$, the contact forces can then be determined in a unified manner for normal/tangential contact and sticking/sliding by including the global contact potential 
\eqb{l}
\Pi_\mrc = \ds\int_{\partial\sB_{0k}}  \ds  W (\hat{\bg}) \,\dif A~,
\label{e:Pic}
\eqe
in the principle of virtual work. Eq~\eqref{e:Pic} can be seen as the surface potential in the framework of \citet{spbc}, but here we have extended it to frictional contact. 

%
By inserting Eq.~\eqref{e:Wx} into Eq.~\eqref{e:Pic}, the variation of $\Pi_\mrc$ reads
\eqb{l}
\delta\Pi_\mrc = \ds\int_{\partial\sB_{0k}} \ds  \bT\cdot\delta \hat{\bg}  \,\dif A~,
\label{e:vPics}
\eqe
where
\eqb{l}
 \bT:=\epsilon\, \hat{\bg}
\eqe
denotes the nominal contact traction. Note that, alternatively, we could also define Eq.~\eqref{e:Wx} per current area. In this case, the resulting contact traction,  denoted as $\bt$, would be the true traction and related to the nominal contact traction by
\eqb{l}
 \bt = J^{-1}\,\bT~.
\eqe

\section{Finite element formulation}\label{s:fem}

This section presents the corresponding finite element formulation of weak form \eqref{e:vPics}. Either isogeometric  analysis \citep{hughes05} or quadratic Hermite interpolation \citep{sauer10b} is employed to obtain smooth contact surfaces. Also, both the full-pass  \citep{laursen93}  and the two-half-pass \citep{spbf} algorithm for frictional contact are discussed. 

\subsection{Finite element discretization}
Contact surfaces $\partial\sB_k$ and $\partial\sB_\ell$ are discretized into \textcolor{black}{$n_\mathrm{sel}$} surface finite elements in total, which are numbered $e = 1,\,...,\, \textcolor{black}{n_\mathrm{sel}}$. $\Gamma^e\subset\partial\sB^h$ denotes the current domain of element $e$. Further, we define $\sE_k$ and $\sE_\ell$ as the sets of element numbers on the slave and master surfaces, respectively. 

The geometry of element $e$ in the current configuration (likewise in the reference configuration) can be interpolated from the positions of the elemental nodes (or control points) $\mx_e$ as
\eqb{l}
 \bx = \mN_e(\bxi)\,\mx_e~,\quad \bxi\in\sP~,
 \label{e:discrex}
\eqe
where \textcolor{black}{$\mN_e:= [N_1\,\bone,~ N_2\,\bone,~..., N_{n_e}\,\bone]$}  denotes the element shape function array, and \textcolor{black}{$n_e$ is the number of nodes in a contact element.} With this, the tangent vectors are
 \eqb{l}
\ba_\alpha = \mN_{e,\alpha}(\bxi) \,\mx_e~.
\label{e:tangenv}
\eqe
The variation of $\bx$ and $\ba_\alpha$, considering $\bxi$ fixed, follows as
 \eqb{lll}
 \delta\bx \is \mN_e\, \delta\mx_e~\\[2mm]
\delta\ba_\alpha \is \mN_{e,\alpha}\, \delta\mx_e~.
\label{e:varxa}
\eqe
In the examples of this paper, the bulk of $\sB$ is discretized by linear elements for efficiency, while for accuracy, the contact surface  is either discretized by non-uniform rational B-Splines (NURBS) interpolation (see e.g.~\citet{hughes05}), using the 3D enrichment approach of \citet{nece,nece2}, or  discretized by quadratic Hermite interpolation, using the 2D enrichment approach of \citet{sauer10b}. 


For NURBS interpolation, the NURBS basis function can be computed in an element-wise manner -- as is usually done in finite element analysis -- by employing the  B{\'e}zier extraction operator $\mC^e$ of \citet{borden11}. The shape function of control point $A$ can then be written as
\eqb{lll}
 N_A(\xi^1,\xi^2) = \ds\frac{w_A\,\hat{N}_A^e(\xi^1,\xi^2)}{\sum_{A=1}^n w_A\,\hat{N}_A^e(\xi^1,\xi^2)},
  \label{e:discrexNURBS}
\eqe
where $w_A$ denotes an associated weight, and $\hat{\mN}^e= \{\hat{N}_A^e\}_{A=1}^{n_e}$ contains the B-spline basis functions. $\hat{\mN}^e$ is computed element-wise in terms of $\mC^e$ and $\mB$, the array of Bernstein polynomials, as
\eqb{lll}
\hat{\mN}^e(\xi^1,\xi^2) = \mC^e_{\xi^1}\,\mB(\xi^1)\,\otimes\,\mC_{\xi^2}^e\,\mB(\xi^2).
\eqe
For quadratic Hermite interpolation in 2D,  the position $\bx$ on the contact surface is interpolated by
\eqb{l}
\bx = \ds \sum_{A=1}^{2} ( N_A \, \bx_A + H_A\,\bx_{A,\xi})~,\quad \xi\in[-1,1]~,
\eqe
instead of Eq.~\eqref{e:discrex}. Here $N_A=N_A(\xi)$ and $H_A=H_A(\xi)$ are the Hermite shape functions for the nodal position $\bx_A$ and the nodal derivative dof $\bx_{A,\xi}$. The tangent vector then follows as
\eqb{l}
\ba = \ds\sum_A^2\left( \pa{N_A}{\xi} \bx_A + \pa{H_A}{\xi} \bx_{A,\xi}\right)~,
\eqe
while the variations are
\eqb{l}
\delta\bx =\ds \sum_{A=1}^{2} ( N_A \, \delta\bx_A + H_A\,\delta\bx_{A,\xi})~,
\eqe
and
\eqb{l}
\delta\ba = \ds\sum_A^2 \left( \pa{N_A}{\xi} \delta\bx_A + \pa{H_A}{\xi} \delta\bx_{A,\xi}\right)~.
\eqe
This surface description is then combined with standard Lagrange interpolation in the bulk following  \citet{sauer10b}.



\subsection{Finite element forces}
Next, the finite element contact forces are derived for the full-pass approach of \citet{laursen93} and  the two-half-pass approach of \citet{spbc, spbf}. 

Applying Eq.~\eqref{e:vargapm} to Eq.~\eqref{e:vPics} and taking Eq.~\eqref{e:varxa} into account, we get the \textit{full-pass} contact formulation as
\eqb{l}
\delta\Pi_\mrc =   \ds\sum_{e\in\sE_k}\,( \delta \mx_e\cdot\mf^e_\mrc +  \delta \mx_{\hat{e}}\cdot\mf_\mrc^{\hat{e}} )~.
\label{e:disvirc}
\eqe
where $\hat{e}\in\sE_\ell$ denotes the master elements  that contain the interacting point $\hat{\bxi}$ emanating from $\bx_k$,  and $\mf^e_\mrc$ and $ \mf^{\hat{e}}_\mrc$ denote the finite element forces acting on slave and master surfaces, respectively.  They are given by
\eqb{l}
\mf^e_\mrc := \ds\int_{\Gamma_0^e} \mN_e^\mrT \,  \bT\, \dif A~, \quad $and$\quad \mf^{\hat{e}}_\mrc:= - \ds\int_{\Gamma_0^e} \mN^\mrT_{\hat{e}}(\hat{\bxi})\, \bT\, \dif A~.
\label{e:deltaPice}
\eqe
Here for simplification, we have neglected the contribution of $\delta\bxi_\mrm$ since $\bg_\mre\cdot\ba_\alpha\approx 0$  for sufficiently large $\epsilon$. But  $\Delta\bxi_\mrm$ should still be taken into account for the tangent matrices. Note that this simplification results in unsymmetrical tangent matrices as seen in Appendix~\ref{s:globalNewton}.

For the \textit{two-half-pass} formulation, we have
\eqb{l}
\delta\Pi_\mrc =   \ds\sum_{e\,\in\,\sE_k \cup \sE_\ell }\, \delta \mx_e\cdot\mf_\mrc^e~,
\label{e:deltaPice2hp}
\eqe
where $\mf^e_\mrc$ is computed by Eq.~(\ref{e:deltaPice}.1).
The linearization of Eq.~\eqref{e:disvirc} and \eqref{e:deltaPice2hp} for the Newton-Raphson method can be found in Appendix~\ref{s:globalNewton}.


\subsection{Implementation}
Tab.~\ref{t:algo} provides an algorithm for the finite element formulation presented above. With this, the implementation of friction can be simply extended from an existing code for frictionless contact, since the only difference is that the closest projection point $\bx_\mrp$ is now replaced by the interacting point $\hat{\bx}=\bx_\ell(\hat{\bxi})$. For the frictionless case, i.e.~$\mu=0$, the interacting point $\hat{\bx}=\bx_\ell(\hat{\bxi})$ is identical to the closest projection point $\bxi_\mrp$.
\begin{table}[!htp]
\begin{framed}
1. Loading loop:\\
$\bullet$  at each quadrature point: 
 If $\hat{\bxi}{^n}$ is not available, set $\mu= 0$.\\ 
$\bullet$ apply load or time step: $n\rightarrow n+1$\\
$\bullet$ provide initial guess for the nodal displacements.\\
$\bullet$ provide initial guess for the current contact surface configurations $\partial\sB_k^{n+1}$ and $\partial\sB_l^{n+1}$ 
\begin{framed}
2. Global Newton-Raphson loop:\\[-1mm]

    ~~~ 2.1.~Loop over the bulk elements and their quadrature points:\\[1.5mm]
    \indent~~~   $\bullet$  Compute and assemble the internal forces and tangent matrices.\\[-1mm]

      ~~~  2.2.~Loop over the slave contact elements and their quadrature points:\\[1.5mm]
       \indent~~~   $\bullet$  Determine current position $\bx^{n+1}_k$ of the quadrature point.\\[1.5mm]
       \indent~~~   $\bullet$  If $\hat{\bxi}{^n}$ is not available, set $\hat{\bxi}{^n}$ equal to the closest proj. point $\bxi^{n+1}_\mrp\in\partial\sB_l^{n+1}$  of $\bx^{n+1}_k$.
        
      \indent~~~   $\bullet$ Evaluate $\ba^{n+1}_\alpha(\hat{\bxi}{^n})$,  $\bn^{n+1}(\hat{\bxi}{^n})$, $\bg_\mre(\hat{\bxi}{^n})$, $\bg_{\tau}(\hat{\bxi}{^n})$,   and $\bg_{\tau}^{\mathrm{max}}(\hat{\bxi}{^n})$  based on\\[1.5mm]
      \indent~~~~~~ Eqs.~\eqref{e:tangenv}, \eqref{e:normalmsl}, \eqref{e:gap}, (\ref{e:gtau}.2), and \eqref{e:gtmax}, respectively.  \\[1.5mm]
     \indent ~~~   $\bullet$ If   $\norm{\bg_{\tau}(\hat{\bxi}{^n}) } < \norm{\bg_{\tau}^{\mathrm{max}}(\hat{\bxi}{^n})}$ then \textbf{sticking} occurs. In this case:\\[1.5mm]
      \indent ~~~~~~ $\circ$ Compute  $\phi = H(\bg_\mre(\hat{\bxi}{^n})\cdot\bn^{n+1}(\hat{\bxi}{^n}))$ from the Heaviside function H. \\[1.5mm]
       \indent ~~~~~~ $\circ$ Set $\omega = 0$. \\[1.5mm]
      \indent ~~~  $\bullet$ If $\norm{\bg_{\tau}(\hat{\bxi}{^n}) } \geq \norm{\bg_{\tau}^{\mathrm{max}}(\hat{\bxi}{^n})}$ then \textbf{either sticking or sliding} occurs. Then:\\[1.5mm]
    \indent ~~~~~~ $\circ$ Compute sliding point $\bxi^{n+1}_\mrm$ by solving Eq.~\eqref{e:phiTa} with a local N-R method.\\[1.5mm]
        \indent ~~~~~~ $\circ$ Evaluate  $\ba^{n+1}_\alpha(\bxi_\mrm^{n+1})$,  $\bn^{n+1}(\bxi_\mrm^{n+1})$,  and $\bg_{\tau}^{\mathrm{max}}(\bxi_\mrm^{n+1})$ based on\\[1,5mm]
      \indent ~~~~~~~~~  Eqs.~\eqref{e:tangenv}, \eqref{e:normalmsl},  and \eqref{e:gtmax}, respectively.  \\[1.5mm]
     \indent ~~~~~~ $\circ$ Compute $\phi = H(\bg(\bxi_\mrm^{n+1})\cdot\bn^{n+1}(\bxi_\mrm^{n+1}))$ from the Heaviside function H.\\[1.5mm]
       \indent ~~~~~~ $\circ$ If   $\norm{\bg_{\tau}(\hat{\bxi}{^n}) } \leq \norm{\bg_{\tau}^{\mathrm{max}}(\bxi_\mrm^{n+1})}$ then $\omega=0$, else $\omega=1$.  \\[1.5mm]   
        \indent~~~   $\bullet$ If $\phi = 1$,  compute $\hat{\bxi}{^{n+1}}$ and  $\hat{\bg}{^{n+1}}$ based on Eqs.~\eqref{e:interacxi} and \eqref{e:interacgap}, respectively.\\[1.5mm]
          \indent~~~~~~ $\circ$ Compute contact forces \eqref{e:deltaPice} and their  tangent matrices \eqref{e:tangentStick} and \eqref{e:tangentSlip}.\\[1.5mm]
      \indent~~~~~~ $\circ$ Assemble contact forces and tangent matrices.\\[1.5mm]
       \indent~~~~~~ $\circ$ Store interacting point $\hat{\bxi}{^{n+1}}$.\\[1.5mm]
         \indent~~~   $\bullet$ If $\phi = 0$,  clear interacting point $\hat{\bxi}{^{n+1}}$.\\

      ~~~  2.3.~Apply boundary conditions. \\[-1mm]
      
      ~~~  2.4.~Solve linear system of equations for the nodal displacements.\\[-1mm]
      
      ~~~  2.5.~Update current configuration and evaluate error norm.\\[-1mm]
      
       ~~~  2.6.~Check for the convergence of the global Newton-Raphson loop.\\[-1mm]
      \end{framed}
\end{framed} 
\caption{The \textit{full-pass algorithm} for the proposed frictional contact formulation. 
For the  \textit{two-half-pass algorithm}, loop 2.2 is employed on both surfaces $(k = 1, 2)$ and  the contact force vector $\mf^e_\mrc$ is evaluated on the two surfaces while force vector $\mf_\mrc^{\hat e}$ is disregarded.}
\label{t:algo}
\end{table}

%
%
%

\section{Numerical examples} \label{s:examples}

This section presents several numerical examples in order to assess the accuracy and robustness of the proposed formulation.  The first is a simple two dimensional block sliding on a rigid plane that is used in order to compare with the existing formulation of \citet{wriggers-contact}. 
Next,  some of the challenging examples presented in \cite{spbf} are reproduced here and compared with the proposed formulation. In the examples, a Neo-Hookean material model (see e.g.~\citet{ogden}) is used with Young's modulus $E = E_0$ and Poisson's ratio $\nu=0.3$.

\subsection {2D sliding on a rigid plane}{\label{s:example1}
The first example examines a rubber block with dimension $L_0\times L_0$ in contact with a rigid plane. The two corners of the block are rounded by the fillet radius $0.1\,L_0$ as is shown in Fig.~\ref{f:block_comp}a in order to avoid singular contact pressures there. The block is first pressed onto the rigid plane with vertical displacement $u_y$ and then moved horizontally by the vertical displacement $u_x$. The prescribed displacement is applied on the upper boundary of the block. For all simulations in this example,  penalty parameters $\epsilon_\mrn= 1000~E_0/L_0$ and  $\epsilon_\tau = 100~E_0/L_0$ are used for normal and tangential contact, respectively. Friction coefficient $\mu$ is considered during both the pressing and sliding phases.   Since the master surface is a straight line in this case, the sliding point $\bx_\mrm$ can be found analytically (see Eq.~\eqref{e:slidepoint2D}) and the formulation simplifies significantly. 


\begin{figure}[htp]
\begin{center} \unitlength1cm
\begin{picture}(0,12)

\put(-8.0,0){\includegraphics[width=1\textwidth]{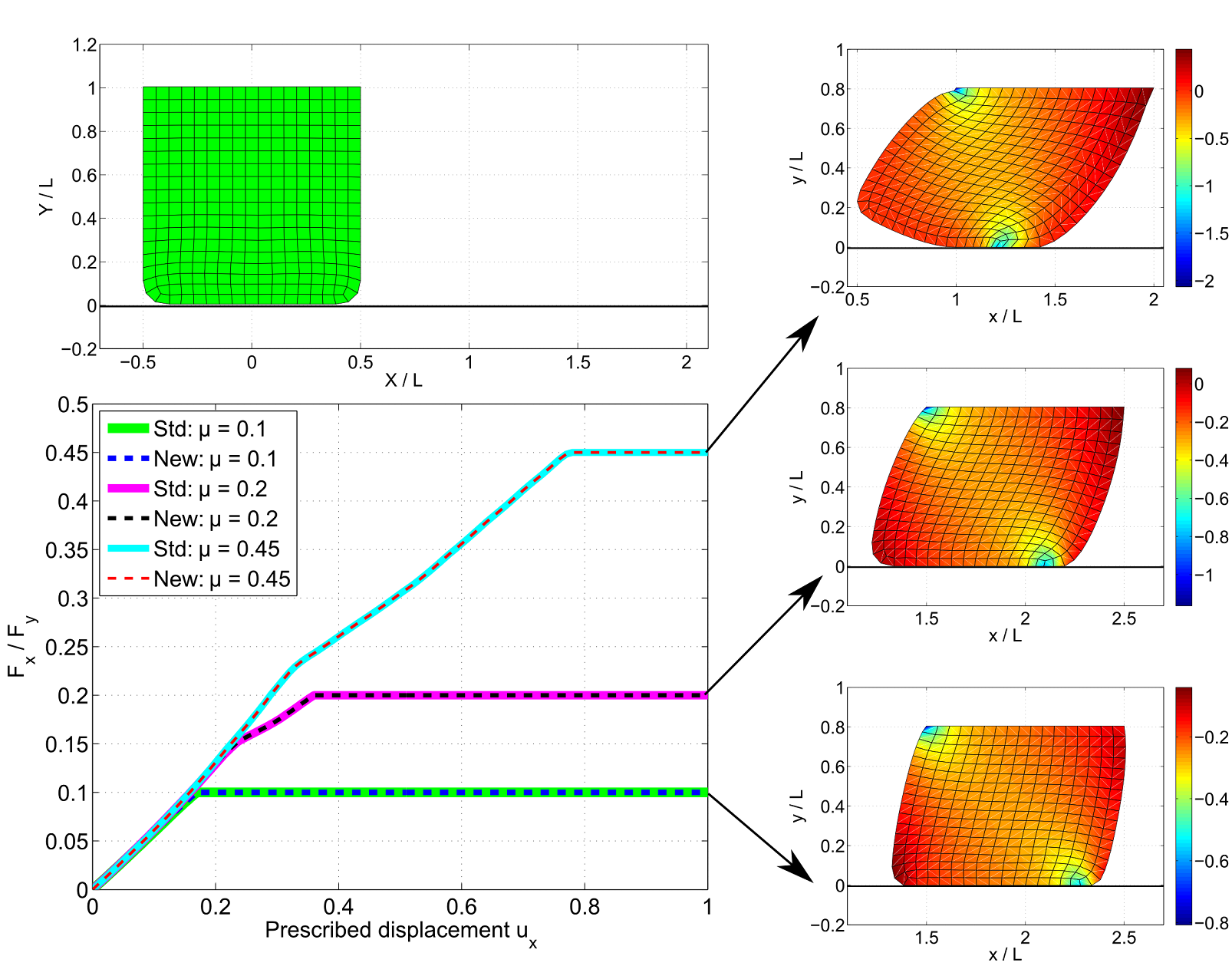}}

\put(-7.6, 7.8){a. }
\put(-7.6,0.5){b. }
\put(2.1,8.5){c. }
\put(2.1,5.1){d. }
\put(2.1,0.8){e. }

\end{picture}
\caption{2D sliding on a rigid plane: a.~Initial configuration discretized by $304$ linear elements. b.~Comparison of the ratio of vertical to horizontal reaction forces  for the standard \citep{wriggers-contact} and the proposed formulation. c-e.~Deformed configurations colored by the stress invariant $I_1=\tr\bsig$ for various $\mu$. \textcolor{black}{Here, the load step size $0.01\,L_0$ is used}. See also the supplementary movie at {\url{https://doi.org/10.5446/37885}}
   for the case $\mu = 0.45$.}
\label{f:block_comp}
\end{center}
\end{figure} 

\begin{figure}[!htp]
\begin{center} \unitlength1cm
\unitlength1cm
\begin{picture}(0,6.2)
\put(-8.0,0){\includegraphics[width=0.49\textwidth]{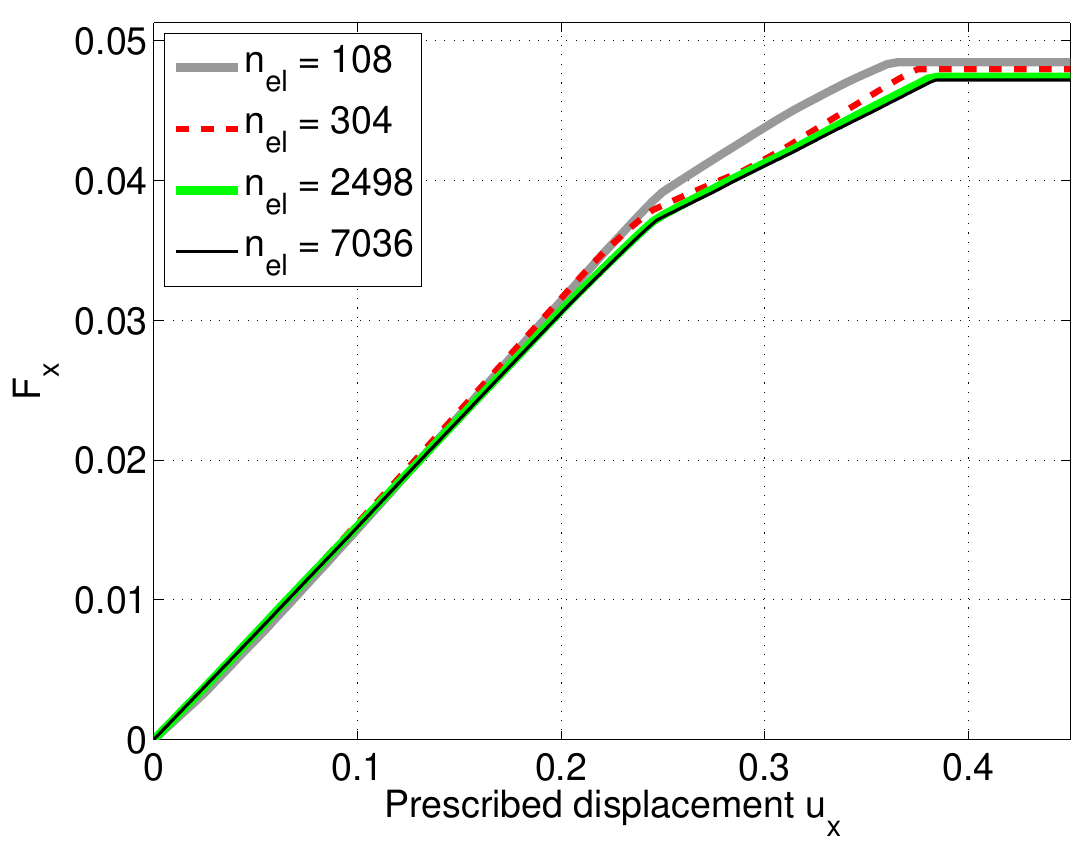}}
\put(0.2,0){\includegraphics[width=0.49\textwidth]{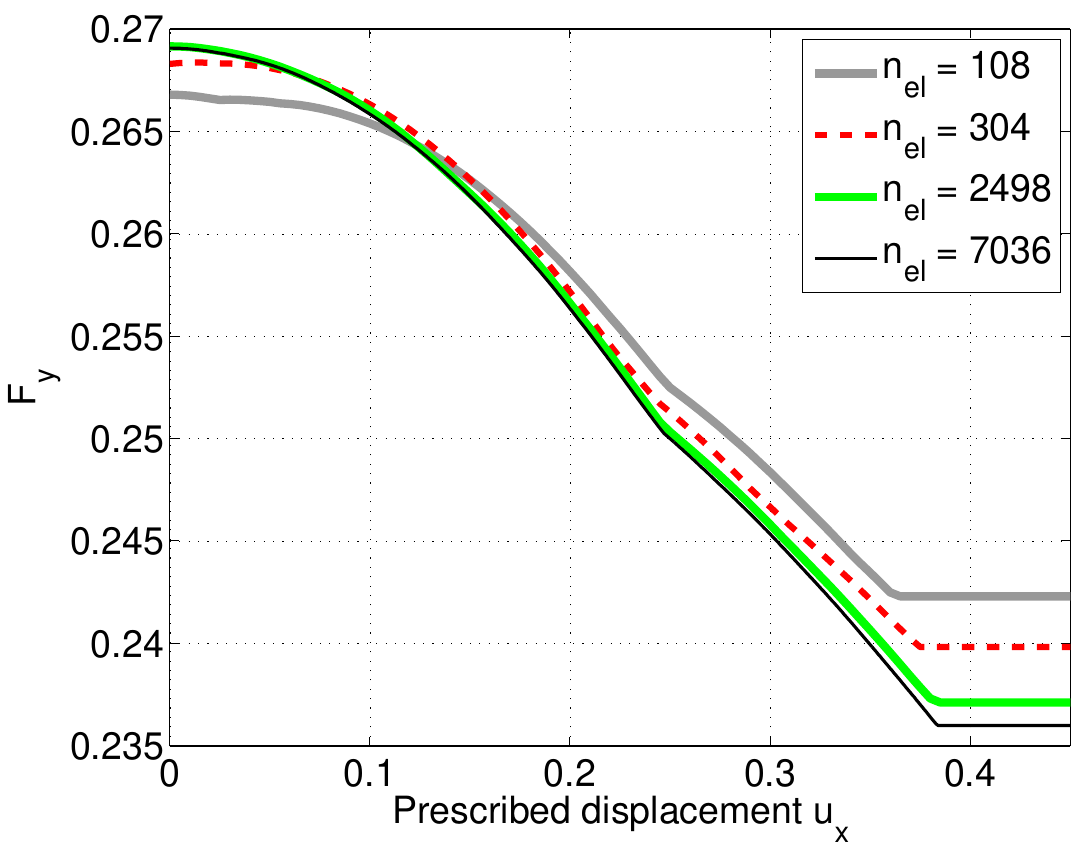}}

\end{picture}
\caption{2D sliding on a rigid plane: horizontal and vertical reaction forces of the proposed formulation considering $\mu=0.2$ and various element numbers $n_\mathrm{el}$. \textcolor{black}{Here, the load step size is $0.001\,L_0$ for the finest mesh, and $0.005\,L_0$ for the other meshes.}}
\label{f:iron2d_conv}
\end{center}
\end{figure}

To verify our formulation, the simulation results are compared in Figs.~\ref{f:block_comp}b-e with the Gauss-point-to-segment formulation of \cite{wriggers-contact} considering various friction coefficients. As expected, the simulation results of both formulations are identical since sliding direction $\btau$, in case of planar contact, is identical in the two formulations. A mesh convergence study of the proposed formulation is shown in Fig.~\ref{f:iron2d_conv}.


\subsection {Contact between two half-cylinders }
The second example considers frictional contact between two half-cylinders  with radius $L_0$ as shown in Fig.~\ref{f:symmetry}a. The example is used to verify the two-half pass version of the proposed formulation. The present simulation results are compared with those of \citet{spbf}. 
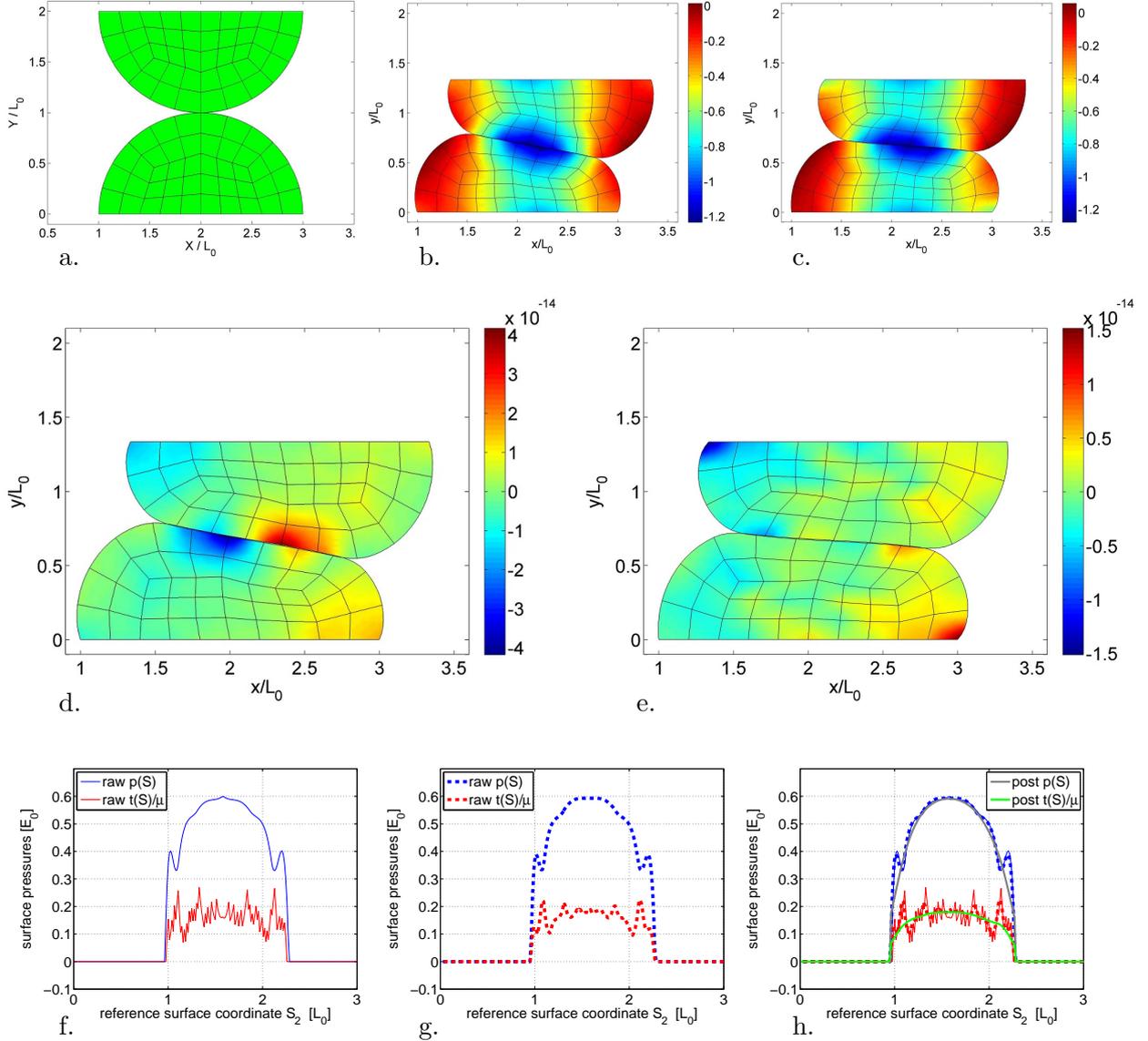
\begin{figure}[!h]
\begin{center} \unitlength1cm
\begin{picture}(0,15.0)

\put(-8.1,11.1){\includegraphics[width=57mm]{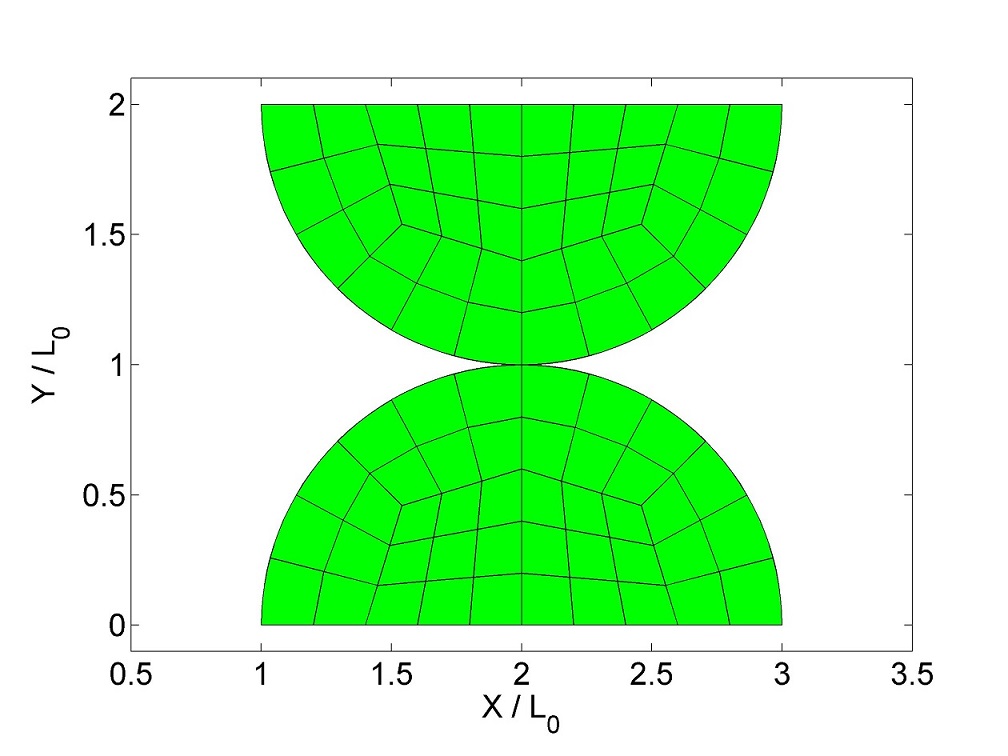}}
\put(-2.9,11.1){\includegraphics[width=57mm]{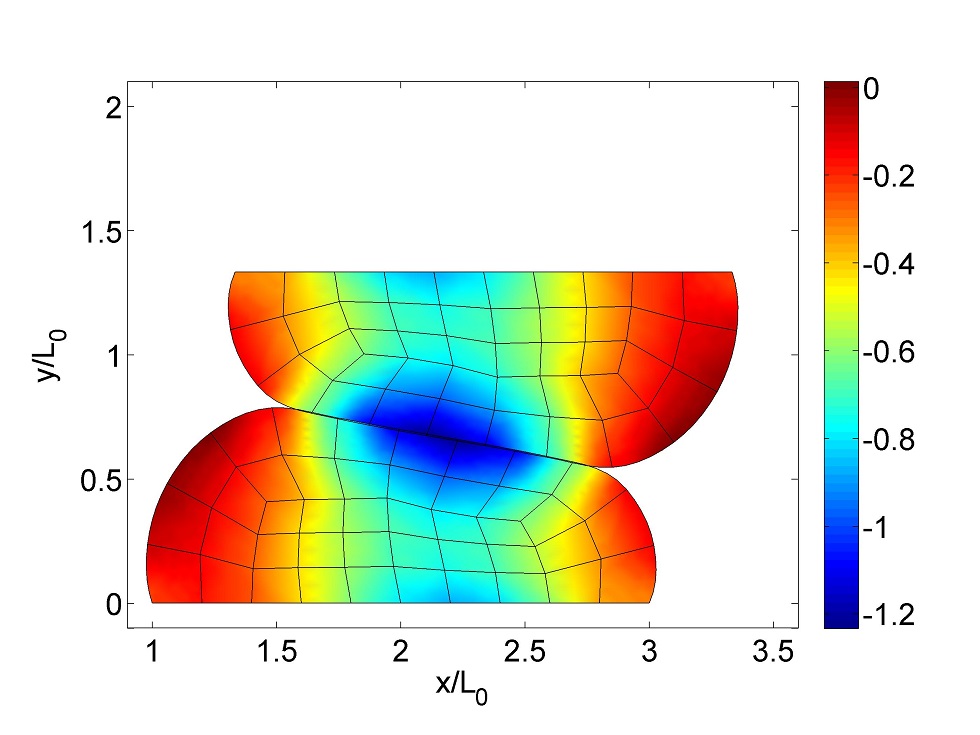}}
\put(2.5,11.1){\includegraphics[width=57mm]{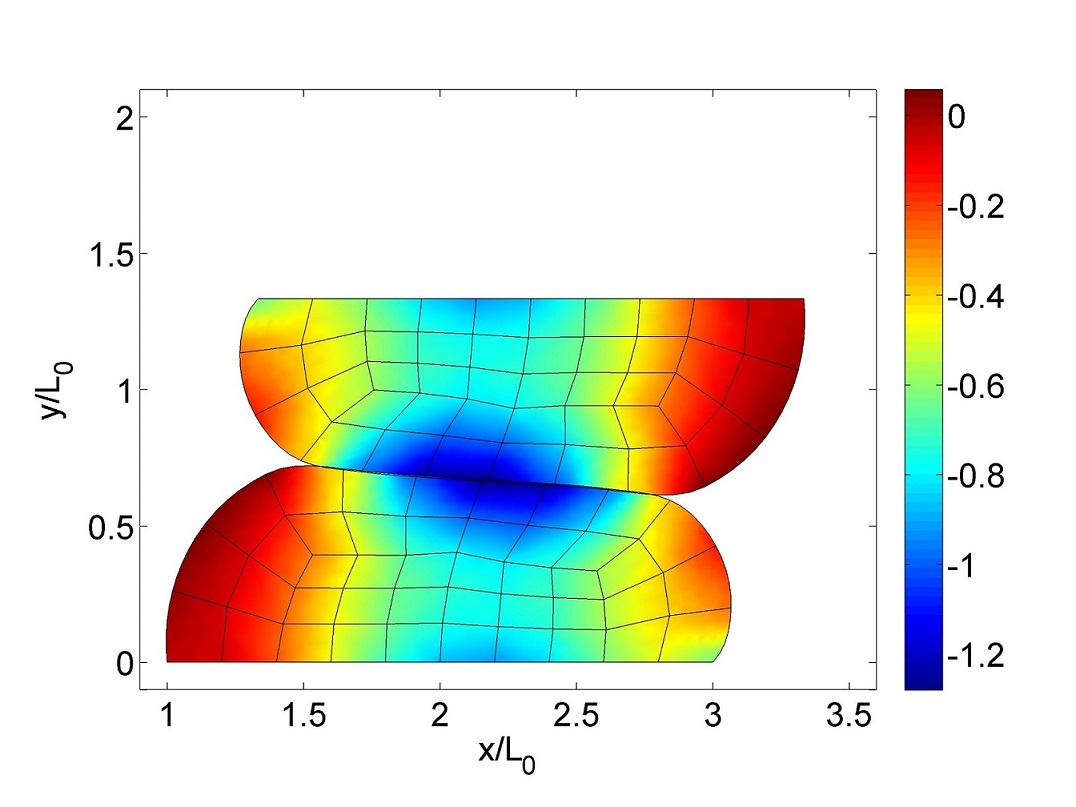}}

\put(-8.2,4.5){\includegraphics[width=85mm]{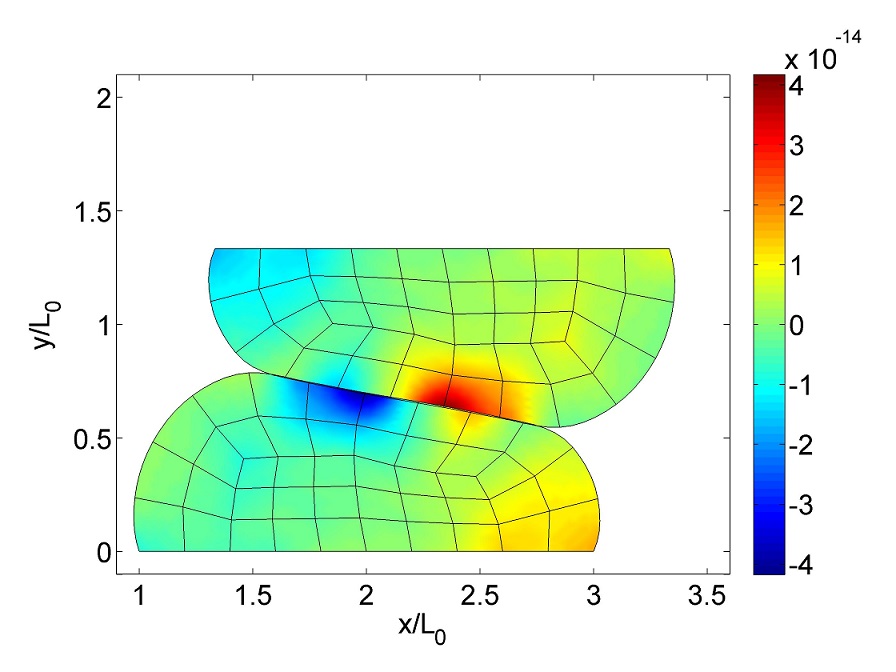}}
\put( 0.15,4.5){\includegraphics[width=85mm]{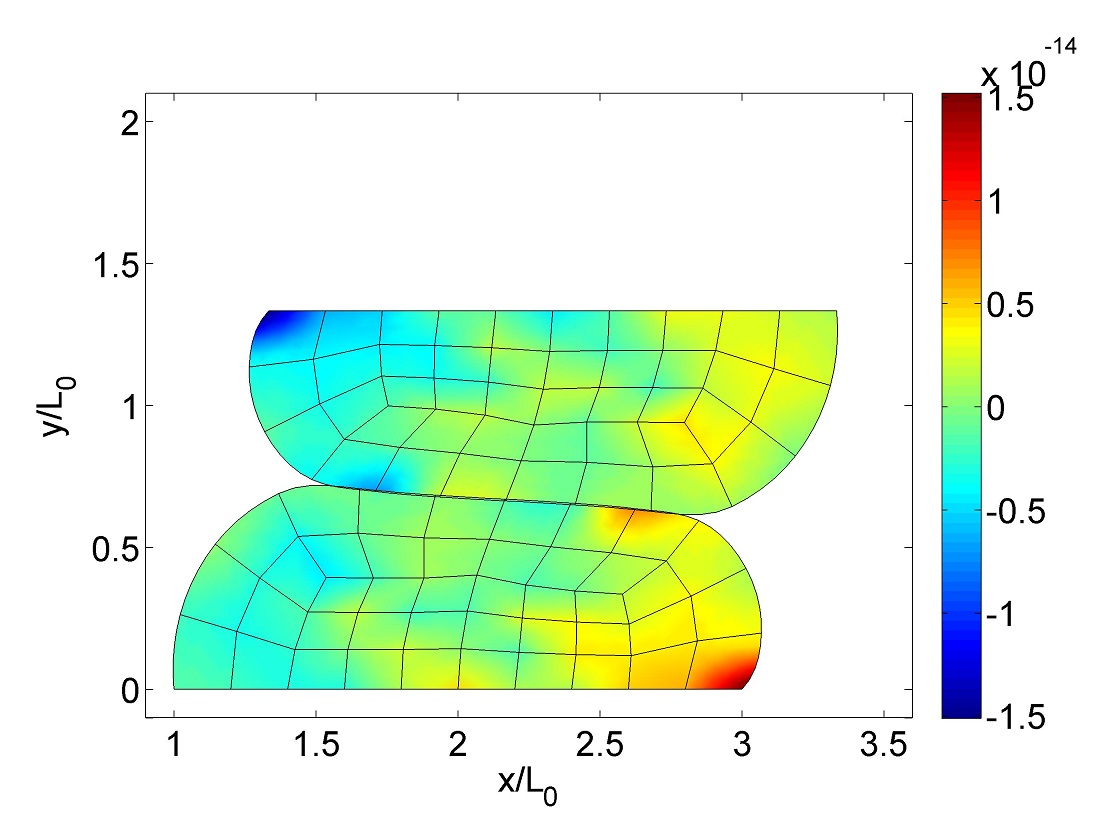}}

\put(-7.8,0){\includegraphics[width=50mm]{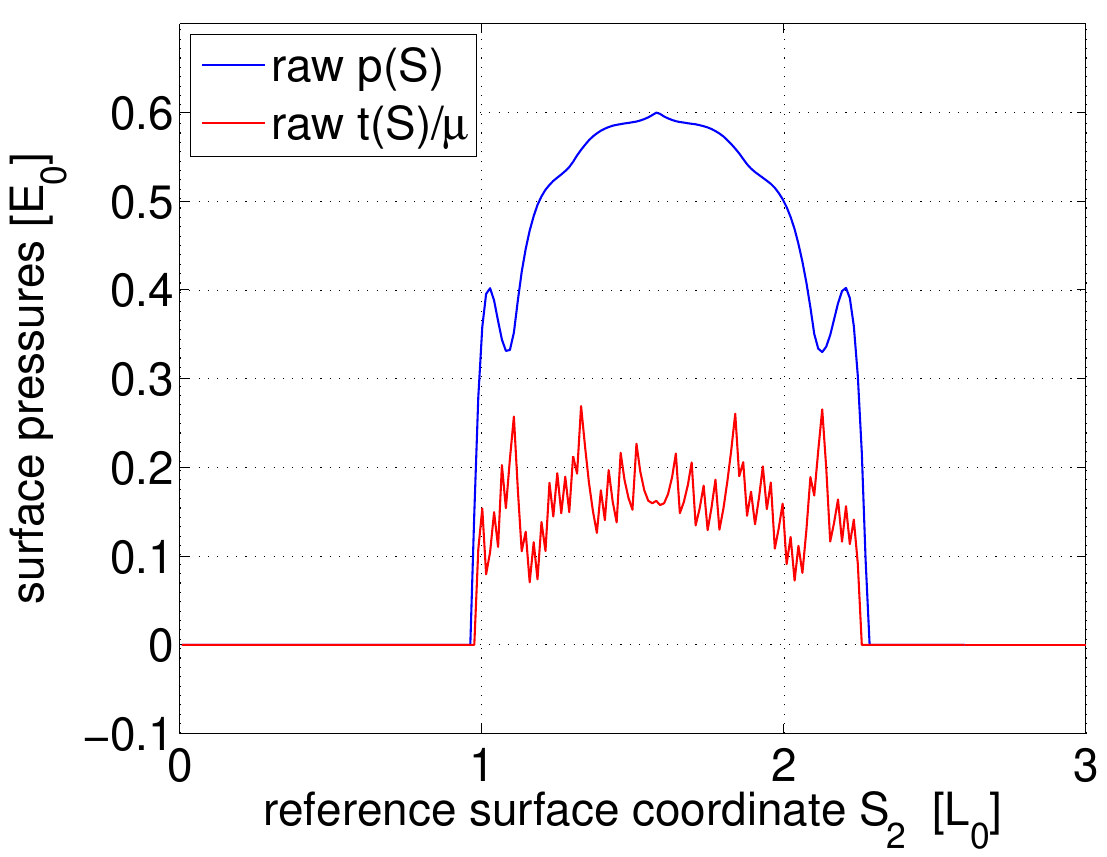}}
\put(-2.5,0){\includegraphics[width=50mm]{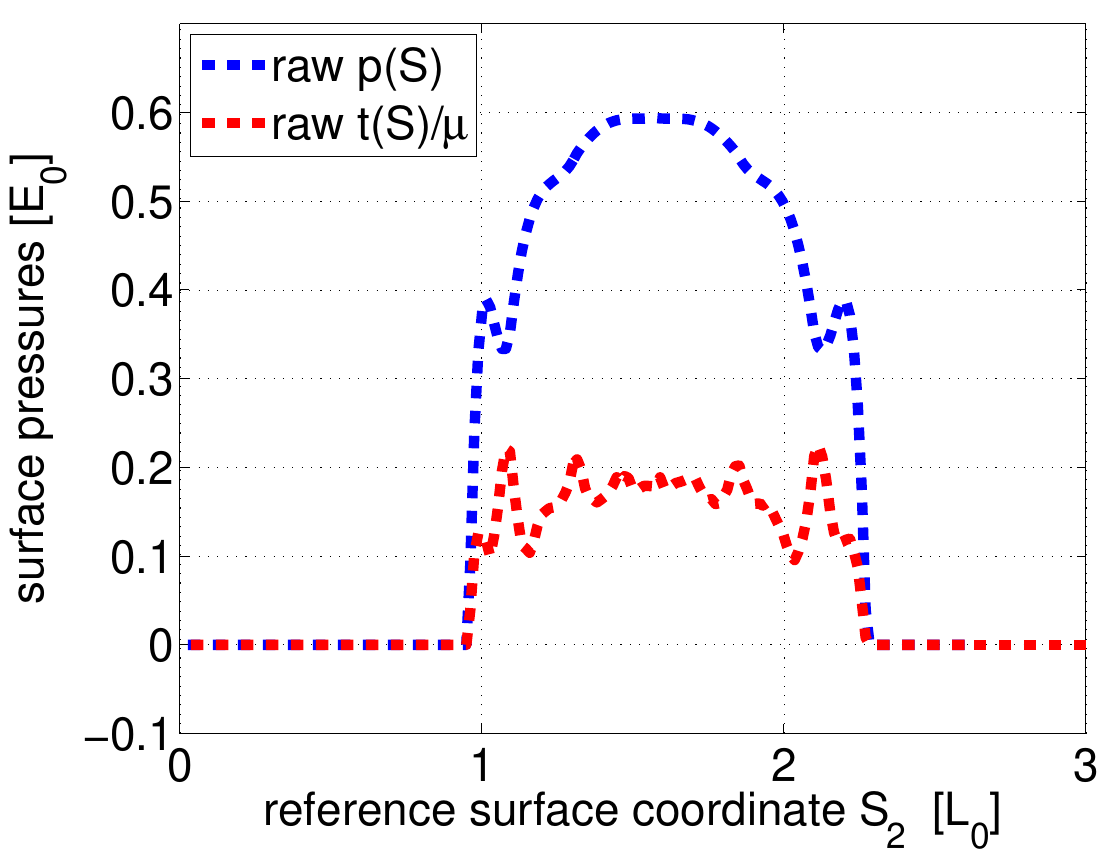}}
\put(2.7,0){\includegraphics[width=50mm]{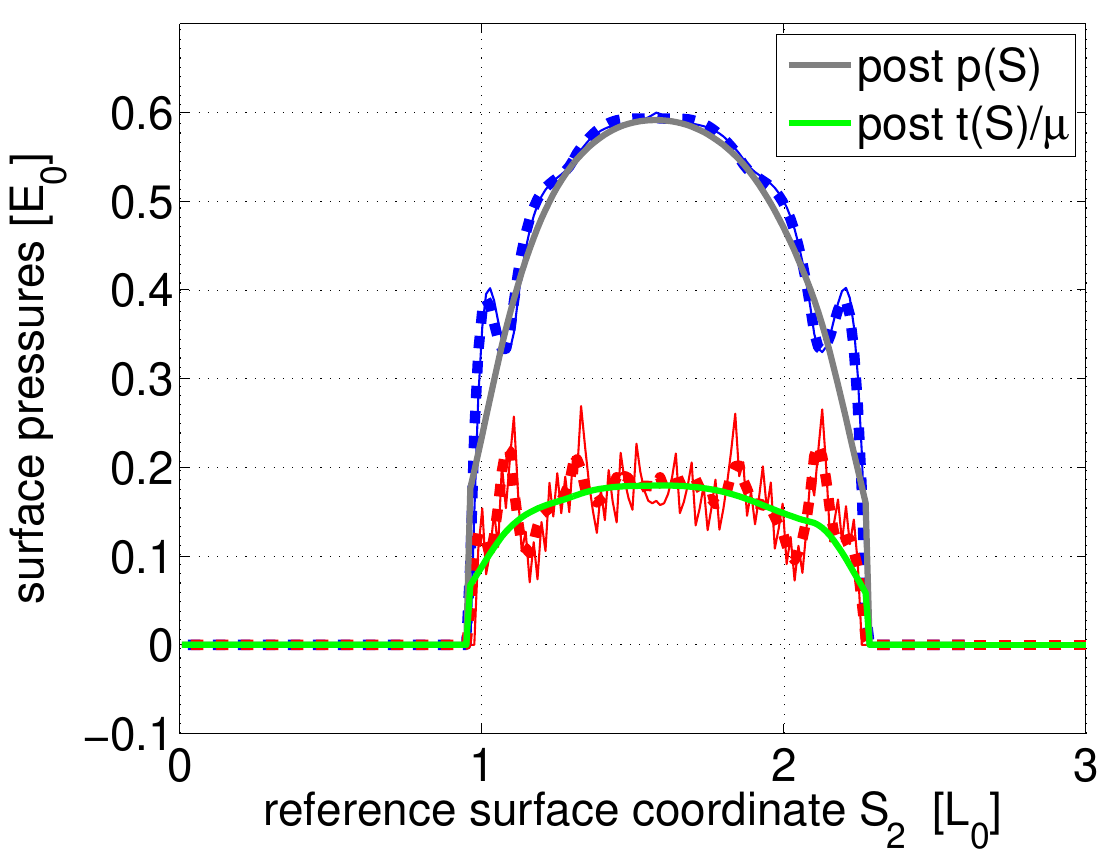}}

\put(-7.2,11.1){a. }
\put(-2.0,11.1){b. }
\put(3.4,11.1){c. }

\put(-7.2,4.6){d. }
\put(1.15,4.6){e. }

\put(-7.2,-0.1){f. }
\put(-2.0,-0.1){g. }
\put(3.4,-0.1){h. }
\end{picture}
\caption{Contact between two half-cylinders: The top row shows the undeformed configuration (a.), and the deformations for b.~$\mu=0.1$ and c.~$\mu=0.6$. The color shows the stress invariant $I_1=\tr\bsig$. The middle row shows the inaccuracy in $I_1$ for d.~$\mu=0.1$ and e.~$\mu=0.6$  considering the two-half-pass algorithm with $n_{u}=20$ \textcolor{black}{load steps}.  The bottom row shows the normal and tangential contact tractions for  $\mu=0.6$ with f.~$n_{u}=20$, g.~$n_{u}=200$, and h.~combining figures f.~and g. using the post-processing scheme of \cite{sauer-ece2}.    Here, $\epsilon = 100 E_0/L$ and $20$ quadrature points are used per contact element. The present simulation results are almost identical to the results reported in \citet{spbf}. }
\label{f:symmetry}
\end{center}
\end{figure}


The two half-cylinders are brought into contact by considering the vertical and horizontal displacements $u_y = 2/3\,L_0$  and  $u_x = 1/3\,L_0$ applied to the top boundary of the upper body. The material parameters, penalty parameter, and discretization are the same as used in \citet{spbf}. That is, $E=E_0$,  $\nu=0.3$, $\epsilon=100\,E_0/L_0$ are used. The half-cylinders are discretized by 4-noded linear finite elements in the bulk, while the contact elements are enriched by quadratic Hermite interpolation on the surface  \citep{sauer10b}.

Figs.~\ref{f:symmetry}b-h show the simulation results computed with the two-half-pass version of the proposed formulation. The results show the deformed configurations (Figs.~\ref{f:symmetry}b-c),  errors in the stress invariant $I_1=\tr\bsig$ (Figs.~\ref{f:symmetry}d-e), and the distribution of the contact tractions  (Figs.~\ref{f:symmetry}f-g). The simulations  consider a low friction coefficient  $\mu=0.1$ versus a high one $\mu=0.6$, as well as a small number of load steps ($n_{{u}}=20$) versus a large one ($n_{{u}}=200$). 

Compared to \citet{spbf}, the present formulation yields the relative difference in the net tangential contact force\footnote{computed by averaging the tangential contact traction over the reference surface.} of $1.5\%$ ($n_{{u}}=20$) and $0.3\%$  ($n_{{u}}=200$) (see Figs.~\ref{f:symmetry}f-g). The computational efficiency of the present formulation is  improved by $1.1\%$ in the contact element routine due to the less complex implementation. Further, from Figs.~\ref{f:symmetry}f-g, the proposed formulation is shown to be less sensitive to the load step size compared to the formulation of \citet{spbf}. This reflects the fact that the sliding direction $\btau$ is chosen here more accurately as shown in Sec.~\ref{s:coulombfr}.

%
%
%
%
%
%
%

%
%
%
%
%

\subsection {2D ironing }
\begin{figure}[!h]
\begin{center} \unitlength1cm
\unitlength1cm
\begin{picture}(0,5.5)

\put(-7.0,0){\includegraphics[width=0.80\textwidth]{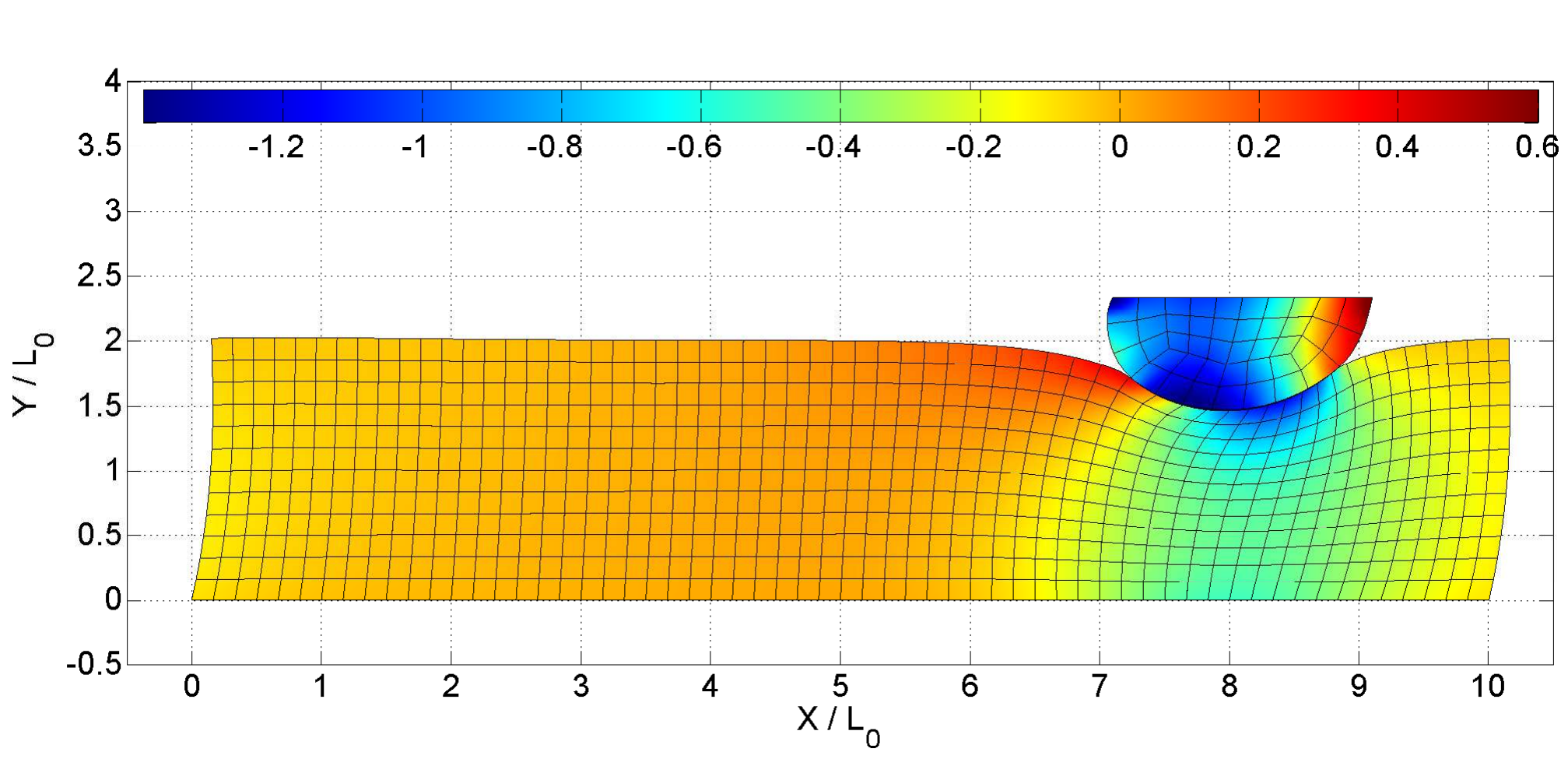}}

\end{picture}
\caption{2D ironing: deformed configuration discretized by $m_1=8$ and $m_2=12$. The color shows the stress invariant $I_1 = \tr \bsig$ normalized by $E_0$.} 
\label{f:iron2d_x}
\end{center}
\end{figure}

\begin{figure}[!h]
\begin{center} \unitlength1cm
\unitlength1cm
\begin{picture}(0,13.2)
\put(-8.0,6.5){\includegraphics[width=0.49\textwidth]{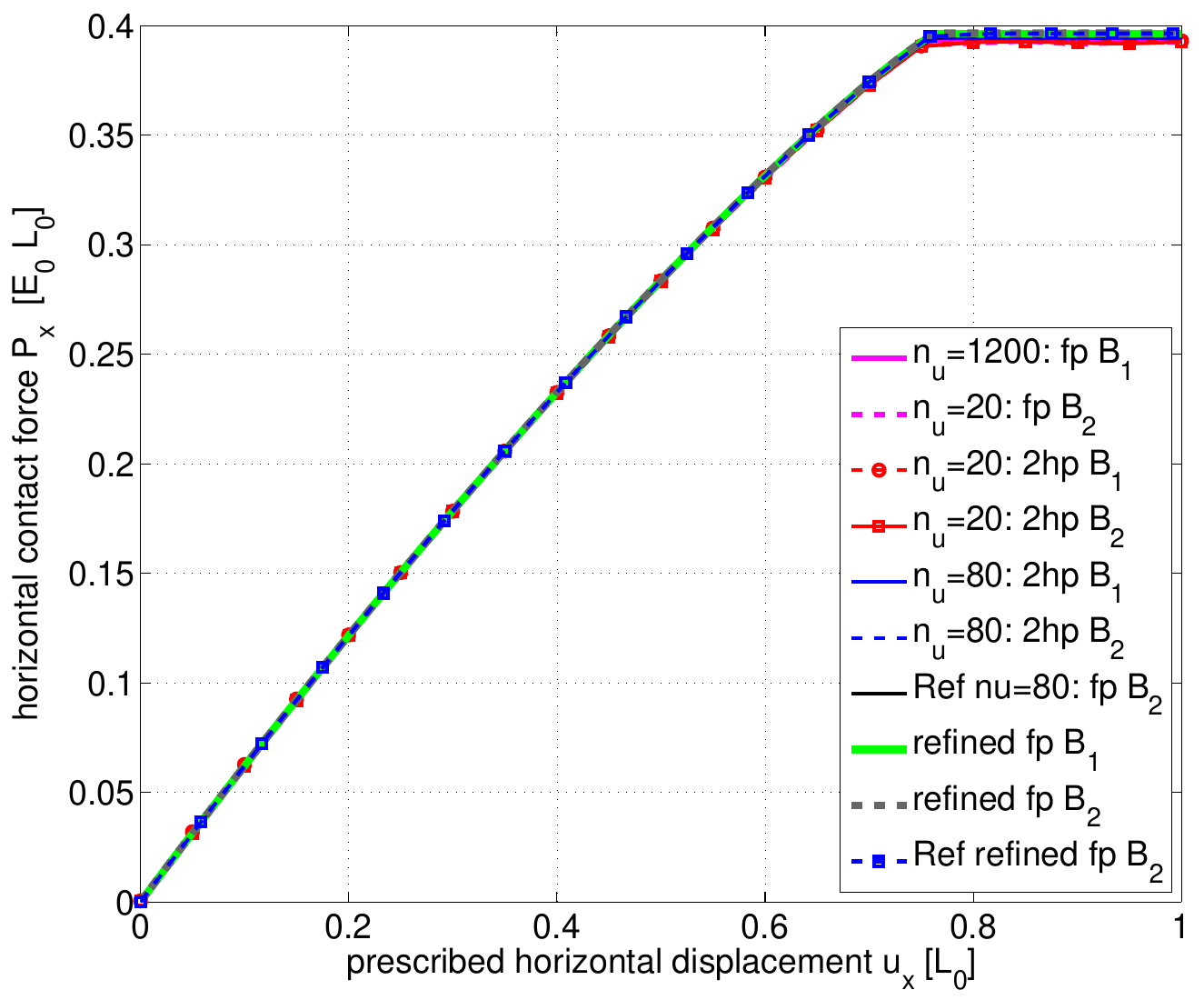}}
\put(0.2,6.5){\includegraphics[width=0.49\textwidth]{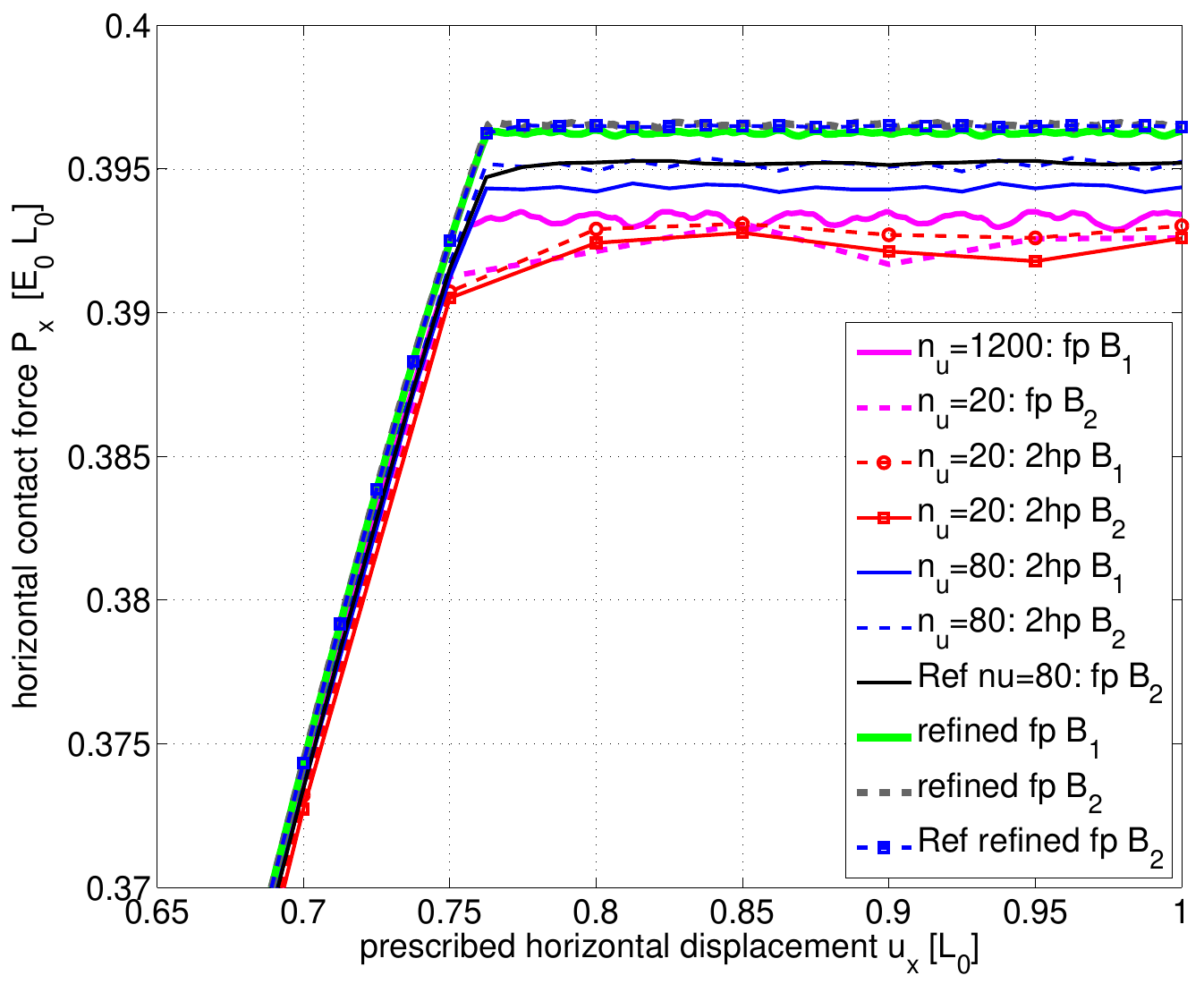}}
\put(-8.0,0){\includegraphics[width=0.49\textwidth]{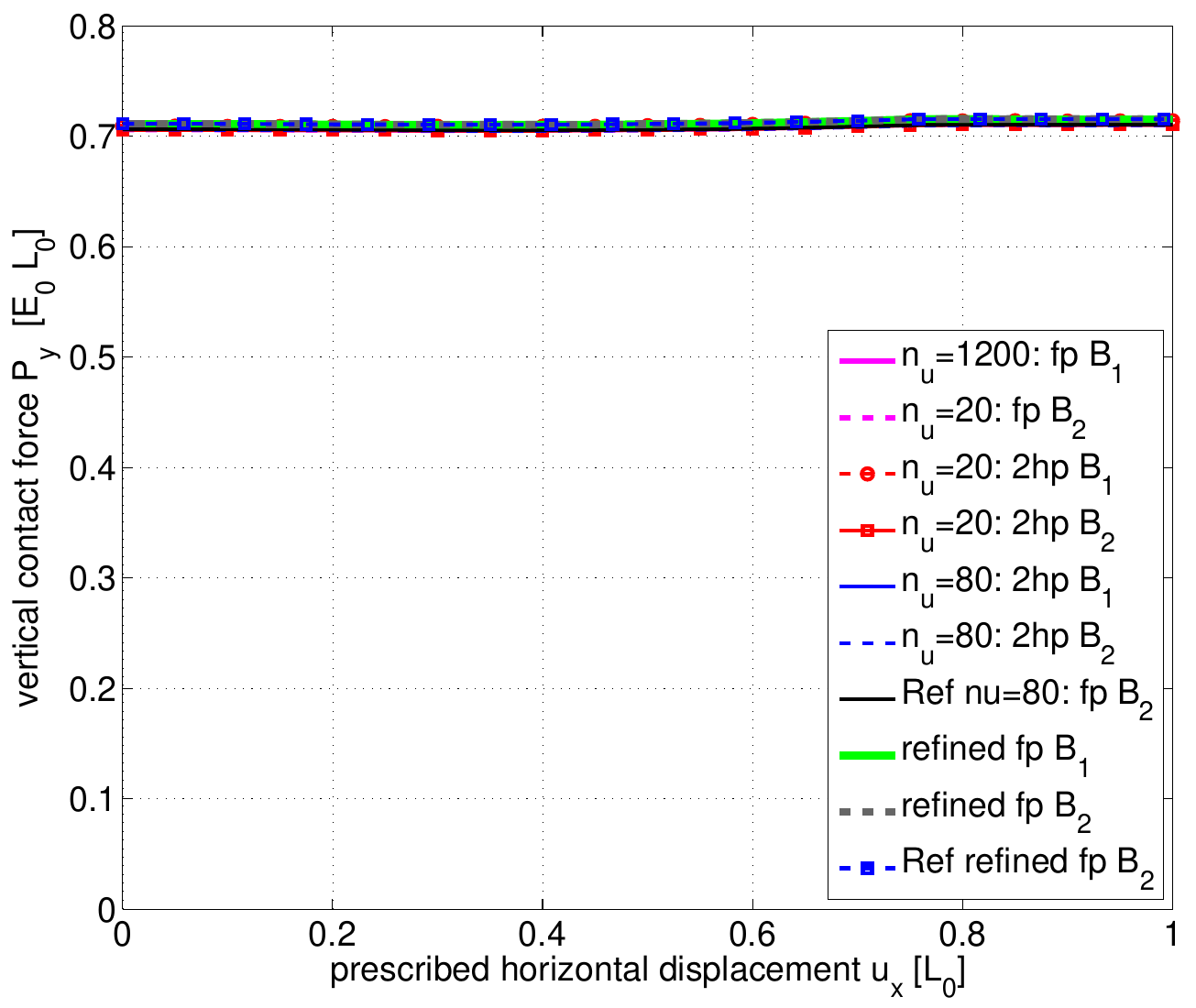}}
\put(0.2,0){\includegraphics[width=0.49\textwidth]{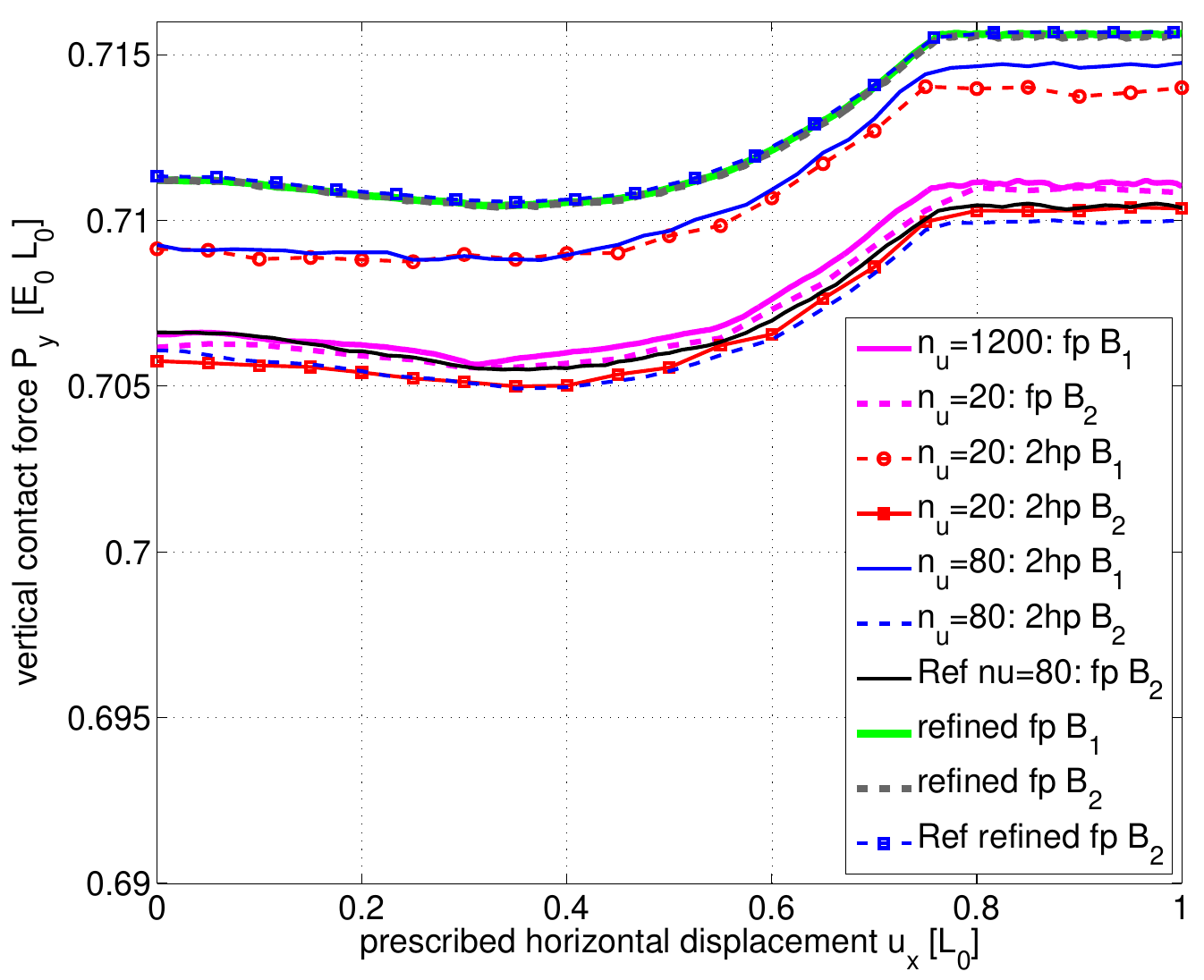}}
\end{picture}
\caption{2D ironing: contact forces shown in overview (left figures) and enlargement (right figures) considering $m_1=m_2=16$, $\mu=0.5$, and $\epsilon = 100~E_0/L_0$, for various $n_u$ and different contact formulations: full-pass (fp) and two-half-pass (2hp). The ``refined'' results are computed with $m_1=m_2=64$, $n_u=1200$, and $\epsilon = 1000~E_0/L_0$. ``Ref'' denotes the simulation results from \citet{spbf}.
} 
\label{f:iron2d_p}
\end{center}
\end{figure}
Next, the 2D ironing problem shown in Fig.~\ref{f:iron2d_x} is considered and also compared with the results of \citet{spbf}. Accordingly, a half-cylinder ($\sB_1$) with radius $L_0$ is pressed and then slid on a slab  ($\sB_2$) with dimension $10\,L_0 \times 2\,L_0$ by prescribing the vertical displacement $u_y = 2/3\,L_0$  and the horizontal displacement $u_x$, respectively, at the top boundary of $\sB_1$. As in the previous example, the bulk is discretized by linear elements while quadratic Hermite enrichment is used for the contact elements. The number of load steps in the simulation is denoted by $n_u$, and  the  the mesh density of $\sB_k$ ($k=1,\,2$) is characterized by the numerical parameter $m_k$. With this, the number of elements of $\sB_1$ and $\sB_2$ becomes $21\,m_1^2/32$ and $5\,m_2^2$, respectively. The material parameters and contact parameters are $E_1 =3\,E_0$ and $E_2=E_0$, $\nu_1=\nu_2=0.3$, $\epsilon = 100\,E_0/L_0$ and $\mu=0.5$.

Fig.~\ref{f:iron2d_p} shows the vertical and horizontal contact forces during the sliding phase considering both the full-pass and the two-half-pass version of the proposed formulation in comparison with the formulation of \citet{spbf}. The influence of the load step size is also shown. 

As seen, the results of both proposed formulations and \citet{spbf} are of the same order and become almost identical when the mesh is refined. However, the two-half-pass version of the proposed formulation is shown to be less sensitive to the number of load steps compared to the results reported in \citet{spbf}.

\subsection {3D twisting}
For general 3D frictional contact, we test our formulation with the twisting example presented in \citet{spbf}. Accordingly, a hollow-hemisphere  ($\sB_1$) with outer radius $L_0$ and thickness $1/3\,L_0$ is pressed and then twisted against a solid block ($\sB_2$) with dimension $L_0\times L_0\times L_0$, as is shown in Fig.~\ref{f:twist_x}. The parameters $E_1 = 5\,E_0$, $E_2 = E_0$, $\nu =0.3$, $\epsilon = 100\,E_0/L_0$ are taken for the simulation. Frictionless contact is assumed during the pressing phase, while frictional contact with friction coefficient $\mu=0.5$ is considered during the twisting phase. To improve both efficiency and accuracy, the bulk is approximated with linear elements while both contact surfaces are discretized with cubic NURBS-enriched surface elements proposed by \citet{nece,nece2}. $5\times 5$ Gaussian quadrature points are used for all contact elements.

The simulation runs without any convergence problems. The vertical reaction force and torque during the twisting phase  are compared  in Fig.~\ref{f:twist_p} with the reference results of \citet{spbf}. As seen, the present simulation results are in good agreement with the ones reported in \citet{spbf}, which confirms the accuracy of the proposed formulation.

\begin{figure}[!htp]
\begin{center} \unitlength1cm
\unitlength1cm
\begin{picture}(0,7.6)

\put(-7.3,4){\includegraphics[height=40mm]{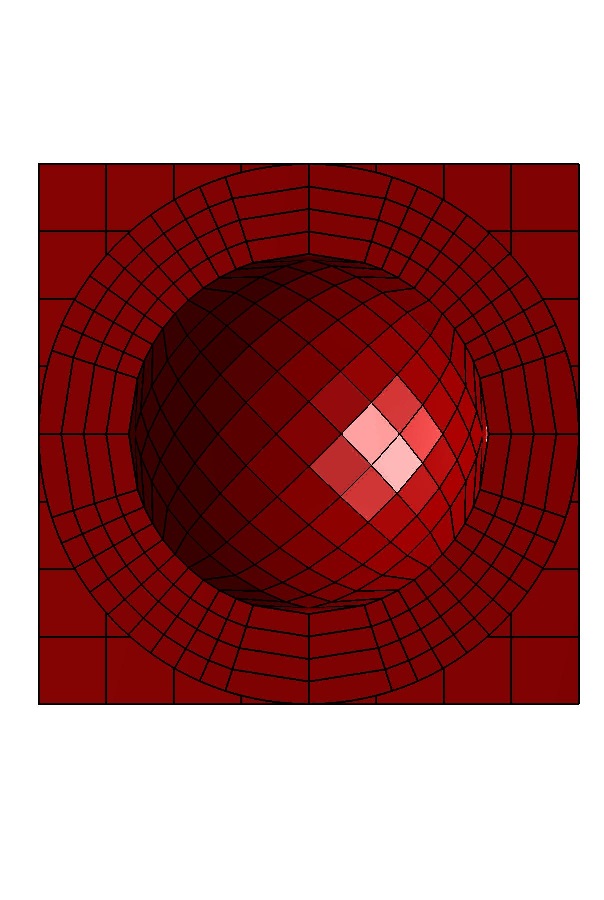}}
\put(-3.7, 4){\includegraphics[height=40mm]{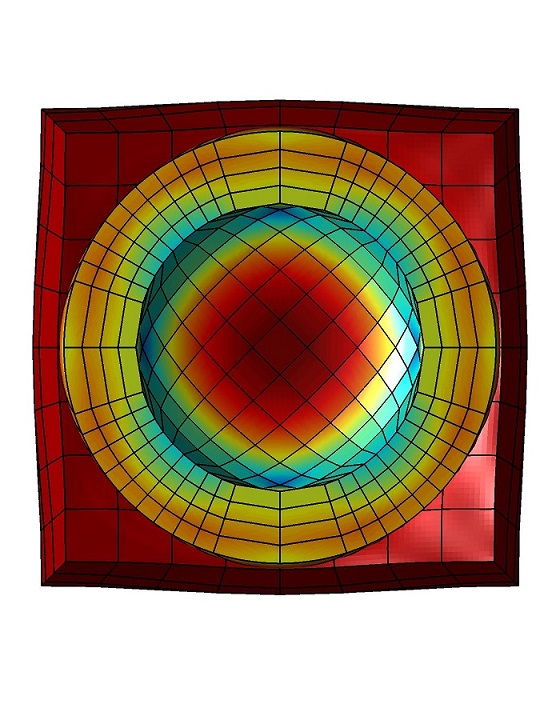}}
\put(-0.25,4){\includegraphics[height=40mm]{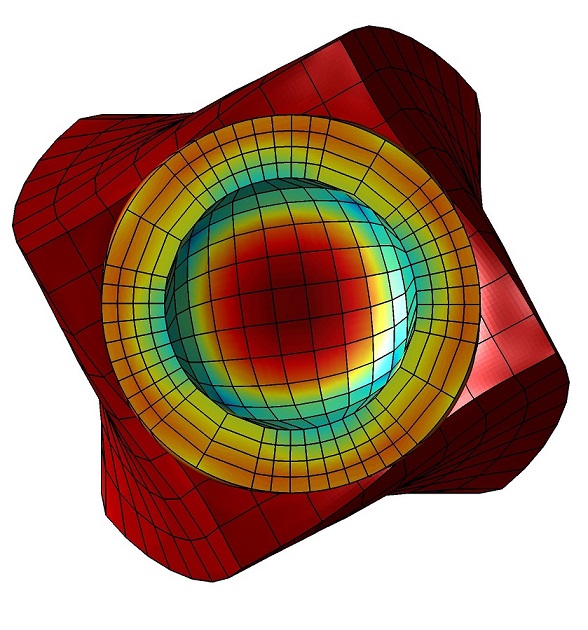}}
\put(3.35,4){\includegraphics[height=40mm]{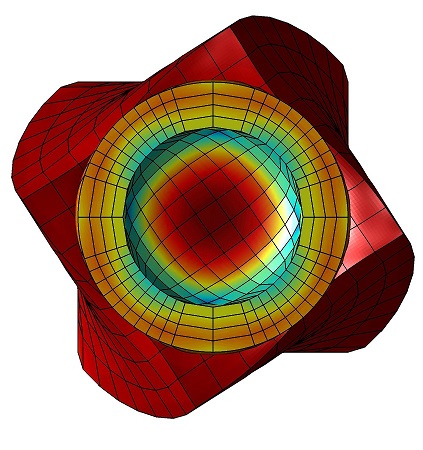}}

\put(-7.7,0){\includegraphics[height=40mm]{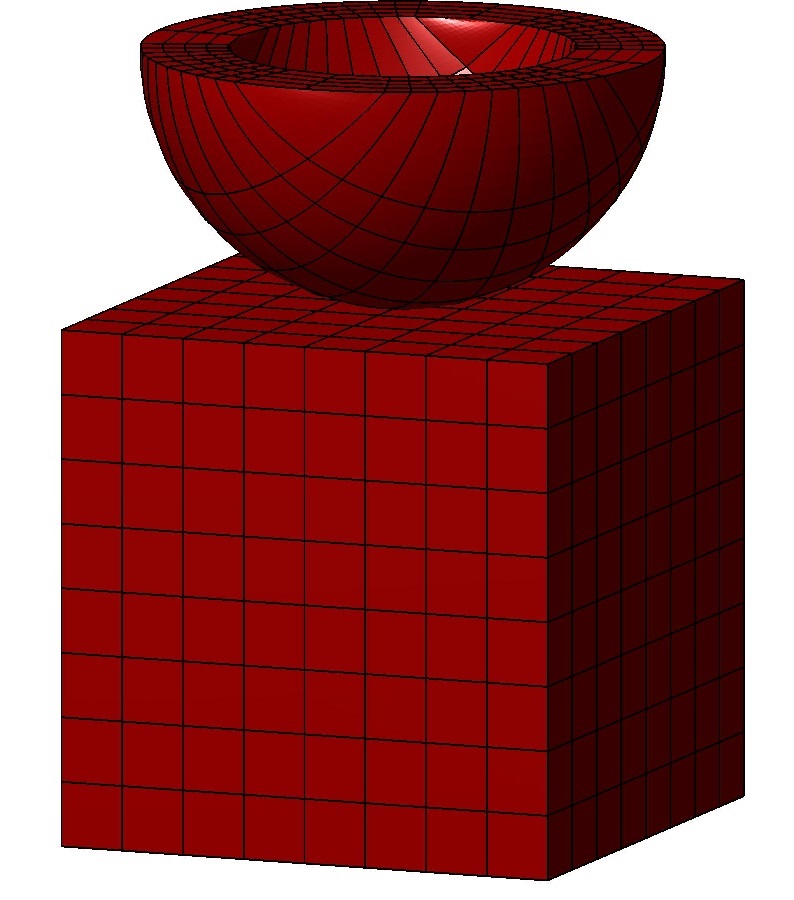}}
\put(-4,0){\includegraphics[height=40mm]{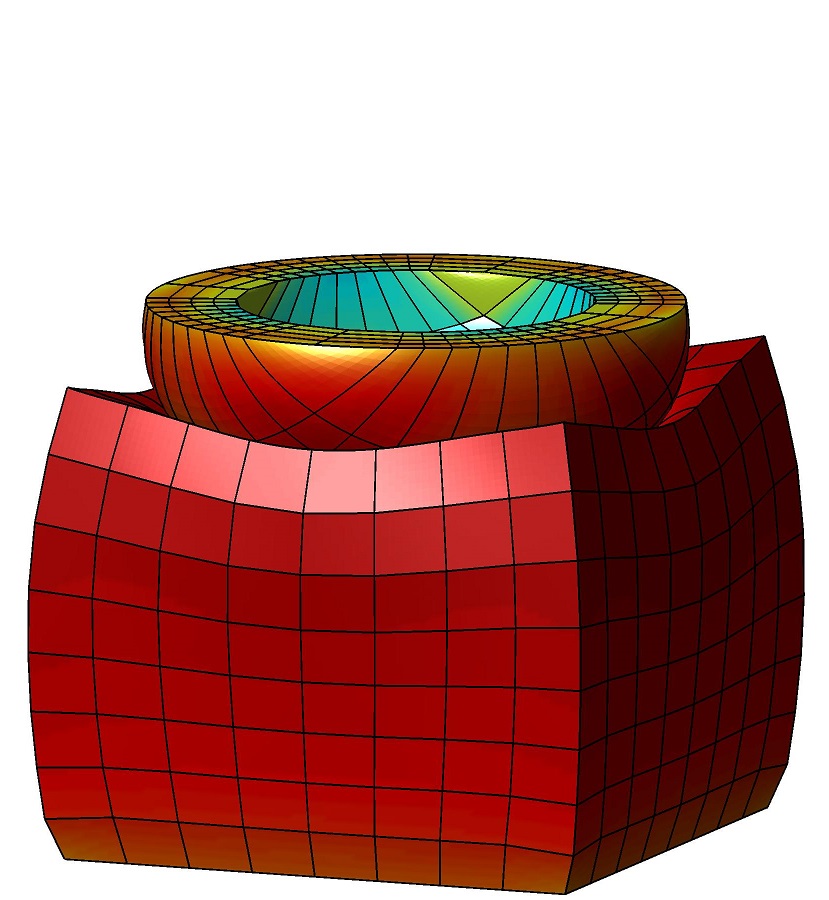}}
\put(0,0){\includegraphics[height=40mm]{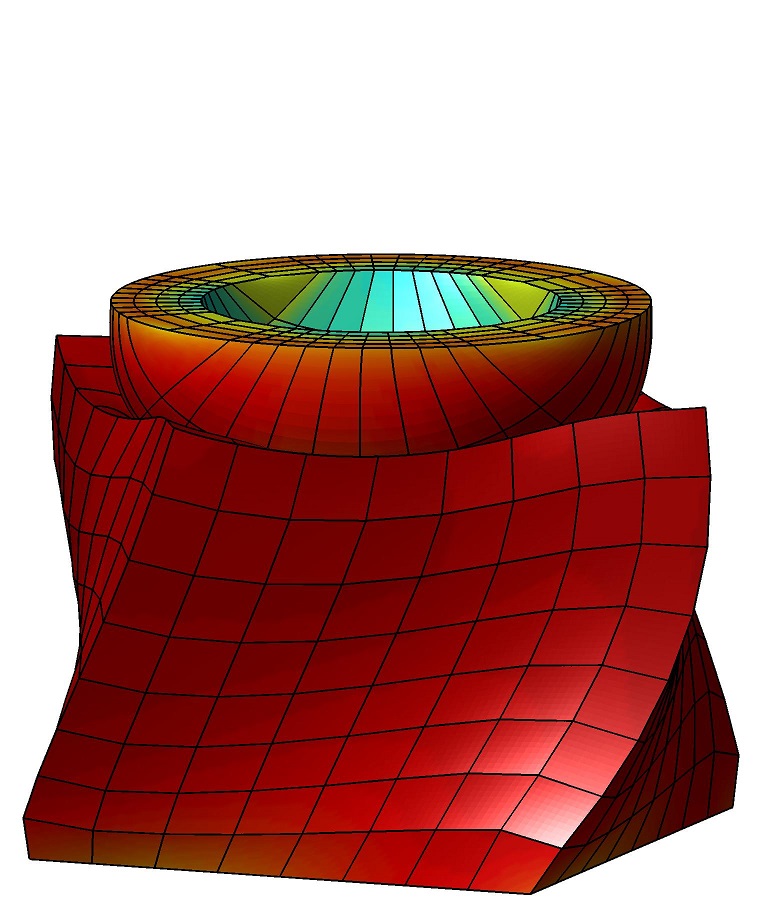}}
\put(3.5,0){\includegraphics[height=40mm]{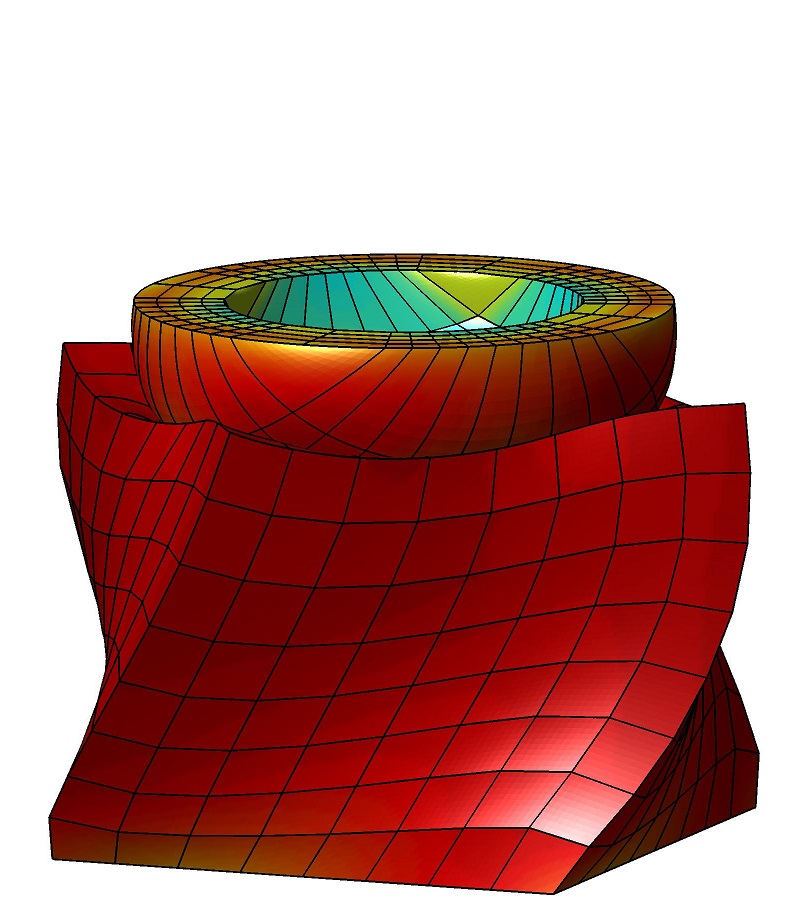}}

\put(-3.7,3.3){\includegraphics[height=7.5mm]{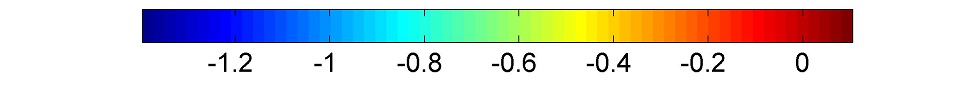}}

\end{picture}
\caption{3D twisting: Plots of the undeformed configuration and the deformed configurations at twisting angles $0^\circ$, $60^\circ$, and $180^\circ$ (from left to right). The color shows the first stress invariant normalized by $E_0$. Here, $\mu=0.5$ and $\epsilon=100\,E_0/L_0$. See also the supplementary movie 
at {\url{https://doi.org/10.5446/37898}}.
}
\label{f:twist_x}
\end{center}

\begin{center} \unitlength1cm
\unitlength1cm
\begin{picture}(0,11.5)
\put(-8.0,6){\includegraphics[width=0.45\textwidth]{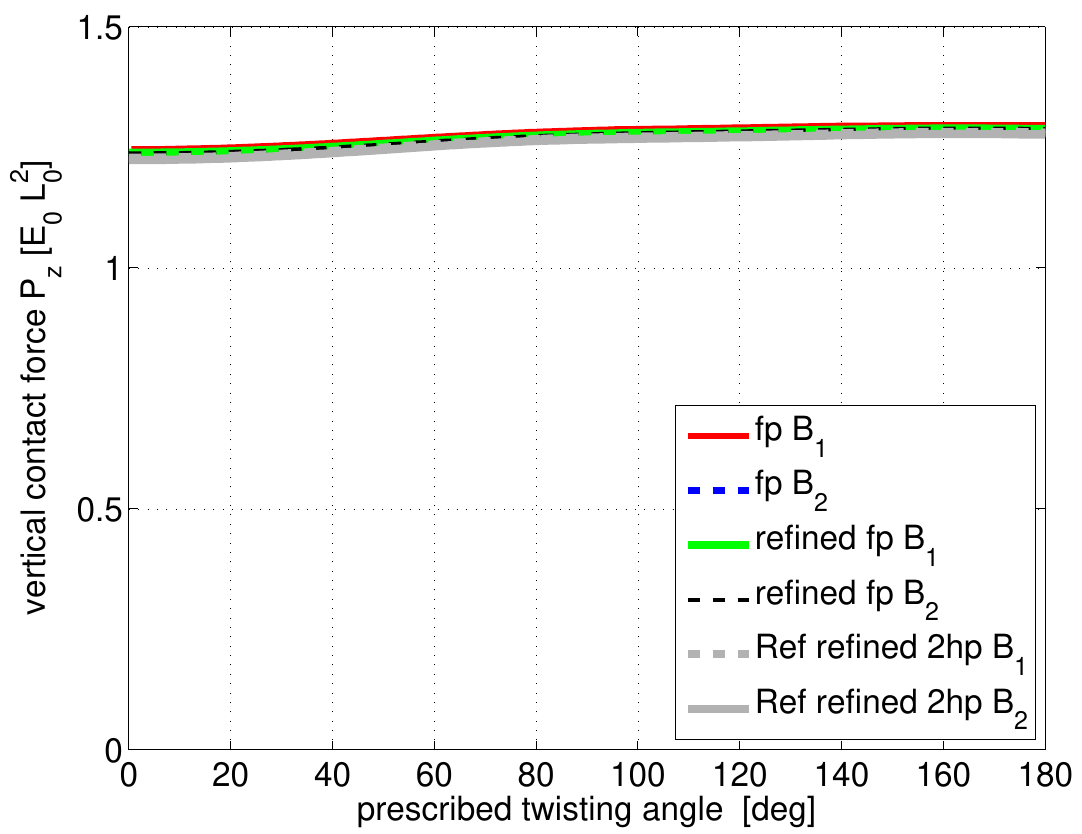}}
\put(0,6){\includegraphics[width=0.45\textwidth]{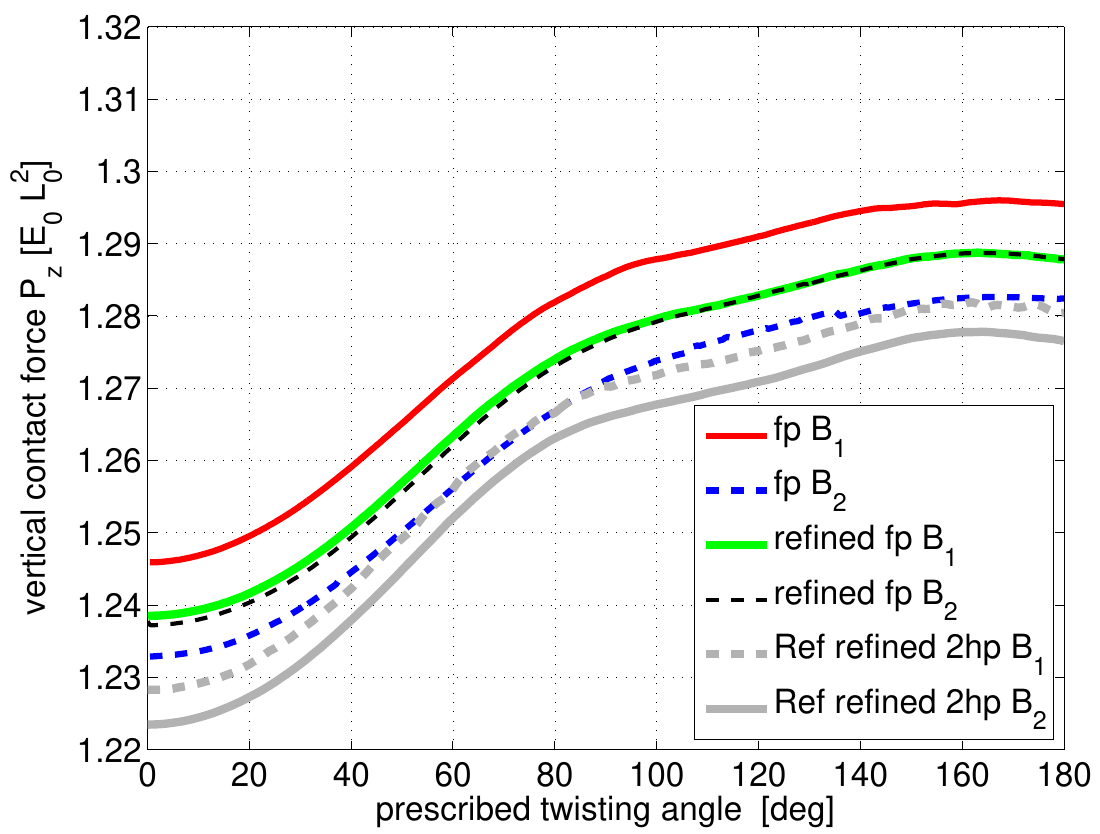}}

\put(-8.0,0){\includegraphics[width=0.45\textwidth]{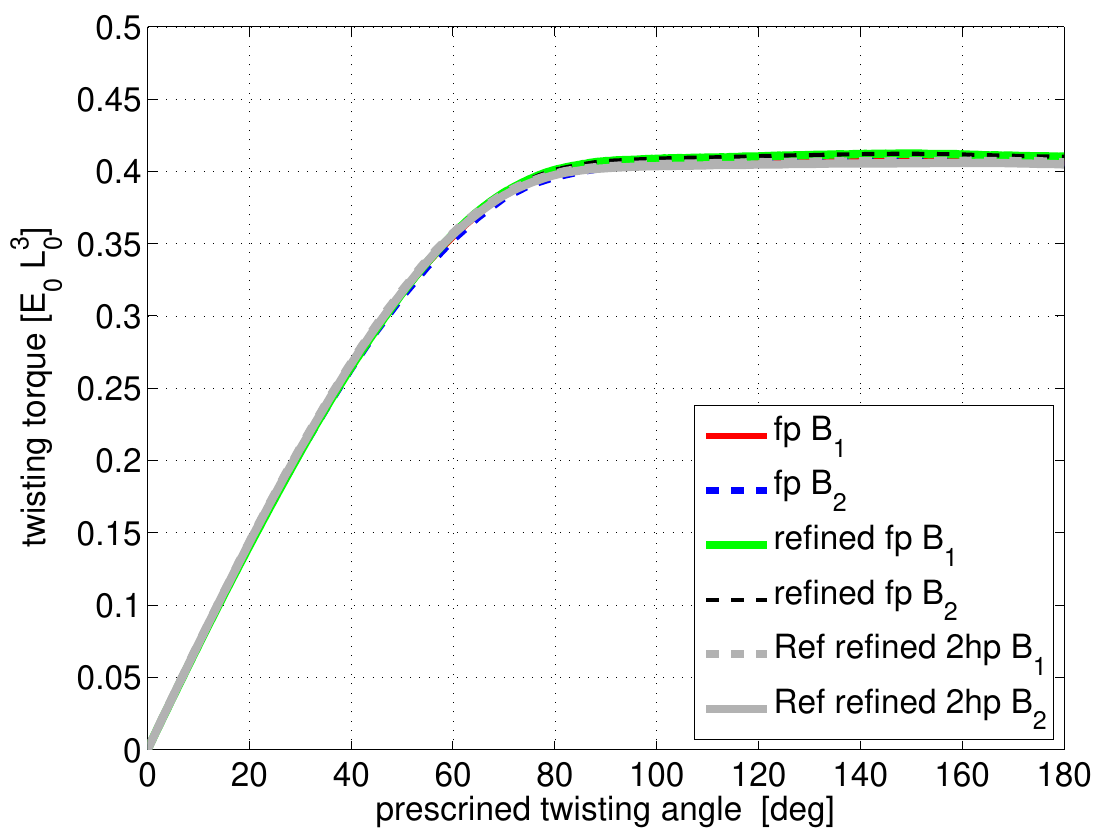}}
\put(0,0){\includegraphics[width=0.45\textwidth]{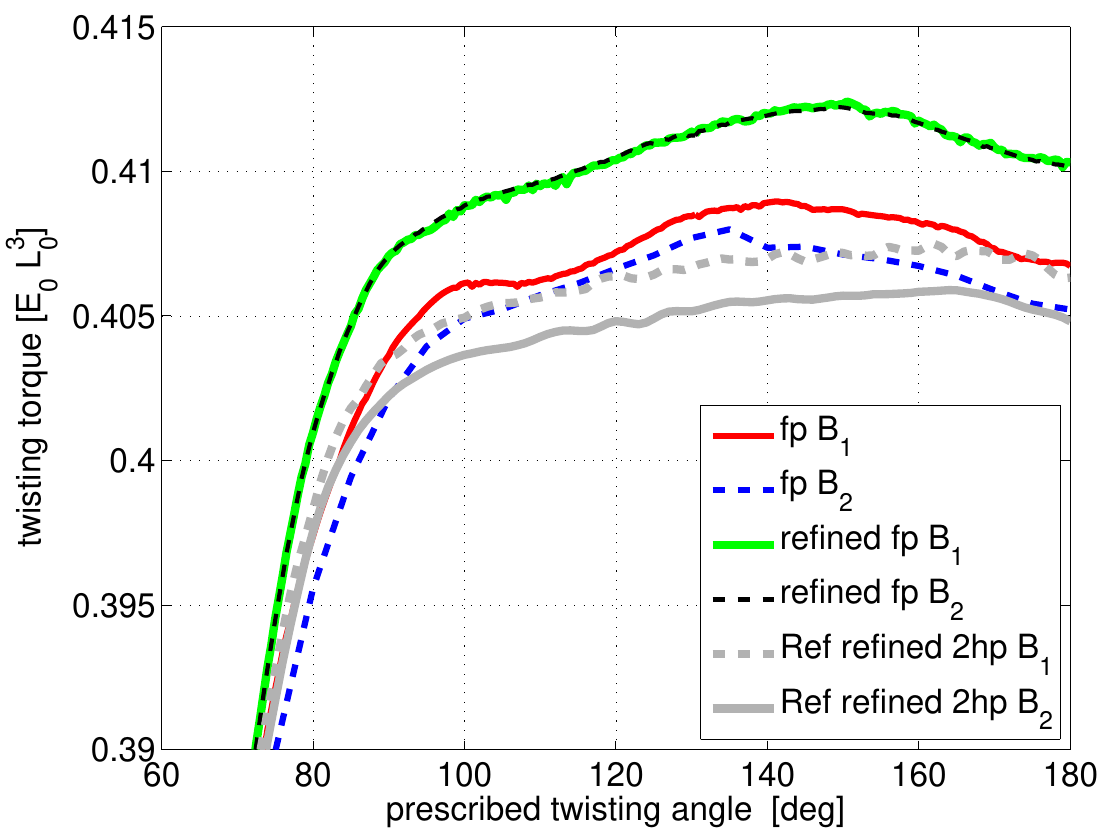}}

\end{picture}
\caption{3D twisting: vertical reaction force $P_z$ and torque $M_z$ shown in overview (left) and enlargement (right), considering $ \mu=0.5$, $\epsilon=100\,E_0/L_0$, and $n_u=360$ for twisting of $180$ degrees.  Both the full-pass (fp) and two-half-pass (2hp) contact formulations are considered. The ``refined'' results are computed with $6\times19^2$ and $16^3$ volume elements for $\sB_1$ and $\sB_2$, respectively, and $\epsilon= 750~E_0/L_0$. ``Ref'' denotes the  simulation results from \citet{spbf}.} 
\label{f:twist_p}
\end{center}
\end{figure}

\subsection {Sliding of two inflated rubber sheets}

\begin{figure}[!htp]

\begin{center} \unitlength1cm
\unitlength1cm
\begin{picture}(0,18.0)

\put(-5.1,13){\includegraphics[width=0.6\textwidth]{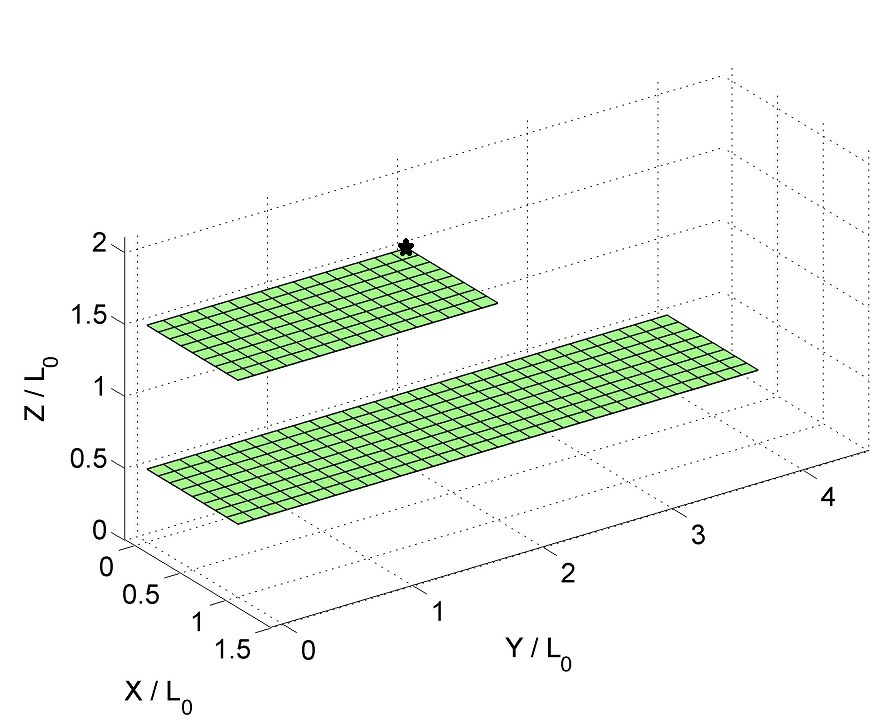}}
\put(-8.0,6.5){\includegraphics[width=0.45\textwidth]{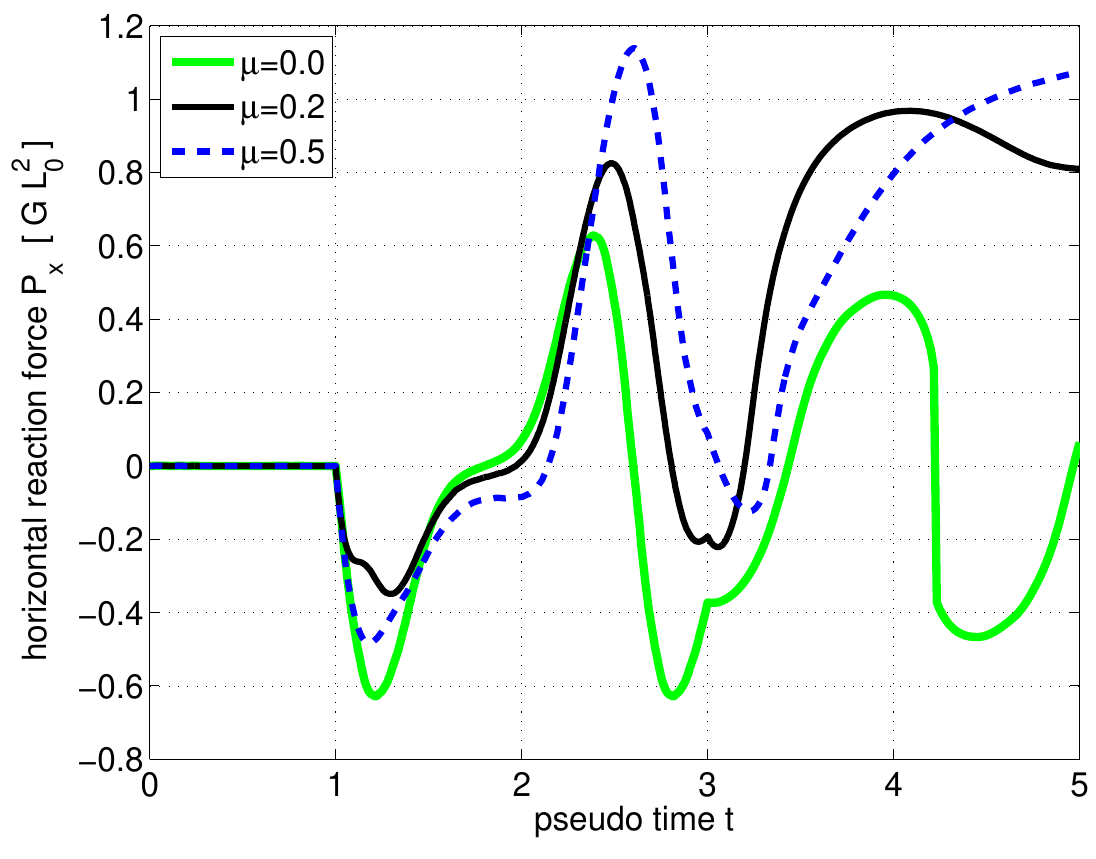}}
\put(0,6.5){\includegraphics[width=0.45\textwidth]{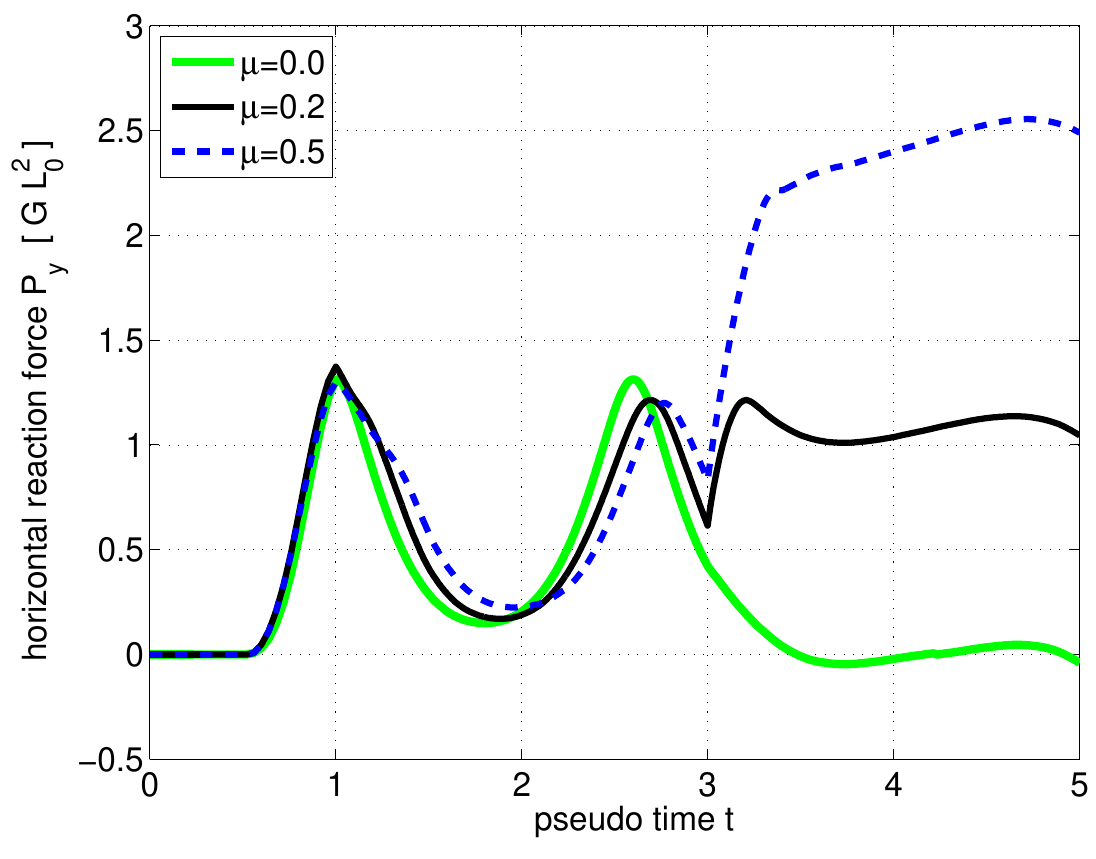}}

\put(-8.0,0){\includegraphics[width=0.45\textwidth]{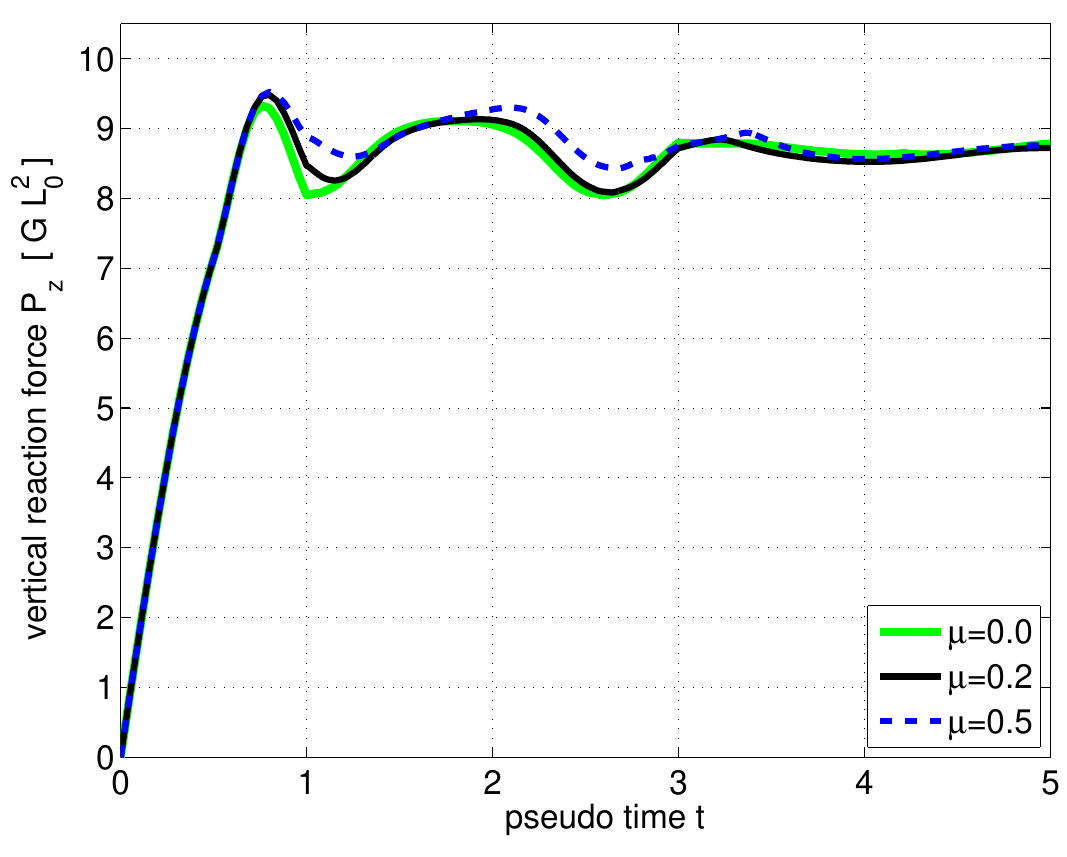}}
\put(0,0){\includegraphics[width=0.45\textwidth]{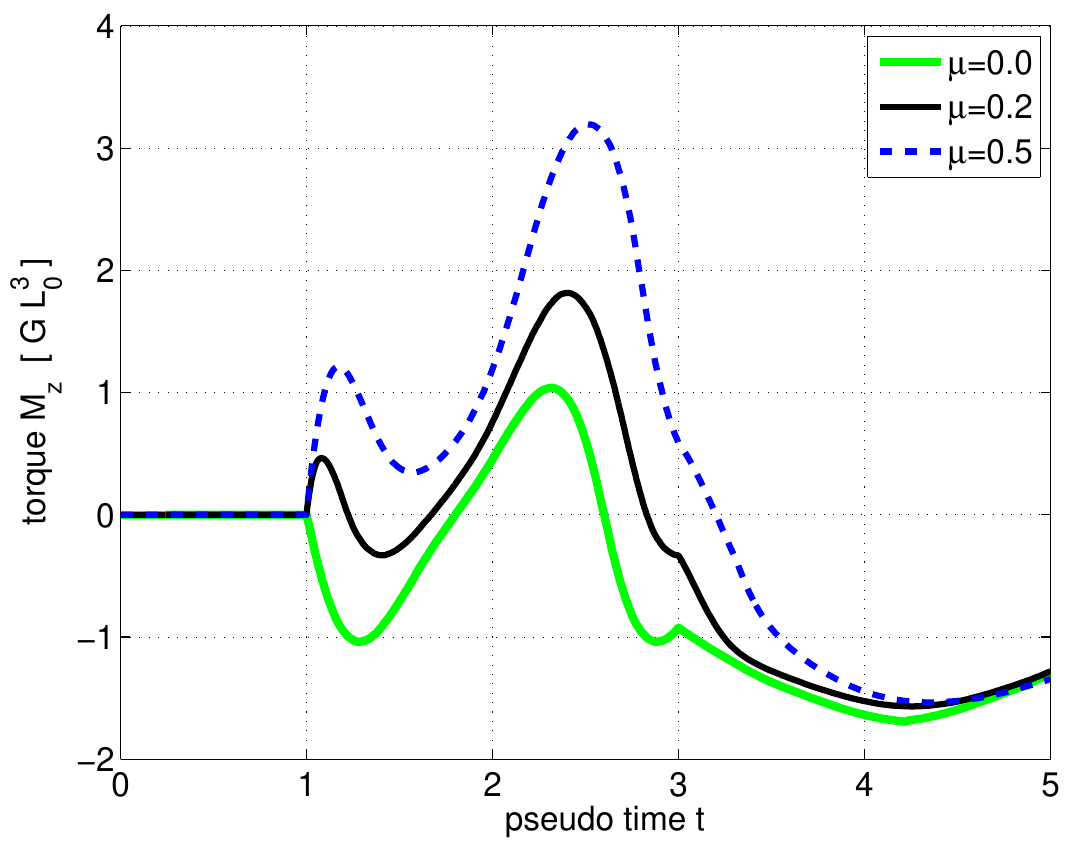}}

\put(-5, 14.5){a. }

\put(-7.8, 6.5){b. }
\put(0.5, 6.5){c. }

\put(-7.8, 0){d. }
\put(0.5, 0){e. }

\end{picture}
\caption{Sliding of two inflated rubber sheets: initial configuration (a), and net reaction forces $P_x$ (b), $P_y$ (c), and $P_z$ (d),  and net torque $M_z$ (e), measured on the boundary of the upper sheet during the inflating phase ($t \in [0,\,1]$), the twisting phase  ($t\in [1,\,3]$), and  the sliding phase  ($t\in [3,\,5]$). Here, $\epsilon=300~G/L_0$ is used for all loading phases. } 
\label{f:inflatedsheet}
\end{center}
\end{figure}
The last example examines  contact between two inflated rubber sheets. The initial configuration is shown in Fig.~\ref{f:inflatedsheet} (left). The upper sheet with size $L_0\times 2\,\! L_0$ is discretized by $8\times 16$ cubic NURBS elements. The lower sheet is twice longer in the $Y$ direction  than the upper one and is discretized by $8\times 32$  cubic NURBS elements.  In order to avoid a compressive stress state, both sheets are pre-stretched by $\lambda =1.5$ and all boundaries are fixed. The membrane formulation of \cite{membrane} is used  for the sheet, and the incompressible Neo-Hookean material model
\eqb{l}
\sigma^{\alpha\beta} = \ds\frac{G}{J}\left(A^{\alpha\beta} - \frac{a^{\alpha\beta}}{J^2}\right)~,
\eqe 
is considered, where $\sigma^{\alpha\beta}$ and $J$ denote the components of the Cauchy stress tensor and the surface stretch, respectively, and $G$ is a material constant. Here,  $G$ of the lower sheet is set five times larger than the upper one. Contact is simulated with the full-pass algorithm using the friction coefficient $\mu=0.5$. $4\times 4$ Gauss points (per element) are used for the quadrature of both membrane and contact elements.

Initially, the sheets are aligned in the $X$ and $Y$ directions and separated by the gap $L_0$ in the $Z$ direction  as shown in Fig.~\ref{f:inflatedsheet}a. Next, the sheets undergo three consecutive loading phases. From (pseudo) time 0 to $1$, contact between the two sheets is induced by 
increasing the volume enclosed by the sheets from 0 to $19/9$ and  $13/3$ for the upper and lower sheet, respectively.  The deformed configuration at the end of this phase is shown in Fig.~\ref{f:inflatedsheetcgf}a. 
From time $1$ to $3$, the upper sheet is rotated by $225^\circ$ around its center as shown in  Fig.~\ref{f:inflatedsheetcgf}a-d. Finally, from time $3$ to $5$, the upper sheet slides against the lower sheet by moving its boundary by the distance $2.5\,L_0$ in the $Y$ direction  (see Fig.~\ref{f:inflatedsheetcgf}e-f).  In the simulation, $25$, $225$, and $125$ loads steps are used for the inflating, twisting, and sliding phases, respectively. 

Selected snapshots during the simulation are shown in Fig.~\ref{f:inflatedsheetcgf}. 
As Fig.~\ref{f:inflatedsheet}{b-e} shows, the net torque and  the reaction forces\footnote{i.e~the resultant of the contact force and the surface force due to the volume constraint.} vary strongly during the three loading phases. 
A vertical reaction force appears during the inflating phase due to the volume constraint. Also, a net force $P_y$ appears during the inflating phase mainly due to the re-distribution of the inflated volumes during contact. During the sliding phase, $P_y$ depends mainly on  friction. The net force $P_x$ and the net torque $M_z$ during the twisting and the sliding phase result from friction in combination with the re-distribution of the volume of the upper sheet from one side to the other, as is seen in Fig.~\ref{f:inflatedsheetcgf}a-f. The successful simulation of this example demonstrates the robustness of the proposed formulation for large sliding contact problems.

\begin{figure}[!htp]
\begin{center} \unitlength1cm
\unitlength1cm
\begin{picture}(0,21.8)

\put(0.5, 20.9){\includegraphics[width=0.47\textwidth]{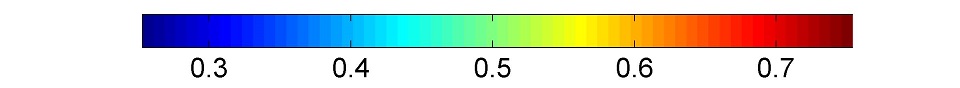}}

\put(-8.5,18.2){\includegraphics[width=0.52\textwidth]{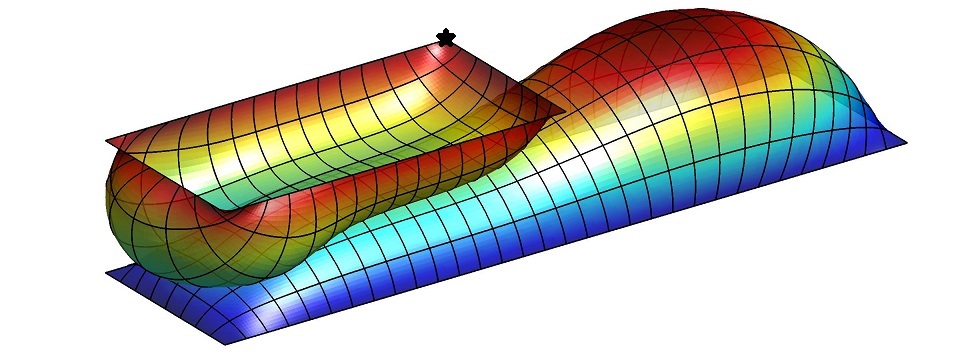}}
\put(0, 18.7){\includegraphics[width=0.47\textwidth]{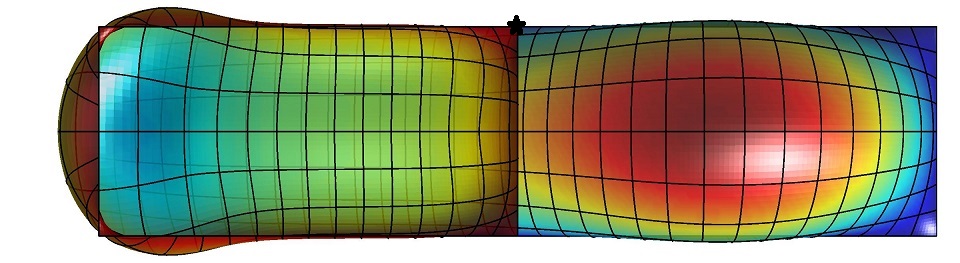}}

\put(-8.5, 14.6){\includegraphics[width=0.52\textwidth]{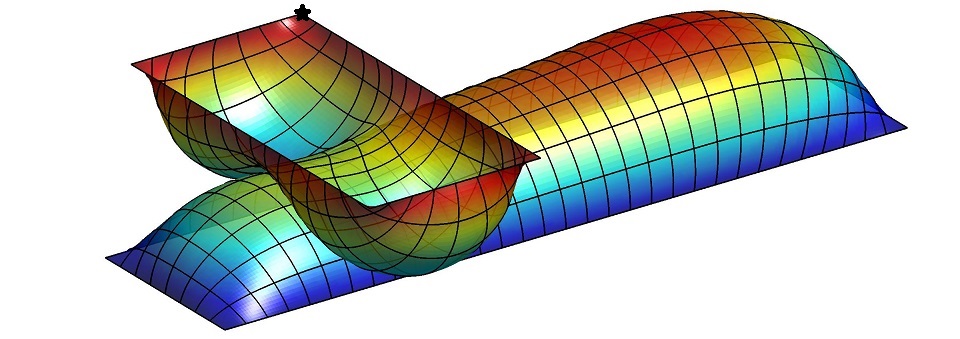}}
\put(0, 14.4){\includegraphics[width=0.47\textwidth]{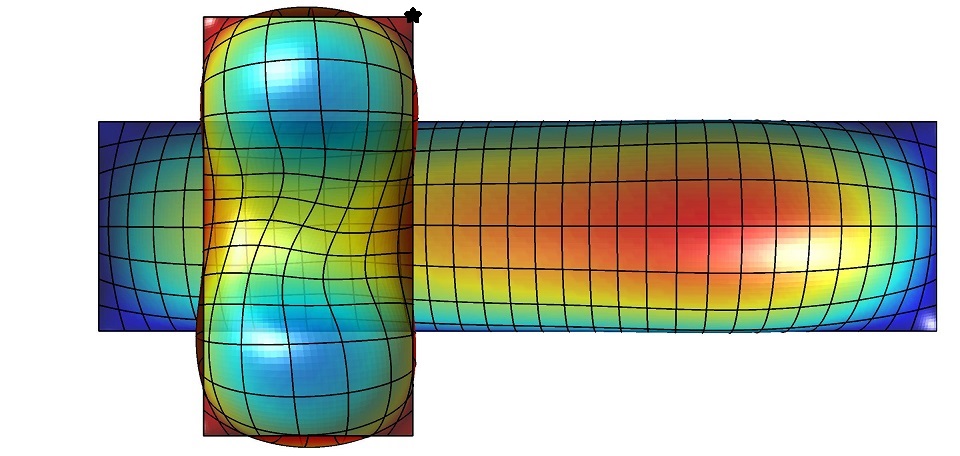}}

\put(-8.5, 11.0){\includegraphics[width=0.52\textwidth]{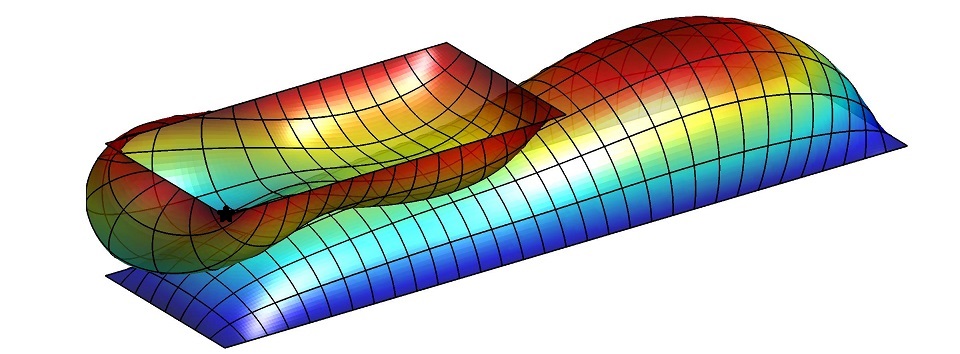}}
\put(0, 11.6){\includegraphics[width=0.47\textwidth]{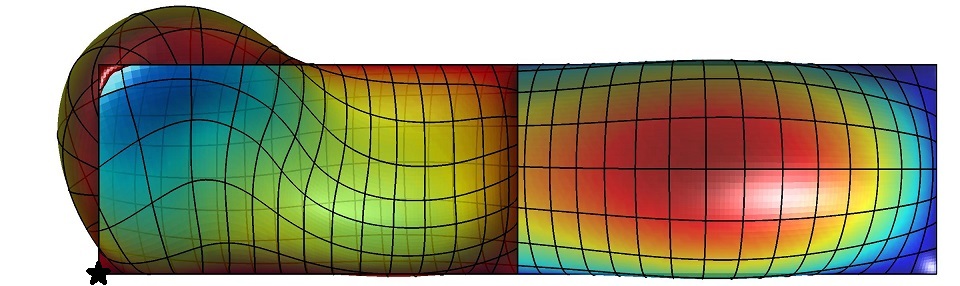}}

\put(-8.5, 7.4){\includegraphics[width=0.52\textwidth]{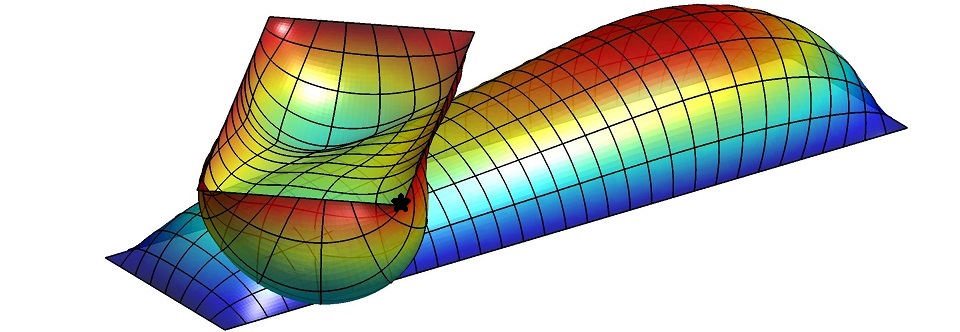}}
\put(0,7.2){\includegraphics[width=0.47\textwidth]{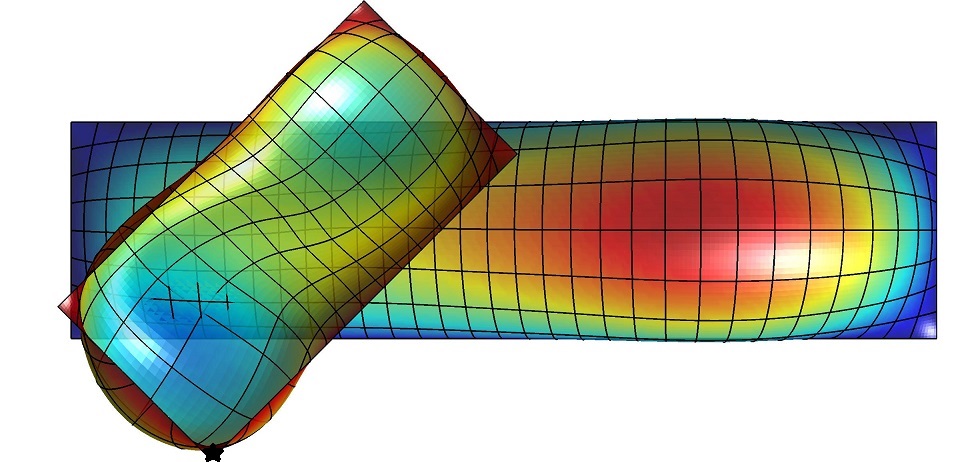}}

\put(-8.5, 3.7){\includegraphics[width=0.52\textwidth]{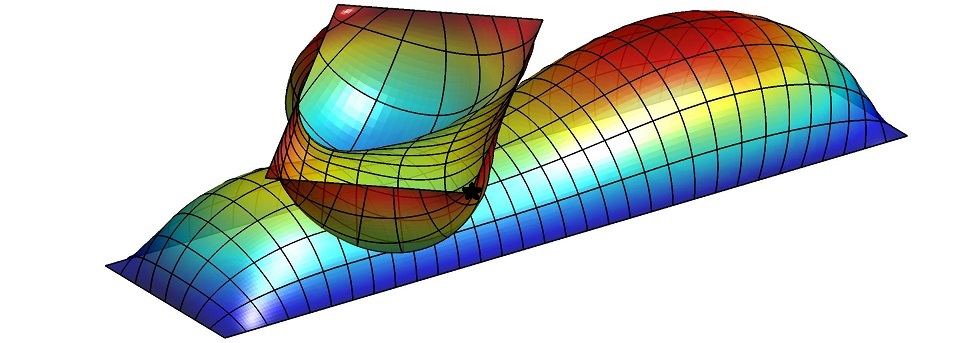}}
\put(0, 3.5){\includegraphics[width=0.47\textwidth]{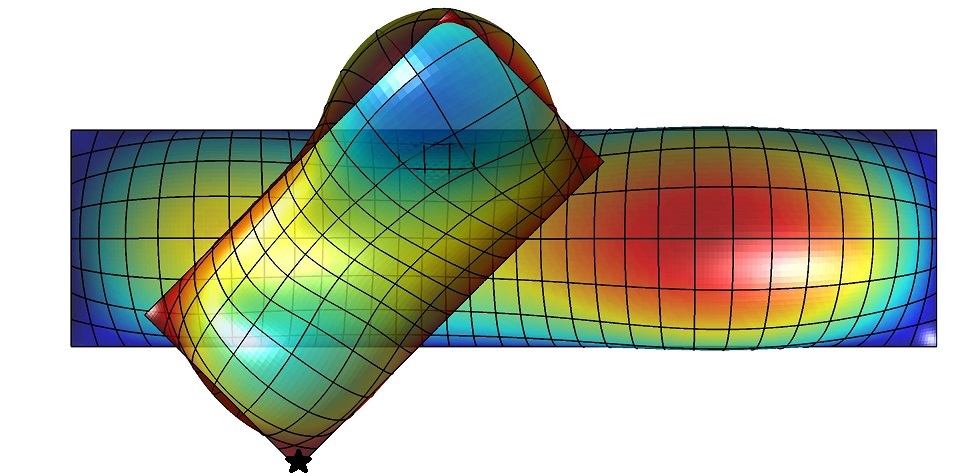}}

\put(-8.5, 0){\includegraphics[width=0.52\textwidth]{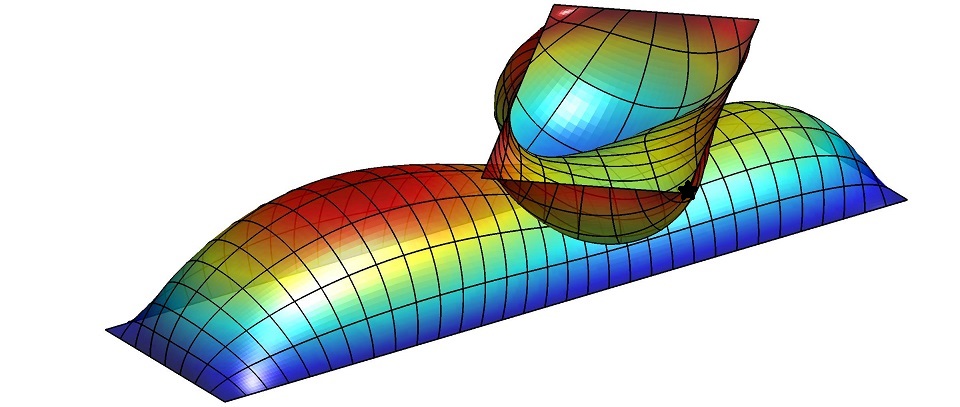}}
\put(0, -0.2){\includegraphics[width=0.47\textwidth]{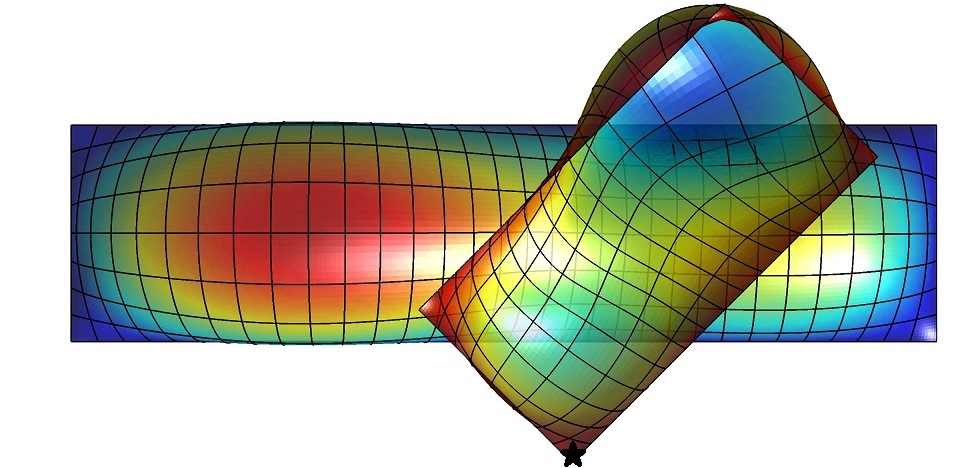}}

\put(-0.9, 19.2){a. }
\put(-0.9, 15.7){b. }

\put(-0.9, 12){c. }

\put(-0.9, 8.4){d. }

\put(-0.9, 4.7){e. }

\put(-0.9, 1.0){f. }

\end{picture}

\caption{Sliding of two inflated rubber sheets: deformed configurations in 3D view (left) and  corresponding top view (right) for the twisting phase at $\theta =  0^\circ$ (a),  $90^\circ$ (b), $180^\circ$ (c), and $225^\circ$ (d), and  the sliding phase at $u_y = 0.6\,L_0$ (e) and $2.5\,L_0$ (f). The color shows the stress invariant $I_1=\tr\bsig$ normalized by $G$. Here, $\mu=0.5$ and $\epsilon=300~G/L_0$. See also the supplementary movie at {\url{https://doi.org/10.5446/37899}}.
 }
\label{f:inflatedsheetcgf}
\end{center}

\end{figure}
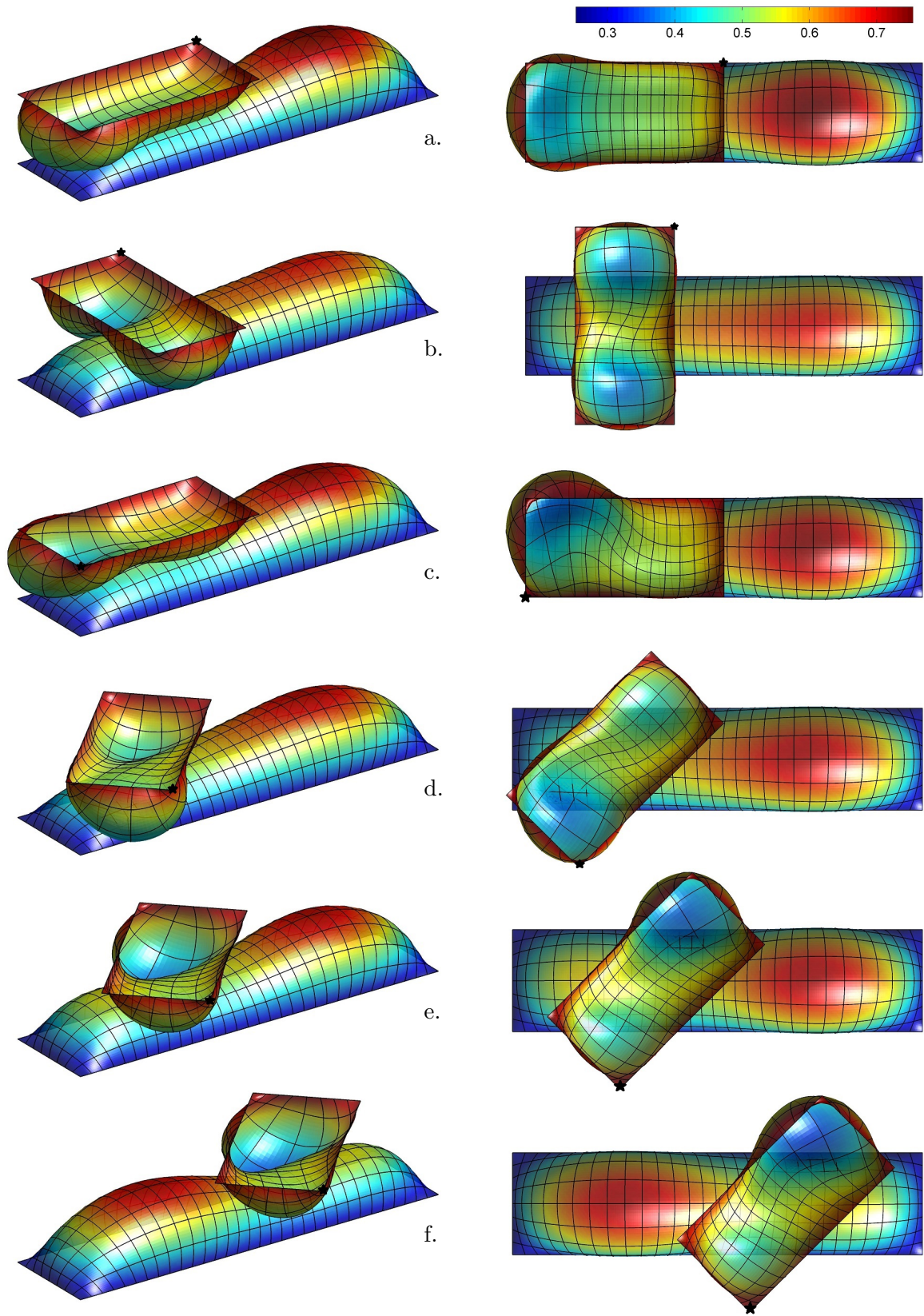

\section {Conclusion}\label{s:conclude}

This paper presents the extension of the surface potential theory of \citet{spbc} to friction for the case of point interactions. The basic equations for friction are first derived for a simple 1D example using the first and the second laws of thermodynamic.  The so-called \textit{interacting gap} is defined as a kinematic variable, which unifies both normal/tangential and sticking/sliding contact.  With this, the computational contact formulation for 3D friction is constructed based on a purely kinematic constraint function.  

We further employ the direct elimination approach on the constraint function, which is then identified to be equivalent to the moving friction cone concept of \cite{Wriggers2003}. The corresponding finite element formulation for quasi-static computations is presented for both the two-half-pass and the full-pass algorithms. The robustness is further enhanced by employing  smooth isogeometric discretization \citep{hughes05}, which further facilitates a more accurate choice of the tangential sliding direction. Consequently, as the numerical examples show, the proposed formulation exhibits lower sensitivity to the load step size than previous formulations.
\begin{table}[!htp]
\begin{small}
\begin{center}
\def\arraystretch{1.5}\tabcolsep=3pt
\resizebox{\textwidth}{!}{%
\begin{tabular}{|l|c|c|c|c|}
\hline 
 Item &Standard & MFC & Present formulation\\ 
\hline 
Treatment of normal and tangential gap&split & unified  & unified  \\ 
\hline 
Underlying contact theory&numerical constraint & numerical constraint & surface potential-based \\ 
\hline 
Interpretation of frictional sliding contact&plasticity theory & plasticity theory&  kinematical constraint\\ 
\hline 
Computation method for the sliding point& predictor-corrector  & moving friction cone  & direct elimination \\ 
\hline
Direction of the sliding traction& secant  & secant & tangent  \\ 
\hline
\end{tabular}}
\caption{Comparison between the standard, MFC, and the proposed formulations.}
\label{t:compare}  
\end{center} 
\end{small}
 \end{table}  
 
Tab.~\ref{t:compare} compares the proposed formulation with the standard formulation (e.g.~as considered by \cite{krstulovic02,laursen,spbf,neto16}) and the moving friction cone formulation \citep{Wriggers2003}.   
In comparison with the standard formulation}, the implementation of the proposed formulation is much easier, since its theory is more concise even though it is still consistent with the surface potential-based contact  theory of \citet{spbc}. An advantage of the surface potential-based contact theory is that it provides a unified framework for both numerical constraint formulations, like the penalty and Lagrange multiplier methods, and physically motivated contact interactions like van-der-Waals adhesion, electrostatic interactions, or cohesive-zone models.

The current friction formulation focuses exclusively on penalty-based constraint enforcement. However, since the theory also allows for surface potentials, those can for example be constructed from the homogenization of atomistic interaction potentials. This will be considered in future work.

\appendix

\section{Linearization of the kinematical constraint} \label{s:localNewton}
Given the current position of $\bx_k$ and $\bx_{\ell}(\hat\bxi^n)$, the sliding point $\bxi_\mrm$ is determined by solving nonlinear Eq.~\eqref{e:phiTa} with the Newton-Raphson method. Accordingly, the Taylor series of $f_\alpha(\xi^\beta)$ about the point $\xi^\beta+ \Delta\xi^\beta$ is given by
\eqb{l}
f_\alpha(\xi^\beta+ \Delta\xi^\beta) \approx f_\alpha(\xi^\beta) + \ds\pa{f_\alpha}{\xi^\beta}\,\Delta\xi^\beta~.
\eqe
With this,  the increment $\Delta\xi^\beta$ for the iterative procedure is determined from setting $f_\alpha(\xi^\beta+ \Delta\xi^\beta)=0$, giving
\eqb{l}
\Delta\xi^\beta = c^{\alpha\beta}\,f_\alpha(\xi^\beta)~,
\eqe
where $c^{\alpha\beta}$ are the components of the matrix 
\eqb{l}
 [c^{\alpha\beta}] = \ds \left[\pa{f_\alpha}{\xi^\beta}\right]^{-1}~.
\eqe
Here, following from Eq.~\eqref{e:phiTa}, we have
\eqb{l}
\ds\pa{f_\alpha}{\xi^\beta} = -\bc_\alpha\cdot\ba_\beta +  (\bg  - \bg^\mathrm{max}_{\tau})\cdot \ba_{\alpha,\beta}  - \bd_\alpha^\gamma\cdot\ba_{\gamma,\beta}~,
\eqe
where we have denoted
\eqb{lll}
\bc_\alpha \dis \ba_\alpha - \mu\,\sign(g_\mathrm{n})\tau_\alpha\,\bn~,\\[2mm]
\bd_\alpha^\beta \dis \ds \mu\,\frac{\norm{\bg_{\mrn}}}{\norm{\hat{\bg}^n_{\tau}}}(\delta_\alpha^\beta - \tau_\alpha^\beta)\,\hat{\bg}_{\mrn}^n - \mu\,\sign(g_\mathrm{n})\,\tau_\alpha\,\bn\,g^\beta~,
\eqe
with $g_\mathrm{n}:=\bg_\mathrm{e}\cdot\bn$, $g^\alpha:=\bg_\mathrm{e}\cdot\ba^\alpha$, $\tau_\alpha:=\btau\cdot\ba_\alpha$, $\tau_{\alpha\beta}:=\tau_\alpha\,\tau_\beta$, and $\tau_{\alpha}^{\beta}:=\tau_{\alpha\gamma}\,a^{\gamma\beta}$.

\section{Tangent matrices} \label{s:globalNewton}
The tangent matrices for the full-pass algorithm follow from the linearization of Eq.~\eqref{e:vPics}. In general, we have
\eqb{l}
\Delta\delta\Pi_\mrc = \ds\int_{\partial\sB_{0k}}  (\Delta\bT\cdot\delta \hat{\bg} +\bT\cdot \Delta\delta \hat{\bg} )  \,\dif A~,
\label{e:LinvPics}
\eqe
which includes both sticking and sliding.  However,  when sticking occurs,  $\hat{\bg}$ becomes $\hat{\bg}{^n}$ since  $\omega=0$ in Eq.~\eqref{e:interacgap}. Eq.~\eqref{e:LinvPics} then reduces to
\eqb{l}
\Delta\delta\Pi_\mrc = \ds  \delta\mx_e^\mrT\, \mk_{kk} \,\Delta\mx_e  +  \delta\mx_e^\mrT\, \mk_{k\bar{\ell}} \,\Delta\mx_{\bar{e}}  + \delta\mx_{\bar{e}}^\mrT\, \mk_{\bar{\ell} k} \,\Delta\mx_{e} + \delta\mx_{\bar{e}}^\mrT\, \mk_{\bar{\ell}\bar{\ell}} \,\Delta\mx_{\bar{e}}~,
\label{e:LinvPicsStick}
\eqe
where  $\bar{e}\in\sE_\ell$ denotes the master elements  that contain the previous interacting point $\hat{\bxi}{^n}$, and 
\eqb{lll}
\mk_{kk} \dis \ds  \int_{\partial\sB_{0k}}  \mN_e^T\,\epsilon\, \mN_e  \,\dif A~,\\[3mm]
\mk_{k\bar{\ell}} \dis \ds  -\int_{\partial\sB_{0k}}  \mN_e^T\,\epsilon\, \mN_{\bar{e}}(\hat{\bxi}{^n})  \,\dif A~,\\[3mm]
\mk_{\bar{\ell} k} \dis \ds  -\int_{\partial\sB_{0k}}  \mN_{\bar{e}}^T(\hat{\bxi}{^n}) \,\epsilon\, \mN_{{e}}  \,\dif A~,\\[3mm]
\mk_{\bar{\ell} \bar{\ell}} \dis \ds  \int_{\partial\sB_{0k}}  \mN_{\bar{e}}^T(\hat{\bxi}{^n}) \,\epsilon\, \mN_{\bar{e}}(\hat{\bxi}{^n})   \,\dif A~,
\label{e:tangentStick}
\eqe
denote the tangent matrices. When sliding occurs, i.e.~$\omega=1$, Eq.~\eqref{e:LinvPics} becomes
\eqb{llll}
\Delta\delta\Pi_\mrc \is \ds  \delta\mx_e^\mrT\, \mk_{kk} \,\Delta\mx_e  + \delta\mx_e^\mrT\, \mk_{k\hat{\ell}} \,\Delta\mx_{\hat{e}}  +  \delta\mx_e^\mrT\, \mk_{k\bar{\ell}} \,\Delta\mx_{\bar{e}}\\[2mm]
\plus \delta\mx_{\hat{e}}^\mrT\, \mk_{\hat{\ell} k} \,\Delta\mx_{e} + \delta\mx_{\hat{e}}^\mrT\, \mk_{\hat{\ell}\hat{\ell}} \,\Delta\mx_{\hat{e}} + \delta\mx_{\hat{e}}^\mrT\, \mk_{\hat{\ell}\bar{\ell}} \,\Delta\mx_{\bar{e}}~,
\label{e:LinvPicsSlip}
\eqe
where $\hat{e}\in\sE_\ell$ denotes the master elements  that contain the current interacting point $\hat{\bxi}{^{n+1}}$, and the tangent matrices are defined by
\eqb{llll}
\mk_{kk} \dis \ds  \int_{\partial\sB_{0k}}  \mN_e^T\,\epsilon\, (\mN_e - \ba_\alpha\,\mM^\alpha_e) \,\dif A~,\\[3mm]
\mk_{k\hat{\ell}} \dis \ds -\int_{\partial\sB_{0k}}  \mN_e^T\,\epsilon\, (\mN_{\hat{e}} + \ba_\alpha\,\mM^\alpha_{\hat{e}}) \,\dif A~,\\[3mm]
\mk_{k\bar{\ell}} \dis \ds -\int_{\partial\sB_{0k}}  \mN_e^T\,\epsilon\,  \ba_\alpha\,\mM^\alpha_{\bar{e}} \,\dif A~,\\[3mm]
\mk_{\hat{\ell} k} \dis \ds -\int_{\partial\sB_{0k}} \left[ \mN_{\hat{e}}^T\,\epsilon\, (\mN_e - \ba_\alpha\,\mM^\alpha_e)  - \mN^\mrT_{\hat{e},\alpha}\,\bT\,\mM^\alpha_e     \right]\,\dif A~,\\[3mm]
\mk_{\hat{\ell} \hat{l}} \dis \ds \int_{\partial\sB_{0k}} \left[ \mN_{\hat{e}}^T\,\epsilon\, (\mN_{\hat{e}} + \ba_\alpha\,\mM^\alpha_{\hat{e}} )  -   \mN^\mrT_{\hat{e},\alpha}\,\bT\,\mM^\alpha_{\hat{e}}      \right]\,\dif A~,\\[3mm]
\mk_{\hat{\ell}\bar{\ell}} \dis \ds \int_{\partial\sB_{0k}}\left[  \mN_{\hat{e}}^T\,\epsilon\,  \ba_\alpha\,\mM^\alpha_{\bar{e}} - \mN^\mrT_{\hat{e},\alpha}\,\bT\,\mM^\alpha_{\bar{e}}      \right] \,\dif A~,
\label{e:tangentSlip}
\eqe
with
\eqb{llll}
\mM^\alpha_e \dis  \ds \pa{\xi^\alpha}{\mx_e} = -c^{\alpha\beta}\,(\bc_\beta - \bm_\beta)\cdot\mN_e,\\[3mm]
\mM^\alpha_{\hat{e}} \dis  \ds \pa{\xi^\alpha}{\mx_{\hat{e}}} = -c^{\alpha\beta}\, \Big[  (\bg  - \bg^\mathrm{max}_{\tau})\cdot \mN_{\hat{e},\beta} - \bc_\beta   \cdot\mN_{\hat{e}} - \bd_\beta^\gamma\cdot\mN_{\hat{e},\gamma} \Big]\\[3mm]
\mM^\alpha_{\bar{e}} \dis  \ds \pa{\xi^\alpha}{\mx_{\bar{e}}} = -c^{\alpha\beta}\, \bm_\beta\cdot\mN_{\bar{e}},
\eqe
where 
\eqb{l}
\bm_\alpha := \ds \mu\,\frac{\norm{\bg_{\mrn}}}{\norm{\hat{\bg}^n_{\tau}}}   (\ba_\alpha - \tau_{\alpha\beta}\,\ba^\beta)~.
\eqe
For the two-half-pass algorithm, all the tangent matrices associated with the variation of the master surface, i.e.~$\mk_{\bar{\ell}k}$ and $\mk_{\bar{\ell}\bar{\ell}}$ in Eq.~\eqref{e:LinvPicsStick}; $\mk_{\hat{\ell}k}$,  $\mk_{\hat{\ell}\hat{\ell}}$, and  $\mk_{\hat{\ell}\bar{\ell}}$ in Eq.~\eqref{e:LinvPicsSlip}, are not needed.

\vspace{1cm}
{\Large{\bf Acknowledgements}}

The authors are grateful to the German Research Foundation (DFG)
for supporting this research under grants  GSC 111 and SA1822/8-1. 

\bigskip
\bibliographystyle{apalike}
\bibliography{sauerduong,bibliography}

\begin{thebibliography}{}

\bibitem[Argento et~al., 1997]{argento97}
Argento, C., Jagota, A., and Carter, W.~C. (1997).
\newblock Surface formulation for molecular interactions of macroscopic bodies.
\newblock {\em J. Mech. Phys. Solids}, {\bf 45}(7):1161--1183.

\bibitem[Borden et~al., 2011]{borden11}
Borden, M.~J., Scott, M.~A., Evans, J.~A., and Hughes, T. J.~R. (2011).
\newblock Isogeometric finite element data structures based on bezier
  extraction of {NURBS}.
\newblock {\em Int. J. Numer. Meth. Engng.}, {\bf {87}}:15--47.

\bibitem[Brivadis et~al., 2015]{Buffa15}
Brivadis, E., Buffa, A., Wohlmuth, B., and Wunderlich, L. (2015).
\newblock Isogeometric mortar methods.
\newblock {\em Comput. Methods Appl. Mech. Engrg.}, {\bf 284~}(Supplement
  C):292 -- 319.

\bibitem[Corbett and Sauer, 2014]{nece}
Corbett, C.~J. and Sauer, R.~A. (2014).
\newblock {NURBS}-enriched contact finite elements.
\newblock {\em Comput. Methods Appl. Mech. Engrg.}, {\bf 275}:55--75.

\bibitem[Corbett and Sauer, 2015]{nece2}
Corbett, C.~J. and Sauer, R.~A. (2015).
\newblock Three-dimensional isogeometrically enriched finite elements for
  mixed-mode contact and debonding.
\newblock {\em Comput. Methods Appl. Mech. Engrg.}, {\bf 284}:781--806.

\bibitem[{De~Lorenzis} et~al., 2014]{Laura2014}
{De~Lorenzis}, L., , Wriggers, P., and Hughes, T. J.~R. (2014).
\newblock Isogeometric contact: {A} review.
\newblock {\em GAMM Mitteilungen}, {\bf {37}}:85--123.

\bibitem[{De~Lorenzis} et~al., 2011]{Laura2011}
{De~Lorenzis}, L., Temizer, I., Wriggers, P., and Zavarise, G. (2011).
\newblock A large deformation frictional contact formulation using
  {NURBS}-based isogeometric analysis.
\newblock {\em Int. J. Numer. Meth. Engrg.}, {\bf{87}}:1278--1300.

\bibitem[{De~Lorenzis} et~al., 2012]{Lorenzis2012}
{De~Lorenzis}, L., Wriggers, P., and Zavarise, G. (2012).
\newblock A mortar formulation for {3D} large deformation contact using
  {NURBS}-based isogeometric analysis and the augmented {L}agrangian method.
\newblock {\em Comput. Mech.}, {\bf{49}}:1--20.

\bibitem[{Del~Piero} and Raous, 2010]{delpiero10}
{Del~Piero}, G. and Raous, M. (2010).
\newblock A unified model for adhesive interfaces with damage, viscosity, and
  friction.
\newblock {\em Eur. J. Mech. A-Solid}, {\bf 29}:496--507.

\bibitem[Dimitri and Zavarise, 2017]{dimitri2017}
Dimitri, R. and Zavarise, G. (2017).
\newblock Isogeometric treatment of frictional contact and mixed mode debonding
  problems.
\newblock {\em Comput. Mech.}, {\bf{60}}(2):315--332.

\bibitem[Dittmann et~al., 2014]{Dittmann14}
Dittmann, M., Franke, M., Temizer, I., and Hesch, C. (2014).
\newblock Isogeometric analysis and thermomechanical mortar contact problems.
\newblock {\em Comp. Meth. Appl. Mech. Engrg.}, {\bf{274}}:192--212.

\bibitem[Duong et~al., 2018]{Duong2018}
Duong, T.~X., De~Lorenzis, L., and Sauer, R.~A. (2018).
\newblock A segmentation-free isogeometric extended mortar contact method.
\newblock {\em Comput. Mech.}, DOI: 10.1007/s00466-018-1599-0.

\bibitem[Fischer and Wriggers, 2006]{fischer06}
Fischer, K.~A. and Wriggers, P. (2006).
\newblock Mortar based frictional contact formulation for higher order
  interpolations using the moving friction cone.
\newblock {\em Comput. Methods Appl. Mech. Engrg.}, {\bf 195}:5020--5036.

\bibitem[Gitterle et~al., 2010]{Gitterle10}
Gitterle, M., Popp, A., Gee, M.~W., and Wall, W.~A. (2010).
\newblock Finite deformation frictional mortar contact using a semi-smooth
  newton method with consistent linearization.
\newblock {\em Int. J. Numer. Meth. Engrg.}, {\bf{84}}(5):543--571.

\bibitem[Hiermeier et~al., 2018]{Hiermeier2018}
Hiermeier, M., Wall, W.~A., and Popp, A. (2018).
\newblock A truly variationally consistent and symmetric mortar-based contact
  formulation for finite deformation solid mechanics.
\newblock {\em Comp. Meth. Appl. Mech. Engrg.}, DOI: 10.1016/j.cma.2018.07.020.

\bibitem[Hughes et~al., 2005]{hughes05}
Hughes, T. J.~R., Cottrell, J.~A., and Bazilevs, Y. (2005).
\newblock Isogeometric analysis: {CAD}, finite elements, {NURBS}, exact
  geometry and mesh refinement.
\newblock {\em Comp. Meth. Appl. Mech. Engrg.}, {\bf{194}}:4135--4195.

\bibitem[Khi{\^{e}}m and Itskov, 2017]{Khiem2017}
Khi{\^{e}}m, V.~N. and Itskov, M. (2017).
\newblock {An averaging based tube model for deformation induced anisotropic
  stress softening of filled elastomers}.
\newblock {\em International Journal of Plasticity}, 90:96--115.

\bibitem[Kiliç and Temizer, 2016]{kilic2016}
Kiliç, K. and Temizer, I. (2016).
\newblock Tuning macroscopic sliding friction at soft contact interfaces:
  {Interaction} of bulk and surface heterogeneities.
\newblock {\em Tribol. Int.}, {\bf{104}}:83--97.

\bibitem[Kim and Youn, 2012]{Kim12}
Kim, J.-Y. and Youn, S.-K. (2012).
\newblock Isogeometric contact analysis using mortar method.
\newblock {\em Int. J. Numer. Meth. Engrg.}, {\bf 89}(12):1559--1581.

\bibitem[Krstulovic-Opara et~al., 2002]{krstulovic02}
Krstulovic-Opara, L., Wriggers, P., and Korelc, J. (2002).
\newblock A {$C^1$}-continuous formulation for {3D} finite deformation friction
  contact.
\newblock {\em Comp. Mech.}, {\bf 29}:27--42.

\bibitem[Laursen, 2002]{laursen}
Laursen, T.~A. (2002).
\newblock {\em Computational Contact and Impact Mechanics: Fundamentals of
  modeling interfacial phenomena in nonlinear finite element analysis}.
\newblock Springer-Verlag Berlin Heidelberg.

\bibitem[Laursen and Simo, 1993]{laursen93}
Laursen, T.~A. and Simo, J.~C. (1993).
\newblock A continuum-based finite element formulation for the implicit
  solution of multibody, large deformation frictional contact problems.
\newblock {\em Int. J. Numer. Meth. Engng.}, {\bf 36}:3451--3485.

\bibitem[Lu, 2011]{Lu2011}
Lu, J. (2011).
\newblock Isogeometric contact analysis: Geometric basis and formulation for
  frictionless contact.
\newblock {\em Comp. Meth. Appl. Mech. Engrg.}, {\bf{200}}:726--741.

\bibitem[Mergel et~al., 2018]{adhesionfric}
Mergel, J.~C., Sahli, R., Scheibert, J., and Sauer, R.~A. (2018).
\newblock Continuum contact models for coupled adhesion and friction.
\newblock {\em The Journal of Adhesion}, {\bf 94}:1--33.

\bibitem[Neto et~al., 2016]{neto16}
Neto, D., Oliveira, M., Menezes, L., and Alves, J. (2016).
\newblock A contact smoothing method for arbitrary surface meshes using
  {Nagata} patches.
\newblock {\em Comp. Meth. Appl. Mech. Engrg.}, {\bf{299}}:283 -- 315.

\bibitem[Ogden, 1987]{ogden}
Ogden, R.~W. (1987).
\newblock {\em Non-Linear Elastic Deformations}.
\newblock Dover Edition, Mineola.

\bibitem[Persson, 2000]{persson}
Persson, B. N.~J. (2000).
\newblock {\em Sliding friction: {P}hysical principles and application}.
\newblock Springer-Verlag Berlin Heidelberg, 2$^{\text{nd}}$ edition.

\bibitem[Popp et~al., 2012]{Apop2012}
Popp, A., Wohlmuth, B.~I., Gee, M.~W., and Wall, W.~A. (2012).
\newblock Dual quadratic mortar finite element methods for {3D} finite
  deformation contact.
\newblock {\em SIAM J. Sci. Comput.}, {\bf{34}}:B421--B446.

\bibitem[Puso and Laursen, 2004]{puso04a}
Puso, M.~A. and Laursen, T.~A. (2004).
\newblock A mortar segment-to-segment contact method for large deformation
  solid mechanics.
\newblock {\em Comput. Methods Appl. Mech. Engrg.}, {\bf 193}:601--629.

\bibitem[Raous et~al., 1999]{raous99}
Raous, M., Cang\'{e}mi, L., and Cocu, M. (1999).
\newblock A consistent model coupling adhesion, friction, and unilateral
  contact.
\newblock {\em Comput. Methods Appl. Mech. Engrg.}, {\bf 177}:383--399.

\bibitem[Sauer, 2006]{sauer-phd}
Sauer, R.~A. (2006).
\newblock {\em An atomic interaction based continuum model for computational
  multiscale contact mechanics}.
\newblock PhD thesis, University of California, Berkeley, USA.

\bibitem[Sauer, 2011]{sauer10b}
Sauer, R.~A. (2011).
\newblock Enriched contact finite elements for stable peeling computations.
\newblock {\em Int. J. Numer. Meth. Engrg.}, {\bf 87}:593--616.

\bibitem[Sauer, 2013]{sauer-ece2}
Sauer, R.~A. (2013).
\newblock Local finite element enrichment strategies for {2D} contact
  computations and a corresponding postprocessing scheme.
\newblock {\em Comput. Mech.}, {\bf 52}(2):301--319.

\bibitem[Sauer and {De~Lorenzis}, 2013]{spbc}
Sauer, R.~A. and {De~Lorenzis}, L. (2013).
\newblock A computational contact formulation based on surface potentials.
\newblock {\em Comput. Methods Appl. Mech. Engrg.}, {\bf 253}:369--395.

\bibitem[Sauer and {De~Lorenzis}, 2015]{spbf}
Sauer, R.~A. and {De~Lorenzis}, L. (2015).
\newblock An unbiased computational contact formulation for {3D} friction.
\newblock {\em Int. J. Numer. Meth. Engrg.}, {\bf{101}}:251--280.

\bibitem[Sauer et~al., 2014]{membrane}
Sauer, R.~A., Duong, T.~X., and Corbett, C.~J. (2014).
\newblock A computational formulation for constrained solid and liquid
  membranes considering isogeometric finite elements.
\newblock {\em Comput. Methods Appl. Mech. Engrg.}, {\bf 271}:48--68.

\bibitem[Sauer and Li, 2007]{sauer07b}
Sauer, R.~A. and Li, S. (2007).
\newblock An atomic interaction-based continuum model for adhesive contact
  mechanics.
\newblock {\em Finite Elem. Anal. Des.}, {\bf 43}(5):384--396.

\bibitem[Sauer and Li, 2008]{sauer08b}
Sauer, R.~A. and Li, S. (2008).
\newblock An atomistically enriched continuum model for nanoscale contact
  mechanics and its application to contact scaling.
\newblock {\em J. Nanosci. Nanotech.}, {\bf 8}(7):3757--3773.

\bibitem[Seitz et~al., 2016]{Apop2016}
Seitz, A., Farah, P., Kremheller, J., Wohlmuth, B.~I., Wall, W.~A., and Popp,
  A. (2016).
\newblock Isogeometric dual mortar methods for computational contact mechanics.
\newblock {\em Comput. Methods Appl. Mech. Engrg.}, {\bf{301}}:259--280.

\bibitem[Shadowitz, 1988]{shadowitz}
Shadowitz, A. (1988).
\newblock {\em The electromagnetic field}.
\newblock Dover Publications, New York.

\bibitem[Simo and Ju, 1987]{Simo1987}
Simo, J. and Ju, J. (1987).
\newblock Strain- and stress-based continuum damage models—i. formulation.
\newblock {\em International Journal of Solids and Structures}, 23(7):821 --
  840.

\bibitem[Temizer, 2013]{Temizer2013}
Temizer, I. (2013).
\newblock A mixed formulation of mortar-based contact with friction.
\newblock {\em Comput. Methods Appl. Mech. Engrg.}, {\bf{255}}:183--195.

\bibitem[Temizer, 2016]{Temizer2016}
Temizer, I. (2016).
\newblock Sliding friction across the scales: {Thermomechanical} interactions
  and dissipation partitioning.
\newblock {\em J. Mech. Phys. Solids}, {\bf{89}}:126--148.

\bibitem[Temizer et~al., 2011]{temizer2011}
Temizer, I., Wriggers, P., and Hughes, T. (2011).
\newblock Contact treatment in isogeometric analysis with {NURBS}.
\newblock {\em Comput. Methods Appl. Mech. Engrg.}, {\bf{200}}:1100--1112.

\bibitem[Temizer et~al., 2012]{temizer12}
Temizer, I., Wriggers, P., and Hughes, T. J.~R. (2012).
\newblock Three-dimensional mortar-based frictional contact treatment in
  isogeometric analysis with {NURBS}.
\newblock {\em Comput. Methods Appl. Mech. Engrg.}, {\bf 209-212}:115--128.

\bibitem[Weeger et~al., 2018]{weeger_2018}
Weeger, O., Narayanan, B., and Dunn, M.~L. (2018).
\newblock Isogeometric collocation for nonlinear dynamic analysis of {Cosserat}
  rods with frictional contact.
\newblock {\em Nonlinear Dyn.}, {\bf{91}}(2):1213--1227.

\bibitem[Wriggers, 2006]{wriggers-contact}
Wriggers, P. (2006).
\newblock {\em Computational Contact Mechanics}.
\newblock Springer-Verlag Berlin Heidelberg, 2$^{\text{nd}}$ edition.

\bibitem[Wriggers and Haraldsson, 2003]{Wriggers2003}
Wriggers, P. and Haraldsson, A. (2003).
\newblock A simple formulation for two-dimensional contact problems using a
  moving friction cone.
\newblock {\em Comm. Num. Meth. Engrg.}, {\bf{19}}:285--295.

\bibitem[Wriggers and Krstulovic-Opara, 2004]{Wriggers04}
Wriggers, P. and Krstulovic-Opara, L. (2004).
\newblock The moving friction cone approach for three-dimensional contact
  simulations.
\newblock {\em Int. J. Comput. Methods}, {\bf{01}}(01):105--119.

\bibitem[Yang et~al., 2005]{yang05}
Yang, B., Laursen, T.~A., and Meng, X. (2005).
\newblock Two dimensional mortar contact methods for large deformation
  frictional sliding.
\newblock {\em Int. J. Numer. Meth. Engng}, {\bf 62}:1183--1225.

\end{thebibliography}

\end{document}